\documentclass[twocolumn,nofootinbib,superscriptaddress,preprintnumbers,floatfix
]{revtex4}

\pdfoutput=1
\usepackage{graphicx}
\usepackage{amsmath}
\usepackage{amssymb}
\usepackage{color}
\usepackage{xspace}
\usepackage[small]{subfigure}
\usepackage[hyperfootnotes=false]{hyperref}
\usepackage{multirow}

\newcommand{\eq}[1]{Eq.~\eqref{eq:#1}}
\newcommand{\eqs}[2]{Eqs.~\eqref{eq:#1} and \eqref{eq:#2}}
\renewcommand{\sec}[1]{Sec.~\ref{sec:#1}}

\newcommand{\OMIT}[1]{}

\newcommand{\df}{\mathrm{d}}

\newcommand{\beq}{\begin{equation}}
\newcommand{\eeq}{\end{equation}}
\newcommand{\beqa}{\begin{eqnarray}}
\newcommand{\eeqa}{\end{eqnarray}}

\newcommand{\nn}{\nonumber}

\newcommand{\msbar}{\overline{\textrm{MS}}}

\newcommand{\ntll}{\mbox{N${}^3$LL}\xspace}
\newcommand{\muj}{\mu_J}
\newcommand{\muh}{\mu_H}
\newcommand{\mus}{\mu_S}

  \newcount\hour \newcount\minute
  \hour=\time \divide \hour by 60
  \minute=\time
  \newcommand{\mydate}{\ \today \ - \number\hour :\ifnum \minute<10 0\fi 
\number\minute}

\allowdisplaybreaks[4]

\begin{document}

\preprint{\vbox{
\hbox{MIT--CTP
4343}\hbox{IFIC/12-02}\hbox{UWThPh-2012-9}\hbox{LPN12-023}\hbox{INT-PUB-12-006}
}}

\title{\boldmath 
Precision Thrust Cumulant Moments at N${}^3$LL
\vspace{0.1cm}
}

\author{Riccardo Abbate}
\affiliation{Center for Theoretical Physics, Massachusetts Institute of
Technology, Cambridge, MA 02139\vspace{0.cm}}

\author{Michael Fickinger}
\affiliation{Department of Physics, University of Arizona, Tucson, AZ 85721 
}

\author{Andr\'e H. Hoang}
\affiliation{University of Vienna, Faculty of Physics, Boltzmanngasse 5,
1090 Vienna, Austria\vspace{0.cm}}

\author{Vicent Mateu} 
\affiliation{Center for Theoretical Physics, Massachusetts Institute of
Technology, Cambridge, MA 02139\vspace{0.cm}}
\affiliation{IFIC, UVEG - CSIC, Apartado de Correos 22085, E-46071, 
        Valencia, Spain\vspace{0.cm}}

\author{Iain W.\ Stewart\vspace{0.5cm}}
\affiliation{Center for Theoretical Physics, Massachusetts Institute of
Technology, Cambridge, MA 02139\vspace{0.cm}}

\begin{abstract}
  
  We consider cumulant moments (cumulants) of the thrust distribution using
  predictions of the full spectrum for thrust including ${\cal O}(\alpha_s^3)$
  fixed order results, resummation of singular N$^3$LL logarithmic
  contributions, and a class of leading power corrections in a renormalon-free
  scheme. From a global fit to the first thrust moment we extract the strong
  coupling and the leading power correction matrix element $\Omega_1$. We obtain
  $\alpha_s(m_Z) = 0.1140 \,\pm\, (0.0004)_{\rm exp} \,\pm\, (0.0013)_{\rm hadr}
  \,\pm \, (0.0007)_{\rm pert}$, where the $1$-$\sigma$ uncertainties are
  experimental, from hadronization (related to $\Omega_1$) and perturbative,
  respectively, and $\Omega_1=0.377 \,\pm\, (0.044)_{\rm exp} \,\pm\,
  (0.039)_{\rm pert}\,{\rm GeV}$.  The $n$-th thrust cumulants for $n\ge 2$ are
  completely insensitive to $\Omega_1$, and therefore a good instrument for
  extracting information on higher order power corrections,
  $\Omega_n^\prime/Q^n$, from moment data. We find
  $(\tilde\Omega_2^\prime)^{1/2} = 0.74 \,\pm\, (0.11)_{\rm exp} \,\pm\,
  (0.09)_{\rm pert}\,{\rm GeV}$.

\end{abstract}

\maketitle

\section{Introduction}
\label{sec:intro}

The process $e^+e^-\to {\rm jets}$ plays an important role in precise
determinations of $\alpha_s(m_Z)$, as well as for probing the nonperturbative
dynamics of hadronization in jet production. A wealth of high precision data
with percent level uncertainties, is available for jet production in $e^+e^-$
collisions at the Z-pole, $Q=m_Z$, and with somewhat larger uncertainties at
both lower and higher energies $Q$. For a review of classic work on
$\alpha_s(m_Z)$ determinations using event shapes and other jet observables, the
reader is referred to~\cite{Kluth:2006bw}. Accurate predictions
for event shapes are now available which include ${\cal O}(\alpha_s^3)$
corrections \cite{GehrmannDeRidder:2007bj,
  GehrmannDeRidder:2007hr,Weinzierl:2008iv,Weinzierl:2009ms}, a
next-to-next-to-next-to-leading-log (N$^3$LL) resummation of large
logarithms~\cite{Becher:2008cf,Chien:2010kc}, and a high precision method
developed for simultaneously incorporating field theory matrix elements for the
power corrections~\cite{Abbate:2010xh}.

The majority of fits for $\alpha_s(m_Z)$ from event shapes $e$ make use of 
cross section distributions $\df\sigma/\df e$, in a region where
nonperturbative effects enter as power corrections in $1/Q$ and the theoretical
description is the most accurate. In our recent analysis \cite{Abbate:2010xh}
for
the event-shape variable thrust $\tau=1-T$~\cite{Farhi:1977sg},
\begin{align}
\label{eq:Tdef}
T & \, = \,\mbox{max}_{\hat {\bf t}}\frac{\sum_i|\hat {\bf t}\cdot\vec{p}_i|}
{\sum_i|\vec{p}_i|}
\,,
\end{align}
we obtained a precise determination of $\alpha_s(m_Z)$.  Our theoretical
description is based on Soft-Collinear Effective Theory
(SCET)~\cite{Bauer:2000ew, Bauer:2000yr, Bauer:2001ct, Bauer:2001yt,
  Bauer:2002nz}, and has several advanced features, such as:
\begin{enumerate}
\item Matrix elements and nonsingular terms at order $\alpha_s^3$ using results
  from~\cite{GehrmannDeRidder:2007bj}.  Non-logarithmic terms in the hard
  function are included at order $\alpha_s^3$ as well.
\item Resummation of the singular logarithmic terms to all orders in $\alpha_s$
  up to N${}^3$LL order.
\item Profile functions ($\tau$-dependent scales $\mu_J$, $\mu_S$, $R$,
  $\mu_{\rm ns}$) that correctly treat the peak region and account for the
  multijet boundary condition to ensure that predictions converge properly
  into the known fixed order result in the multijet endpoint region. They allow
  an accurate theoretical description over the entire range $\tau \in [0,0.5]$.
\item Description of nonperturbative effects with field theory and a fit to a
  single nonperturbative matrix element of Wilson lines $\Omega_1$ in the tail
  region (where power corrections are described by an OPE). 
\item Definition of $\Omega_1$ in a more stable Rgap scheme \cite{Hoang:2007vb,Hoang:2008fs} rather than in
  $\overline{\rm MS}$.  This ensures $\Omega_1$ and the perturbative
  cross section are free of ${\cal O}(\Lambda_{\rm QCD})$ renormalon
  ambiguities.  An RGE is used to sum large logarithms in the perturbative
  renormalon subtractions \cite{Hoang:2008yj,Hoang:2009yr}.  The fit gives $\Omega_1$ with an accuracy of $16\%$.
\item QED final state corrections at ${\cal O}(\alpha)$ and NNLL (counting
$\alpha\sim  \alpha_s^2$); bottom mass corrections are included using a factorization
  theorem with log resummation; ${\cal O}(\alpha_s^2)$ axial-singlet terms
arising from the large top-bottom mass splitting are included as well.
\end{enumerate}
A two-parameter global fit in the tail of the thrust distribution
gives~\cite{Abbate:2010xh} $\alpha_s(m_Z) \, = \, 0.1135 \,\pm\, (0.0002)_{\rm
  exp} \,\pm\, (0.0005)_{\rm hadr} \,\pm \, (0.0009)_{\rm pert}$ as well as
$\Omega_1=0.323 \,\pm\, (0.009)_{\rm exp}\,\pm\,
(0.013)_{\rm \Omega_2}\pm\, (0.020)_{\rm \alpha_s(m_Z)} \,\pm \, (0.045)_{\rm
  pert}$~GeV where $\Omega_1\equiv\Omega_1(R_\Delta,\mu_\Delta)$ is defined in the Rgap scheme at the scales
$R_\Delta=\mu_\Delta=2$\,GeV.  For $\alpha_s$ the three uncertainties are the
experimental uncertainty, hadronization uncertainty coming mainly from the
determination of $\Omega_1$, and the perturbative theoretical uncertainty. This
result for $\alpha_s$ is one of the most precise in the literature. It is also
one of the lowest, being $3.9\,\sigma$ away from the 2009 world
average~\cite{Bethke:2009jm} and $4.0\,\sigma$ from the 2011 world
average~\cite{PDG:2012}.  For a detailed discussion of $\alpha_s(m_Z)$ determinations
see Ref.~\cite{Bethke:2011tr}. The small value of $\alpha_s(m_Z)$ is directly
connected to the non-negligible correction from $\Omega_1$~\cite{Abbate:2010xh},
whose fit value is of natural size $\Omega_1\sim \Lambda_{\rm QCD}$.  Given the
discrepancy, further tests of the theoretical predictions for event shapes are
warranted.  In this paper we will do so using experimental moments involving the
thrust variable.

The property of the N$^3$LL$\,+\,{\cal O}(\alpha_s^3)$ predictions for
${\rm d}\sigma/{\rm d}\tau$ in Ref.~\cite{Abbate:2010xh} that we will exploit is
that they are valid in both the dijet and tail regions, where singular and large
logarithmic terms in need of resummation arise, and in the multijet region,
where fixed order results without log resummation should be used. That is, they
are valid for all values of $\tau$ (an improvement over earlier results at this
order). Important ingredients are: the inclusion of the nonsingular terms,
important away from the peak region; the use of profile functions that turn off
resummation in the far-tail region; and the inclusion of a soft function, which
is necessary to describe the peak in the dijet region, where nonperturbative
effects are ${\cal O}(1)$.

We will use the full $\tau$ range results to analyze moments $M_n$ of the thrust
distribution in $e^+e^-\to {\rm jets}$,
\begin{align} \label{eq:Mndef}
M_n =  \dfrac{1}{\sigma} \int_0^{\tau_{\rm max}=1/2}\! {\rm d}\tau \ \tau^n\ 
\frac{{\rm d}\sigma}{{\rm d}\tau}\,.
\end{align}
Unlike for tail fits, the entire physical $\tau$ range contributes, providing
sensitivity to a different region of the spectrum. Experimental results are
available for many values of $Q$, and the analysis of systematic uncertainties
is to a large extent independent from that for the binned distributions.  Thus
the outcome for a fit of data for the first moment $M_1$ to $\alpha_s(m_Z)$ and
$\Omega_1$ serves as an important cross check of the results obtained in
Ref.~\cite{Abbate:2010xh}. The $M_n$ moments are also not sensitive to large
logarithms, and hence provide a non-trivial check on whether the
N$^3$LL$\,+\,{\cal
  O}(\alpha_s^3)$ full spectrum results, which contain a summation of logarithms
of $\tau$
with a substantial numerical effect for small $\tau$ values, can reproduce
this property. We explore this issue both for central values and for theory
uncertainty estimates. 

The second purpose of this work is to discuss the structure of higher order
power corrections in thrust moments. We find that cumulant moments $M_n^\prime$
(cumulants) are very useful, since they allow for a cleaner separation of the
subleading nonperturbative matrix elements compared to the $M_n$
moments of Eq.~(\ref{eq:Mndef}). Cumulants 
include the variance $M_2^\prime$ and skewness $M_3^\prime$, and we will
consider the first five:
\begin{align}\label{eq:variance-skewness}
M_1^\prime\,=&\,\,M_1\,, \\
M_2^\prime\,=&\,\,M_2\,-\,M_1^2\,, \nonumber \\
M_3^\prime\,=&\,\,M_3\,-\,3\,M_2\,M_1\,+\,2\,M_1^3\,,\nonumber\\
M_4^\prime\,=&\,\,M_4-4\,M_3\,M_1-3\,M_2^2+12\,M_2\,M_1^2-6\,M_1^4\,,
\nonumber\\
M_5^\prime\,=&\,\,M_5-5\,M_4\,M_1-10\,M_3\,M_2+20\,M_3\,M_1^2\nonumber\\
&+30\,M_2^2M_1-60\,M_1^3\,M_2+24\,M_1^5\,.\nonumber
\end{align}
In the leading order thrust factorization theorem the power correction matrix
elements for the moments $M_n$ are called $\Omega_m$ while for the cumulants
$M^\prime_n$ they are called $\Omega^\prime_m$. (\,The $\Omega^\prime_m$ are
also related to the $\Omega_m$ by \eq{variance-skewness} with $M_n\to
\Omega_n$.\,) In particular, the invariance of the cumulants to shifts in $\tau$
implies that the $M_{n\ge 2}^\prime$ moments are completely insensitive to the
leading thrust power correction parameter $\Omega_1$, and hence can provide
non-trivial information on the higher order power corrections which enter as
$\Omega_n^\prime/Q^n$ and as $1/Q^2$ power corrections from terms beyond the
leading factorization theorem. In contrast, for each $M_{n\ge 2}$ there is a
term $\sim \alpha_s\Omega_1/Q$ that for larger $Q$s dominates over the
$\Omega_m/Q^m$ terms.\footnote{The cumulants begin to differ for $n\ge 4$
  from the so-called central moments, $\langle (\tau-M_1)^n\rangle$. Both
  cumulants and central moments are shift independent, but the cumulants are
  slightly preferred because they are only sensitive to a single moment of the
  leading order soft function in the thrust factorization theorem.}

\subsection{Review of Experiments and Earlier Literature}

Dedicated experimental analyses of thrust moments have been reported by various
experiments: JADE~\cite{MovillaFernandez:1997fr} measured the first moment at
$Q=35,\,44$ GeV, and in~\cite{Pahl:2008uc} reported measurements of the first
five moments at $Q=14$, $22$, $34.6$, $35$, $38.3$, $43.8$ GeV; OPAL
\cite{Abbiendi:2004qz} measured the first five moments at $Q=91$, $133$, $177$,
$197$ GeV, and there is an additional measurement of the first moment at $Q=161$
GeV \cite{Ackerstaff:1997kk}; ALEPH \cite{Heister:2003aj} measured the first
moment at $Q=91.2$, $133$, $161$, $172$, $183$, $189$, $196$, $200$, $206$ GeV;
DELPHI~\cite{Abdallah:2003xz} has measurements of the first moment at $Q=45.2$,
$66$, $76.3$ GeV, measurements of the first three moments at $Q=183$, $189$,
$192$, $196$, $200$, $202$, $205$, $207$ GeV~\cite{Abdallah:2004xe}, and at
$Q=91.2$, $133$, $161$, $172$, $183$ GeV~\cite{Abreu:1999rc}; L3
\cite{Acciarri:2000hm} measured the first two moments at $Q=91.2$ GeV and other
center of mass energies which are superseded by the ones in \cite{Achard:2004sv}
at $Q=41.4$, $55.3$, $65.4$, $75.7$, $82.3$, $85.1$, $130.1$, $136.1$, $161.3$,
$172.3$, $182.8$, $188.6$, $194.4$, $200.2$, $206.2$ GeV; TASSO measured the
first moment at $Q=14$, $22$, $35$, $44$ GeV~\cite{Braunschweig:1990yd}; and AMY
measured the first moment at $Q=55.2$ GeV~\cite{Li:1989sn}. Finally, the
variance and skewness have been explicitly measured by DELPHI
\cite{Abreu:1999rc} at $Q=133$, $161$, $172$, $183$ GeV; and OPAL
\cite{Ackerstaff:1997kk} at $Q=161$ GeV.  All of the experimental moments will
be used in our fits, with the exception of the results in
Ref.~\cite{Pahl:2008uc} and data with $Q\le 22\,{\rm GeV}$ where our
treatment of $b$-quark mass effects may not suffice.

In principle the JADE results in Ref.~\cite{Pahl:2008uc} supersede the earlier
analysis of this data reported in Ref.~\cite{MovillaFernandez:1997fr}. In the
more recent analysis the contribution of primary $b\bar{b}$ events has been
subtracted using Monte Carlo generators.\footnote{We thank C.~Pahl for
  clarifying precisely how this was done.} Since the theoretical precision of
these generators is significantly worse than our N$^3$LL$\,+\,{\cal
  O}(\alpha_s^3)$ treatment of massless quark effects and our NNLL\,+\,${\cal
  O}(\alpha_s)$ treatment of $m_b$-dependent corrections, it is not clear how
our code should be modified consistently to account for these subtractions.
Comparing the old versus new JADE data at $Q=44\,{\rm GeV}$ one finds
$M_1=0.0860 \pm 0.0014$ versus $M_1=0.0807\pm 0.0016$. This corresponds to a
$3.4\,\sigma$ change assuming 100\% correlated uncertainties (or a $2.6\,\sigma$
change with uncorrelated uncertainties). In our analysis we find that the older
JADE data provides more consistent results when employed in a combined fit with
data from the other experiments (related to smaller $\chi^2$ values). For this
reason our default dataset incorporates only the older JADE moment data. We will
report on the change that would be induced by using the new JADE data if we
simply ignore the fact that the $b\bar b$ events were removed.

Event shape moments have also been extensively studied in the theoretical
literature. The ${\cal O}(\alpha_s^3)$ QCD corrections for event shape moments
have been calculated in Ref.~\cite{GehrmannDeRidder:2009dp,Weinzierl:2009yz}.
The leading $\Lambda/Q$ power correction to the first moment of event shape
distributions were first studied
in~\cite{Dokshitzer:1995zt,Akhoury:1995sp,Akhoury:1995fb,Nason:1995hd} often
with the study of renormalons (see~\cite{Korchemsky:1994is}, and
\cite{Beneke:1998ui} for a review).  Ref.~\cite{Gardi:2000yh} made a renormalon
analysis of the second moment of the thrust distribution, finding that the leading
renormalon contribution is not $1/Q^2$ but rather $1/Q^3$. Hadronization effects have also been
frequently considered in the framework of the dispersive model for the strong
coupling~\cite{Dokshitzer:1995zt,Dokshitzer:1995qm,Dokshitzer:1998pt}\,\footnote{Another
approach to hadronization corrections to moments of event shapes distributions
based on renormalons is that of Gardi and Grunberg \cite{Gardi:1999dq}.}. In this
approach an IR cutoff $\mu_I$ is introduced and the strong coupling constant
below the scale $\mu_I$ is replaced by an effective coupling $\alpha_{\rm eff}$
such that perturbative infrared effects coming from scales below $\mu_I$ are
subtracted. In the dispersive model the term $\mu_I\alpha_0$ is the analog of
the QCD matrix element $\Omega_1$ that is derived from the operator product
expansion (OPE).  Since in the dispersive model there is only one
nonperturbative parameter, it does not contain analogs of the independent
nonperturbative QCD matrix elements $\Omega_{n\ge 2}$ of the operator product
expansion. Thus measurements of $\Omega^\prime_{n\ge 2}$ can be used as a test
for additional nonperturbative physics that go beyond this framework.

The dispersive model has been used in Refs.~\cite{Biebel:2001dm,
  Abbiendi:2004qz, Pahl:2009aa} together with ${\cal O}(\alpha_s^2)$ fixed order
results to analyze event shape moments, fitting simultaneously to
$\alpha_s(m_Z)$ and $\alpha_0$. Recently these analyses have been extended to
${\cal O}(\alpha_s^3)$ in Ref.~\cite{Gehrmann:2009eh}, based on code for $n_f=5$
massless quark flavors, using data from \cite{Abbiendi:2004qz,Pahl:2008uc} and
fitting to the first five moments for several event-shape variables. Our
numerical analysis only considers thrust moments, but with a global dataset from all
available experiments. A detailed comparison with Ref.~\cite{Gehrmann:2009eh}
will be made at appropriate points in the paper.  Theoretically our analysis
goes beyond their work by using a formalism that has no large logarithms in the
renormalon subtraction, includes the analog of the ``Milan
factor''~\cite{Dokshitzer:1997iz,Dokshitzer:1998pt} in our framework at ${\cal
  O}(\alpha_s^3)$ (one higher order than~\cite{Gehrmann:2009eh}), and
incorporates higher order power corrections beyond the leading shift from
$\Omega_1$. We also test the effect of including resummation.

\subsection{Outline}

This article is organized as follows: We start out by defining moments and
cumulants of distributions, and their respective generating functions in
Sec.~\ref{sec:bsg}, where we also discuss the leading and subleading power
corrections of thrust moments in an OPE framework. In Sec.~\ref{sec:results} we
present and discuss our main results for $\alpha_s(m_Z)$ from fits to the first
thrust moment $M_1$.  In Sec.~\ref{sec:comparison} we analyze higher moments
$M_{n\ge 2}$.  Sec.~\ref{sec:power-data} contains an analysis of subleading
power corrections from fits to cumulants $M_{n\ge 2}^\prime$ obtained from the moment data.
Our conclusions are presented in Sec.~\ref{eq:conclusions}.

\section{Formalism}
\label{sec:bsg}

\subsection{Various Moments of a Distribution}

The moments of a probability distribution function $p(k)$ are given by
\begin{align}
  M_n=\langle k^n \rangle=\int\! {\rm d}k\: p(k)\,k^n.
\end{align}
The characteristic function is the generator of these moments and is defined as
the Fourier transform 
\begin{align}
  \tilde p(y)=\langle e^{-iky} \rangle=\!\int\! {\rm
d}k\,p(k)\,e^{-iky}=\sum_{n=0}^\infty\,\frac{(-iy)^n}{n!}\,M_n,
\end{align}
with $M_0=1$. The logarithm of $\tilde p(y)$ generates the cumulants (or
connected moments) $M^\prime_n$ of the distribution
\begin{align}
  \ln\,\tilde p(y)=\sum_{n=1}^{\infty}\frac{(-iy)^n}{n!}M^\prime_n\,,
\label{eq:cumulantdef}
\end{align}
and is called the cumulant generating function.  For $n\ge 2$ the cumulants have the
property of being invariant under shifts of the distribution. Replacing $p(k)\to
p(k-k_0)$ takes $\tilde p(y) \to e^{-iy k_0}\, \tilde p(y)$, which shifts
$M_1^\prime\to M_1^\prime + k_0$ while leaving all $M_{n\ge 2}^\prime$
unchanged.  Writing
\begin{align} \label{eq:relateM}
  \sum_{N=0}^\infty\,\frac{(-iy)^N}{N!}\,M_N
&=\exp\bigg[\sum_{j=1}^{\infty}\frac{(-iy)^j}{j!}M^\prime_j\bigg]
\nonumber\\
&=\prod_{j=1}^{\infty}\sum_{R=0}^{\infty}\frac{(-iy)^{jR}}{R!}
  \bigg(\frac{M^\prime_j}{j!}\bigg)^R\,,
\end{align}
one can derive an all-$n$ relation between moments and cumulants of a distribution:
\begin{align} \label{eq:MNpartition}
 M_N=N!\sum_{i=1}^{p(N)} \prod_{j=1}^{N}
 \frac{(M_j^\prime)^{\kappa_{ij}}}{\kappa_{ij}!\,(j!)^{\kappa_{ij}} }\, .
\end{align}
Here the $\kappa_{ij}$ are non-negative integers which determine a partition of
the integer $N$ through $\sum_{j=1}^N j\, \kappa_{ij} = N$, and $p(N)$ is the
the number of unique partitions of $N$.  (\,A partition of $N$ is a set of
integers which sum to $N$. Here $\kappa_{ij}$ is the number of times the value
$j$ appears as a part in the $i$'th partition, and corresponds to $R$ in
\eq{relateM}.\,) As an example we quote the relation for $N=4$ which has five
partitions, $p(4)=5$, giving
\begin{align}
M_4=&M_4^\prime+4\,M_3^\prime\, M_1^\prime+3\,M_2'^2+6\,M_2^\prime
\,M_1'^2+M_1'^4\,.
\end{align}
In the fourth partition, $i=4$, we have $\kappa_{41}=2$, $\kappa_{42}=1$, and
$\kappa_{43}=\kappa_{44}=0$, and the factorials give the prefactor of $6$.
\eq{MNpartition} gives the moments $M_i$ in terms of the cumulants $M_i^\prime$,
and these relations can be inverted to yield the formulas quoted for the
cumulants in \eq{variance-skewness}.  $M_2^\prime\ge 0$ is the well known
variance of the distribution. Higher order cumulants can be positive or
negative.  The skewness of the distribution $M_3^\prime$ provides a measure of
its asymmetry, and we expect $M_3^\prime>0$ for thrust with its long tail to the
right of the peak.  The kurtosis $M_4^\prime$ provides a measure of the
``peakedness'' of the distribution, where $M_4^\prime>0$ for a sharper peak than a
Gaussian.\footnote{The cumulants of a Gaussian are all zero for $n>2$, and the
  cumulants of a delta function are all zero for $n>1$.}

The shift independence of the cumulants $M_n^\prime$ make them an ideal basis
for studying event shape moments. In particular, since the leading ${\cal
  O}(\Lambda_{\rm QCD}/Q)$ power correction acts similar to a shift to the event
shape distribution~\cite{Dokshitzer:1995zt,Dokshitzer:1995qm,Dokshitzer:1997ew,
Lee:2006fn,Lee:2006nr}, we can
anticipate that $M_{n\ge 2}^\prime$ will be more sensitive to higher order power
corrections. We will quantify this statement in the next section by using
factorization for the thrust distribution to derive factorization formulae for
the thrust cumulants in the form of an operator product expansion.

\subsection{Thrust moments}\label{sec:thrust-moments}

We will first make use of the leading order factorization theorem,
$\df\sigma/\df\tau=\int {\rm d}p\, (\df\hat\sigma/\df\tau)(\tau-p/Q)F_\tau(p)$,
which is valid for all $\tau$. It separates perturbative $\df\hat\sigma/{\rm
  d}\tau$ and nonperturbative $F_\tau(p)$ contributions to all orders in
$\alpha_s$ and $\Lambda_{\rm QCD}/(Q\,\tau)$, but is only valid at leading order
in $\Lambda_{\rm QCD}/Q$. For this factorization theorem we follow
Ref.~\cite{Abbate:2010xh} (except that here we denote the nonperturbative soft
function by $F_\tau$).\footnote{Earlier discussions of shape functions for
  thrust can be found in Refs.~\cite{Korchemsky:1999kt,Korchemsky:2000kp}.}  We
will then extend our analysis to parameterize corrections to all orders in
$\Lambda_{\rm QCD}/Q$.

Taking moments of the leading order $\df\sigma/\df\tau$ gives\footnote{This
  manipulation is valid when the renormalization scales of the jet and soft
  function which implement resummation are $\mu_i=\mu_{i}(\tau-p/Q)$, rather
  than the more standard $\mu_i(\tau)$ used in~\cite{Abbate:2010xh}. Both
  choices are perturbatively valid, and we have checked that the difference is
  $0.4\,\%$ for $M_1$, rising to $0.8\,\%$ for $M_5$, and hence is always well
  within the perturbative uncertainty.}
\begin{align} \label{eq:Mnfull}
  M_n &=\int_0^{\tau_{\rm m}}{\rm d}\tau\,\tau^n\,
  \int_0^{Q\tau}\,{\rm d}p\,\frac{1}{\hat\sigma}\,
  \frac{{\rm d}\hat{\sigma}}{{\rm d}\tau}
  \Big(\tau-\frac{p}{Q}\Big)\,F_{\tau}(p)
  \\
&=\int_0^{\infty} \!\!\!\! {\rm d}\tau\, {\rm d}p\:
 \theta\Big(\tau_m \!-\!\tau\!-\!\frac{p}{Q}\Big)\,
 \Big(\tau+\frac{p}{Q}\Big)^n
  \frac{1}{\hat\sigma}\,\frac{{\rm d}\hat{\sigma}}{{\rm
d}\tau}(\tau)\,F_{\tau}(p)
 \nn  \\
&= \bigg[\sum_{\ell=0}^n\,\binom{n}{\ell}\,\Big(\frac{2}{Q}\Big)^{n-\ell}\,
 \hat{M}_{\ell}\: \Omega_{n-\ell} \bigg] - E_n^{(A)} - E_n^{(B)} \,,
\nn
\end{align}
where $\hat \sigma$ is the perturbative total hadronic cross section and all
hatted quantities are perturbative. In the
last line of \eq{Mnfull} we used $\theta(\tau_m-\tau-p/Q)=\theta(\tau_m-\tau)
[1- \theta(p/Q-\tau_m) -\theta(\tau_m-p/Q)\,\theta(p/Q+\tau-\tau_m)]$ to obtain
the three terms.  In \eq{Mnfull} the term in square brackets is our desired
result containing the perturbative $\hat M_n$ and nonperturbative $\Omega_n$
moments
\begin{align} \label{eq:Mnhat}
\hat{M}_n=&\int_0^{\tau_{\rm m}}{\rm d}\tau\:\tau^n\,\frac{1}{\sigma}\,
  \frac{{\rm d}\hat{\sigma}}{{\rm d}\tau}\,(\tau)
 \,,  
 & \hat M_0 & = 1
 \,,\\
 \Omega_n=&\int_0^\infty\!\! {\rm d}p\ \Big(\frac{p}{2}\Big)^n\,F_{\tau}(p)
  \,, 
 & \Omega_0 & = 1 
 \,. \nn
\end{align}
The small ``error'' terms in \eq{Mnfull} are given by
\begin{align}
 E_n^{(A)} &= \sum_{\ell=0}^n \binom{n}{\ell}
 \Big(\frac{2}{Q}\Big)^{n-\ell} \hat{M}_{\ell}
  \int_{Q\,\tau_m}^{\infty}\!\!\!\! \df p\:
 \Big(\frac{p}{2}\Big)^{n-\ell} F_\tau(p)
 , \\
 E_n^{(B)} &=\!\int_0^{\tau_m}\!\! \df\tau\!
  \int^{Q\tau_m}_{Q(\tau_m-\tau)}\!\!\df p\:  
  \Big(\tau+\frac{p}{Q}\Big)^n 
  \frac{1}{\hat\sigma} \frac{{\rm d}\hat{\sigma}}{{\rm d}\tau}(\tau) F_{\tau}(p)
  \,.\nn
\end{align}
For the contribution $E_n^{(A)}$ the $p$-integral is smaller than $10^{-30}$ for
any $Q$ for the first five moments, and hence $E_n^{(A)}\simeq 0$. This occurs
because $F_{\tau}(p)$ falls off exponentially for $p\gtrsim 2\,\Omega_1 \sim
2\,\Lambda_{\rm QCD}$~\cite{Hoang:2007vb,Ligeti:2008ac}, and hence values $p\ge
Q\,\tau_m=Q/2$ are already far out
on the exponential tail.  The $E_n^{(B)}$ term gives a small contribution
because the integral is suppressed by either $F_{\tau}$ or $\df \hat\sigma/\df
\tau$: near the endpoint $\tau\sim\tau_m - 2\,\Lambda_{\rm QCD}/Q$ the
$p$-integration is not restricted and $F_\tau(p)\sim 1$, but $\df \hat\sigma/\df
\tau$ is highly suppressed. For smaller $\tau$ the $p$-integration is restricted
and the exponential tail of $F_{\tau}(p)$ suppresses the contribution.  We have
checked numerically that at $Q=91.2\,$GeV [$Q=35\,$GeV], for the first moment
the relative contribution of $E_1^{(B)}$ compared to the term in square brackets
in \eq{Mnfull} is $\mathcal{O}(10^{-7})\, [\,\mathcal{O}(10^{-6})\,]$, while for
the fifth moment $E_5^{(B)}$ it is
$\mathcal{O}(10^{-6})\,[\,\mathcal{O}(10^{-4})\,]$. This suppression does not
rely on the model used for $F_\tau(p)$.  Thus $E_n^{(B)}$ can also be safely
neglected. 

Within the theoretical precision we conclude that the leading factorization
theorem for the distribution yields an operator product expansion that
separates perturbative and nonperturbative corrections in the moments
\begin{align} \label{eq:Mnfact}
 M_n = &\sum_{\ell=0}^n\,\binom{n}{\ell}\,\Big(\frac{2}{Q}\Big)^{n-\ell}\,
  \hat{M}_{\ell}\: \Omega_{n-\ell}\,.
\end{align}
For $M_n$ the terms that numerically dominate are $\hat M_n$ and $\hat
M_{n-1}\Omega_1/Q$.  However for the cumulants $M_n^\prime$ there are
cancellations, and \eq{Mnfact} does not suffice due to our neglect so far of
$(\Lambda_{\rm QCD}/Q)^j$ suppressed terms in the factorization expression for
the thrust distribution.

To rectify this we parameterize the $(\Lambda_{\rm QCD}/Q)^j$ power corrections
by a series of power suppressed nonperturbative soft functions, $\Lambda^{j-1}
F_{\tau,j}(p/\Lambda)\sim \Lambda_{\rm QCD}^{j-1}$. Here $\Lambda^{-1}
F_{\tau,0}(p/\Lambda)=F_\tau(p)$ is the leading soft function from \eq{Mnfull}.
We introduced the parameter $\Lambda=400\,{\rm MeV}\sim \Lambda_{\rm QCD}$ to
track the dimension of these subleading soft functions.  This parameterization
is motivated by the fact that subleading factorization results can in principle
be derived with SCET~\cite{Lee:2004ja}, and at each order in the power expansion
will
yield new soft function matrix elements.  

Both the factorization analysis and
calculation of cumulants is simpler in Fourier space, so we let
\begin{align}
  \sigma(y)\equiv&\int\df \tau\,e^{-iy\tau}\,\frac{\df \sigma}{\df\tau}(\tau)
 \,,\\
  F_{\tau,j}(z\,\Lambda)\equiv&\int \frac{\df p}{\Lambda}\,
  e^{-iz p}\,F_{\tau,j}\left( \frac{p}{\Lambda}\right) 
\,,\nonumber
\end{align}
and likewise for the leading power partonic cross section
$\df\hat\sigma/\df\tau(\tau) \to \hat\sigma_0(y)$.   The factorization-based
formula for thrust is then
\begin{align} \label{eq:sublead_fact_thm}
\frac{1}{\sigma}\,\sigma (y)\,&=
\frac{1}{\hat \sigma}\sum_{j=0}^\infty 
\Big(\frac{\Lambda}{Q}\Big)^j\,
\hat{\sigma}_j(y)\,
F_{\tau,j}\Big(\frac{y\,\Lambda}{Q}\Big)\,,
\end{align}
where $\hat \sigma_{j>0}(y)$ accounts for perturbative corrections in the
$(\Lambda_{\rm QCD}/Q)^j$ power correction.  The $j=0$ term is equivalent to the
result used in \eq{Mnfull}, $F_{\tau}(p)=\Lambda F_{\tau,0}(p/\Lambda)$, and the
normalization condition for the leading nonperturbative soft function is
$F_{\tau,0}(z=0)=1$. The terms in \eq{sublead_fact_thm} beyond $j=0$ are
schematic since in reality they may involve convolutions in more variables in
the nonperturbative soft functions (as observed in the subleading $b\to
s\,\gamma$ factorization theorem results
\cite{Bauer:2001mh,Bauer:2002yu,Leibovich:2002ys,Lee:2004ja,Bosch:2004cb,Beneke:2004in}).
Nevertheless the scaling is correct, and \eq{sublead_fact_thm} will suffice for
our analysis where we only seek to classify how various power corrections could
enter higher moments or cumulants.

The identities $\sigma(y=0)/\sigma=1$ and
$\hat\sigma_0(y=0)/\hat\sigma =1$ together with \eq{sublead_fact_thm} imply
\begin{align} \label{eq:propsubleading}
F_{\tau, j}(y=0)=0\,,\qquad {\rm for }\,\, j\geq 1\,.
\end{align}
Using the Fourier-space cross section the moments  are
\begin{align} \label{eq:Mny}
 M_n\,=&\,\,i^n\,\frac{{\rm d}^n}{{\rm d}y^n} 
  \bigg[ \frac{1}{\sigma}\,{\sigma}(y)\bigg]_{y=0}
 \\
  =&\,i^n\,\frac{{\rm d}^n}{{\rm d}y^n}\bigg[
  \frac{1}{\hat \sigma}\sum_{j=0}^\infty \hat{ \sigma}_j(y)\,
  \Big(\frac{\Lambda}{Q}\Big)^j\,{F}_{\tau,j}\Big(\frac{y\,\Lambda}{Q}\Big)
  \bigg]_{y=0}
 \nonumber\\
=&\sum_{j=0}^{\infty}\Big(\frac{1}{Q}\Big)^{j}\,
  \sum_{\ell=0}^n\binom{n}{\ell}\,\hat{M}_{n-\ell,j}\,
  \Big(\frac{2}{Q}\Big)^{\ell}\,\Omega_{\ell,j}
 \,, \nn
\end{align}
which extends the OPE in \eq{Mnfact} to parameterize the $(\Lambda_{\rm
  QCD}/Q)^j$ power corrections.  Here the perturbative and nonperturbative
moments are defined as
\begin{align} \label{eq:MnjOnj}
 \hat M_{n,j}\,=&\,\,i^n\,\frac{{\rm d}^n}{{\rm d}y^n}
 \bigg[ \frac{1}{\hat\sigma}\,\hat {\sigma}_j(y)\bigg]_{y=0}
 \,,\nonumber\\
\Omega_{n,j}\,=&\,\,\frac{i^n}{2^n}\,
  \frac{{\rm d}^n}{{\rm d}z^n}
 \bigg[\Lambda^j\,{F}_{\tau,j}\big(z\,\Lambda\big)\bigg]_{z=0}
 \,,
\end{align}
where $\hat M_{n,j}$ is a dimensionless series in $\alpha_s(\mu)$ and
$\Omega_{n,j}\sim \Lambda_{\rm QCD}^{n+j}$. In order for $\hat M_{n,j}$ to exist
it is crucial that our $\hat \sigma_j(y)$ and its derivatives do not contain $\ln(y)$ dependence in
the $y\to 0$ limit at any order in $\alpha_s$.  In $\tau$-space the perturbative
coefficients have support over a finite range, $\tau\in [0,1/2]$, and
\begin{align} \label{eq:sigyexist}
 \hat\sigma_j(y) &= \int_0^{1/2}\!\! \df\tau\: e^{-i\tau y}\: \hat\sigma_j(\tau)
  \,. 
\end{align}
Therefore the existence of $\int_0^{1/2} \df\tau\: \hat \sigma_j(\tau)$
implies a well defined Taylor series in $y$ under the integrand in \eq{sigyexist},
and hence the existence of
$\hat M_{n,j}$.  This integral is the total perturbative
cross section for $j=0$. From \eq{propsubleading} we have $\Omega_{0,j>0}=0$, and
furthermore $\Omega_{n,0}=\Omega_n$ and ${\hat M}_{n,0}={\hat M}_n$.

For the first moment, \eq{Mny} yields
\begin{align} \label{eq:M1final}
M_1\, &= \hat M_1\: +\: \frac{2\,\Omega_1}{Q} \:
  + \:  \sum_{j=0}^\infty \hat M_{0,1+j} \frac{2\, \Omega_{1,1+j}}{Q^{2+j} } \,,
\end{align}
where the first two terms are determined by the leading order factorization
theorem, while the last term identifies the scaling of contributions from
$(\Lambda_{\rm QCD}/Q)^{2+j}$ power corrections. Two properties of \eq{M1final} will
be relevant for our analysis: first, there is no perturbative Wilson
coefficient for the leading $2\,\Omega_1/Q$ power correction; and second, terms
from beyond the leading factorization theorem only enter at ${\cal
  O}(\Lambda_{\rm QCD}^2/Q^2)$ and beyond. For higher order moments, $n\ge 2$,
we have
\begin{align} \label{eq:MnOPE}
M_n &= \hat M_n+ \frac{2\,n\,\Omega_1}{Q}\,\hat M_{n-1}
 +\frac{n(n-1)\Omega_2}{Q^2}\,\hat M_{n-2}\,\nonumber\\
&+\frac{2\,n\,\Omega_{1,1}}{Q^2}\hat
M_{n-1,1}+\mathcal{O}\Big(\frac{1}{Q^3}\Big)\,.
\end{align}

Next we derive an analogous expression for the $n$-th order cumulants for
$n\ge 2$, which are generated from Fourier space by
\begin{align}
  M^\prime_n\,=&\,\,i^n\,\frac{{\rm d}^n}{{\rm d}y^n}
 \bigg[\ln \frac{\sigma(y)}{\sigma}\bigg]_{y=0}
 \,.
\end{align}
Eq.~(\ref{eq:sublead_fact_thm}) can be conveniently written as the product of
three terms
\begin{align} \label{eq:prod}
 \frac{1}{\sigma}\,{\sigma}(y)\,
  =\,&\frac{1}{\hat\sigma}\,\hat{\sigma}_0(y)\, \times 
 {F}_{\tau,0}\Big(\frac{y\,\Lambda}{Q}\Big)
 \\
  \times &\bigg[1+\sum_{j=1}^\infty\overline{ \sigma}_j(y)\,
  \bigg(\frac{\Lambda}{Q}\bigg)^j\,
  \overline{{F}}_{\tau,j}\bigg(\frac{y\,\Lambda}{Q}\bigg)\bigg]
  \,,\nonumber
\end{align}
where bars indicate the ratios
\begin{align}
\overline{{\sigma}}_j(y)=\frac{\hat{ \sigma}_j(y)}{\hat{ \sigma}_0(y)},\qquad
\overline{{F}}_{\tau,j}(x)=\frac{{F}_{\tau,j}(x)}{{F}_{\tau,0}(x)}.
\end{align}
From \eq{propsubleading} we have ${\overline F}_{\tau,j}(x=0)=0$ for all $j\ge
1$.  Taking the logarithm of \eq{prod} expresses the thrust cumulants by the sum
of three terms
\begin{align} \label{eq:Mnint}
  M^\prime_n &=\,\hat M^\prime_n+\bigg(\frac{2}{Q}\bigg)^n\Omega^\prime_n
 +i^n\,\frac{{\rm d}^n}{{\rm d}y^n}\sum_{k=1}^\infty\dfrac{(-1)^{k+1}}{k}
 \nonumber\\
& \times \bigg[\sum_{j=1}^\infty\overline{\sigma}_j(y)\,
\bigg(\frac{\Lambda}{Q}\bigg)^j\,\overline{{F}}_{\tau,j}\bigg(\frac{y\,\Lambda}{
Q}\bigg)\bigg]^k\bigg|_{y=0}
\,. 
\end{align}
The first two terms involve the perturbative cumulants $\hat M^\prime_n$ and
the cumulants of the leading nonperturbative soft functions
$\Omega_n^\prime$,
\begin{align}
 \hat M^\prime_n\,&=\,\,i^n\,\frac{{\rm d}^n}{{\rm d}y^n}
  \bigg[\ln \frac{1}{\sigma}\,\hat{ \sigma}_0(y)\bigg]_{y=0}
 \,,\\
\Omega^\prime_n\,&=\frac{i^n}{2^n}\,\frac{{\rm d}^n}{{\rm d}z^n}
 \bigg[\ln F_{\tau,0}(z\Lambda)\bigg]_{z=0}
\,.\nonumber
\end{align}
The third term in \eq{Mnint} represents contributions from power-suppressed
terms that are not contained in the leading thrust factorization theorem.
These terms start at ${\cal O}(\Lambda_{\rm QCD}^2/Q^2)$. 
At this order only ${\overline F}_{\tau,1}$ has to be considered. 
The terms ${\overline F}_{\tau,i>2}$ do not contribute due to explicit powers
of $\Lambda_{\rm QCD}/Q$. 
Concerning ${\overline
  F}_{\tau,2}$, it must be hit by at least one derivative because
${\overline  F}_{\tau,2}(0)=0$, and hence
does not contribute as well. Performing the $n$-th
derivative at $y=0$ and keeping only the dominant term from the power
corrections gives the OPE
\begin{align}\label{eq:OPE-subleading}
 M^\prime_n&=\hat M^\prime_n+ \frac{2^n\Omega^\prime_n }{Q^n} + n \,
 {\overline M}_{n-1,1} \dfrac{2\,\Omega_{1,1}}{Q^2}\!
 + {\cal O}\Big(\dfrac{\Lambda^3_{\rm QCD}}{Q^3}\Big) .
\end{align}
Here $\Omega_{1,1}$ is defined in \eq{MnjOnj}.  The perturbative coefficient is
\begin{equation}\label{eq:subleading-coeff}
{\overline M}_{j,1}=\bigg[i^j\dfrac{\df^j}{\df y^j}\,
{\overline \sigma}_1(y)\bigg]_{y=0} 
\end{equation}
and so far unknown. For $n=2$ the absence of a $1/Q$ power correction in
Eq.~(\ref{eq:OPE-subleading}) was discussed in Ref.~\cite{Korchemsky:2000kp}.

The majority of our analysis will focus on $M_1$ where terms beyond the leading
order factorization theorem are power suppressed. For our analysis of $M_{n\ge
  2}$ we consider the impact of both $\alpha_s\Omega_1/Q$ corrections, and power
corrections suppressed by more powers of $1/Q$. When we analyze $M_{n\ge
  2}^\prime$ we will consider both $1/Q^n$ and $1/Q^2$ power corrections in the fits.

\section{Results for $\mathbf{M_1}$}
\label{sec:results}

In this section we present the main results of our analysis, the fits to the
first moment of the thrust distribution and the determination of $\alpha_s(m_Z)$
and $\Omega_1$. Prior to presenting our final numbers in
Sec.~\ref{subsec:finalresult} we discuss various aspects important for their
interpretation. In Sec.~\ref{sub:ingredients} we discuss the role of the
log-resummation contained in our fit code, the perturbative convergence for
different kinds of expansion methods, and we illustrate the numerical impact of
power corrections and the renormalon subtraction. We also briefly discuss the
degeneracy between $\alpha_s(m_Z)$ and $\Omega_1$ that motivates carrying out global
fits to data covering a large range of $Q$ values. In
Sec.~\ref{sec:erroranalysis} we present the 
outcome of the theory parameter scans, on which the estimate of theory
uncertainties in our fits are based, and show the
final results. We also display results for the fits 
at various levels of accuracy. Sec.~\ref{subsec:QED} briefly discusses the effects of QED and bottom mass corrections.
Sec.~\ref{sec:FO} shows the results of a fit in which renormalon subtractions and power corrections are included, but
resummation of logs in the thrust distribution is turned off.

\begin{figure}[t!]
\includegraphics[width=0.485\textwidth]{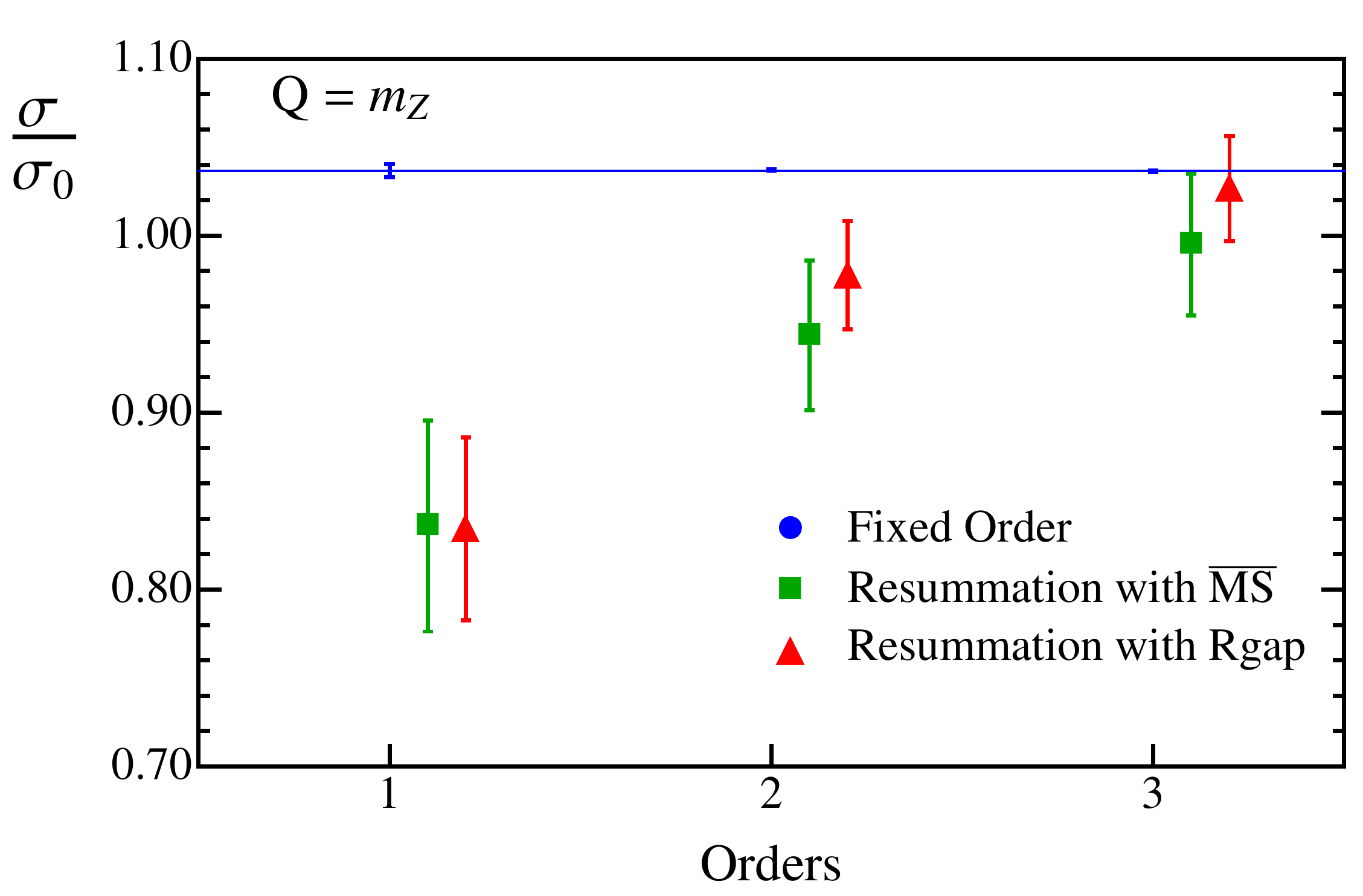}
\vspace{-0.6cm}
\caption{Theoretical computations at various orders in perturbation theory for 
  the total hadronic cross section at the Z-pole normalized to the Born-level 
  cross section $\sigma_0$. Here the small blue points correspond to fixed order
  perturbation theory, green squares to resummation without renormalon 
  subtractions, and red triangles to resummation with renormalon subtractions.
  \label{fig:M0norm} }
\end{figure}

\begin{figure}[t!]
\includegraphics[width=0.485\textwidth]{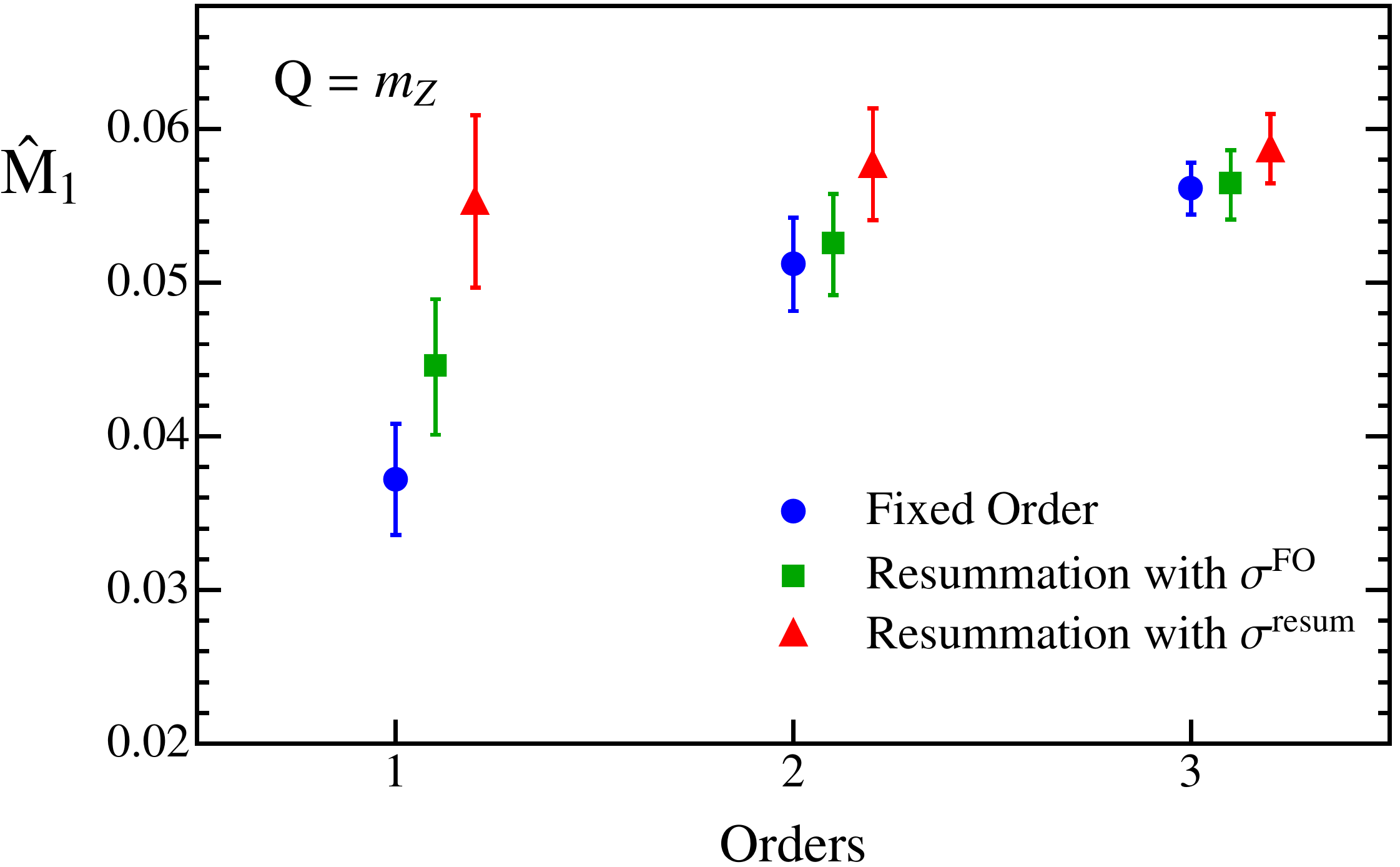}
\includegraphics[width=0.485\textwidth]{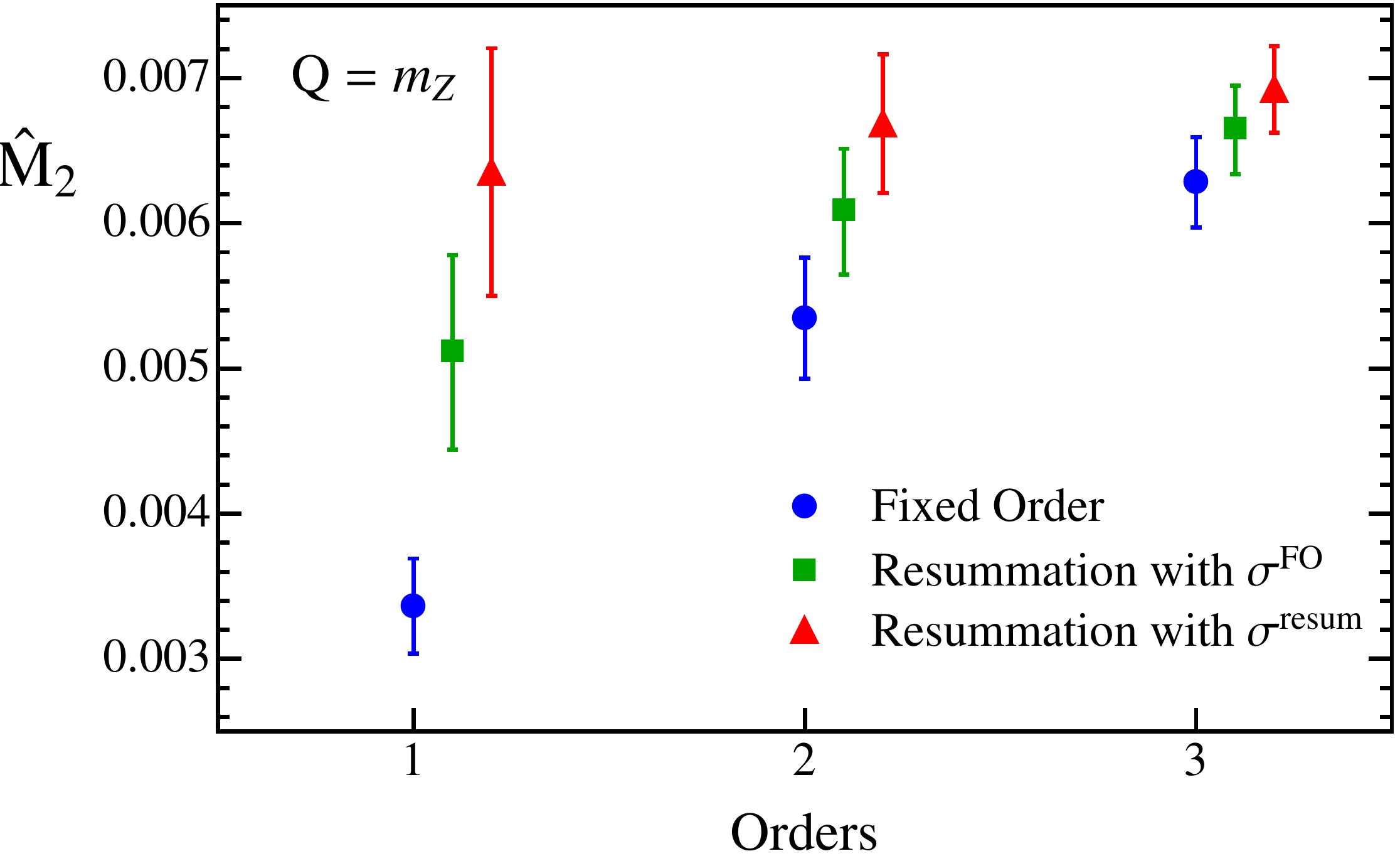}
\includegraphics[width=0.485\textwidth]{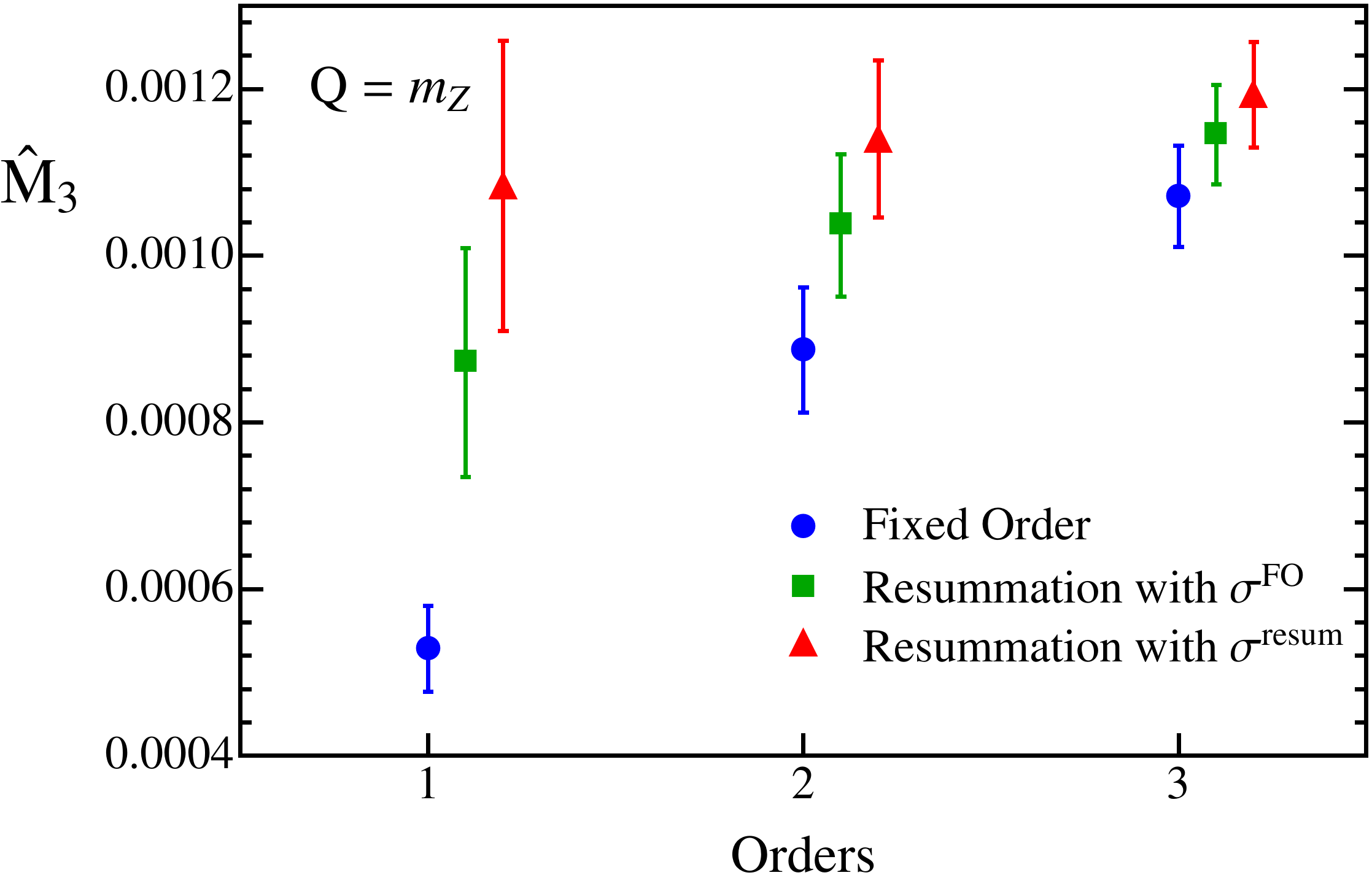}
\vspace{-0.4cm}
\caption{
Theoretical prediction for the first three moments at the Z-pole at various
orders in perturbation theory.  The blue circles correspond to
fixed order perturbation theory (normalized with the total hadronic cross
section) at ${\cal O}(\alpha_s)$, ${\cal O}(\alpha_s^2)$ and ${\cal
  O}(\alpha_s^3)$, green squares correspond to resummed predictions at NLL,
NNLL, and N${}^3$LL normalized with the total hadronic cross section, and 
red triangles correspond to resummation normalized with the norm of the
resummed distribution. For these plots we use $\alpha_s(m_Z)=0.114$.
\label{fig:Mnorm} }
\end{figure}
For our moment analysis we use the thrust distribution code developed in
Ref.~\cite{Abbate:2010xh}, 
where a detailed description of the various ingredients may be found.  We are
able to perform fits with different level of accuracy: fixed order at
$\mathcal{O}(\alpha_s^3)$, resummation of large logarithms to \ntll
accuracy\footnote{Throughout this publication N${}^n$LL corresponds to the same
  order counting as N${}^n$LL${}^\prime$ in Ref.~\cite{Abbate:2010xh}.}, power
corrections, and subtraction of the leading renormalon ambiguity.  Recently the
complete calculation of the ${\cal O}(\alpha_s^2)$ hemisphere soft function has
become available~\cite{Kelley:2011ng,Hornig:2011iu,Monni:2011gb}, so the code is
updated to use the fixed parameter $s_2=-40.6804$ from
Refs.~\cite{Kelley:2011ng,Monni:2011gb}.  A feature of our code is its ability
to describe the thrust distribution in the whole range of thrust values. This is
achieved with the introduction of what we call profile functions, which are
$\tau$-dependent factorization scales.  In the $e^+\,e^-$ annihilation process
there are three relevant scales: hard, jet and soft, associated to the center of
mass energy, the jet mass and the energy of soft radiation, respectively. The
purpose of $\tau$-dependent profile functions for these scales is to smoothly
interpolate between the peak region where we must ensure that $\mu_i >
\Lambda_{\rm QCD}$, the dijet region where the summation of large logs is
crucial, and the multijet region where regular perturbation theory is
appropriate to describe the partonic contribution~\cite{Abbate:2010xh}. The
major part of the higher order perturbative uncertainties are directly related
to the arbitrariness of the profile functions, and are estimated by scanning the
space of parameters that specify them. For details on the profile functions and
the parameter scans we refer the reader to App.~\ref{app:scan}.
We note that our distribution code was designed for $Q$ values above
$22$~GeV.

\subsection{Ingredients}
\label{sub:ingredients}

The theoretical fixed order expression for the thrust moments contain no large
logarithms, so we might not expect that the resummation of logarithms in the
thrust spectrum will play a role in the numerical analysis. We will show that
there is nevertheless some benefit in accounting for the resummation of thrust
logarithms.  This is studied in Figs.~\ref{fig:M0norm} and \ref{fig:Mnorm},
where for $Q=m_Z$ we compare the theoretical value of moments of the thrust
distribution obtained in fixed order with those obtained including resummation.
(The error bars for the fixed order expansion arise from varying the
renormalization scale $\mu$ between $Q/2$ and $2\,Q$ and those for the resummed
results arise from our theory parameter scan method.) 

In Fig.~\ref{fig:M0norm} we show the total hadronic cross section $\sigma$ from
the fixed order $\alpha_s$ expansion (blue points with small uncertainties
sitting on the horizontal line) and
determined from the integral over the log-resummed distribution with/without
renormalon subtractions (red triangles and green squares).  Both expansions are
displayed including fixed order corrections up to order $\alpha_s(m_Z)$,
$\alpha_s^2(m_Z)$ and $\alpha_s^3(m_Z)$, as indicated by the orders 1, 2, 3,
respectively. We immediately notice that the resummed result is not as effective
in reproducing the total cross section as the fixed order expansion. Predictions
that sum large logarithms have a substantial (perturbative) normalization
uncertainty. On the other hand, as shown in Ref.~\cite{Abbate:2010xh}, the
resummation of logarithms combined with the profile function approach leads to a
description of the thrust spectrum that converges nicely over the whole physical
$\tau$ range when the norm of the spectrum is divided out, a property not present
in the spectrum of the fixed order expansion.

\begin{figure*}[t!]
\includegraphics[width=0.48\textwidth]{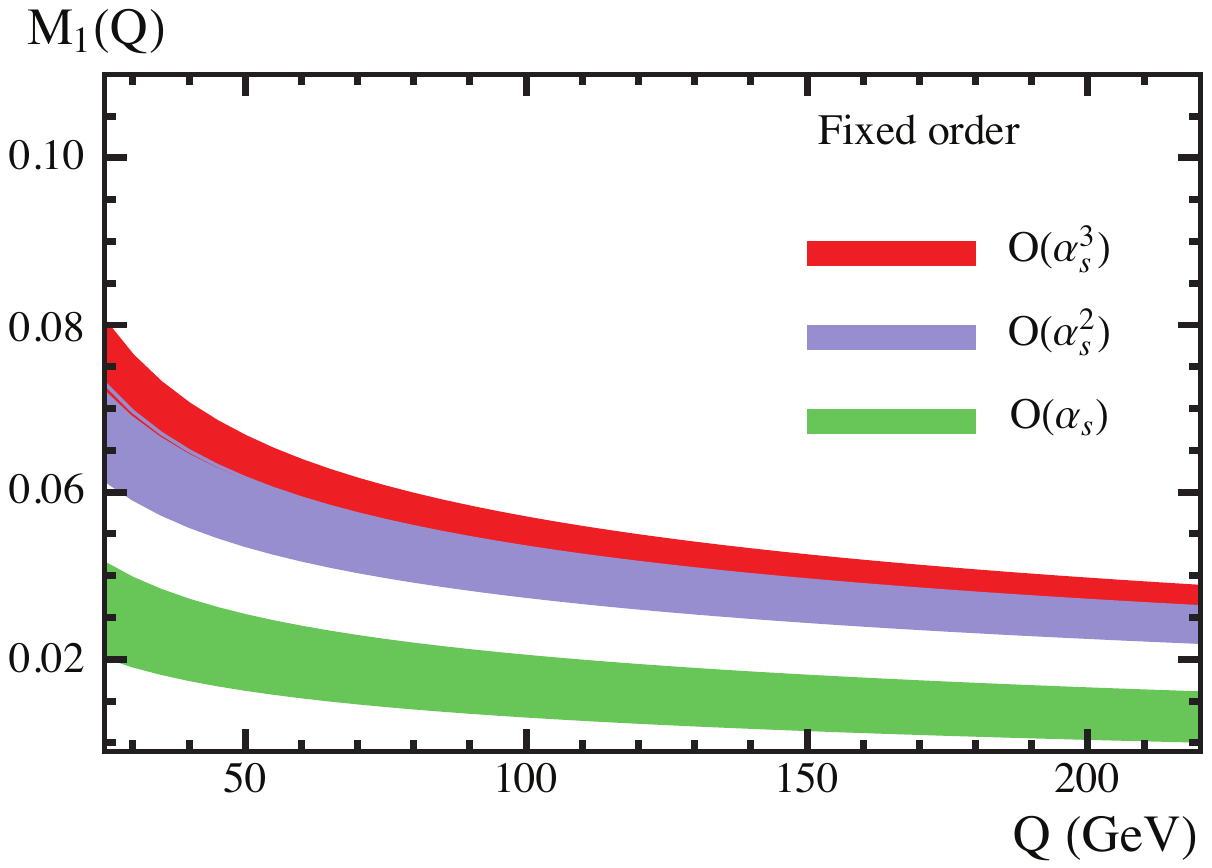}
\includegraphics[width=0.48\textwidth]{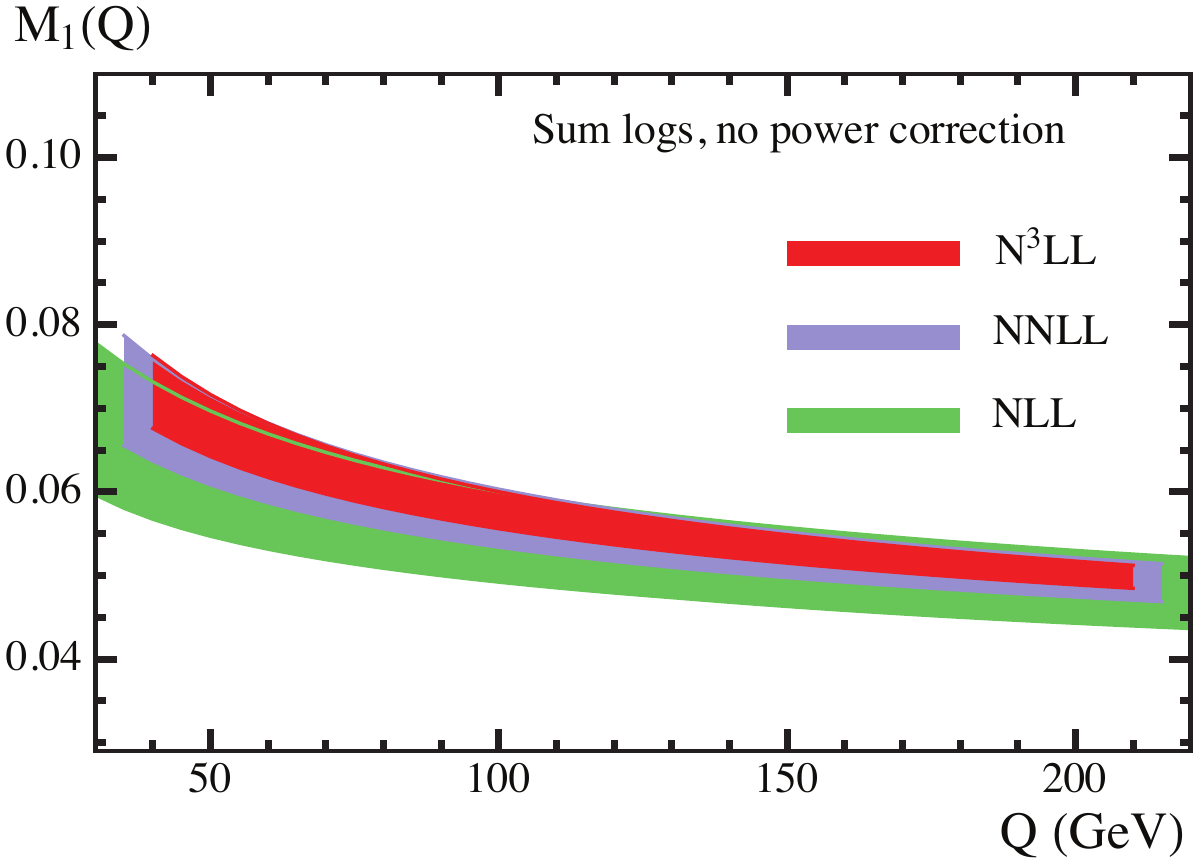}
\includegraphics[width=0.48\textwidth]{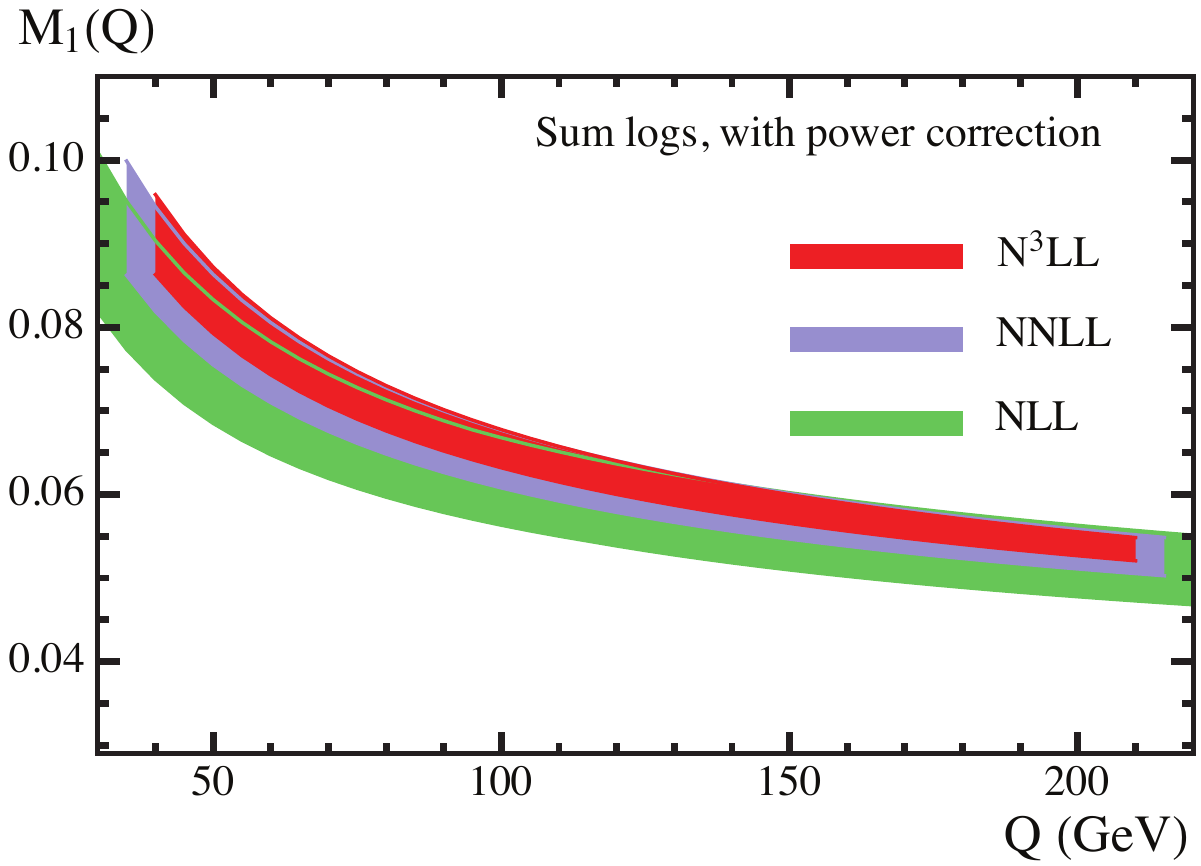}
\includegraphics[width=0.48\textwidth]{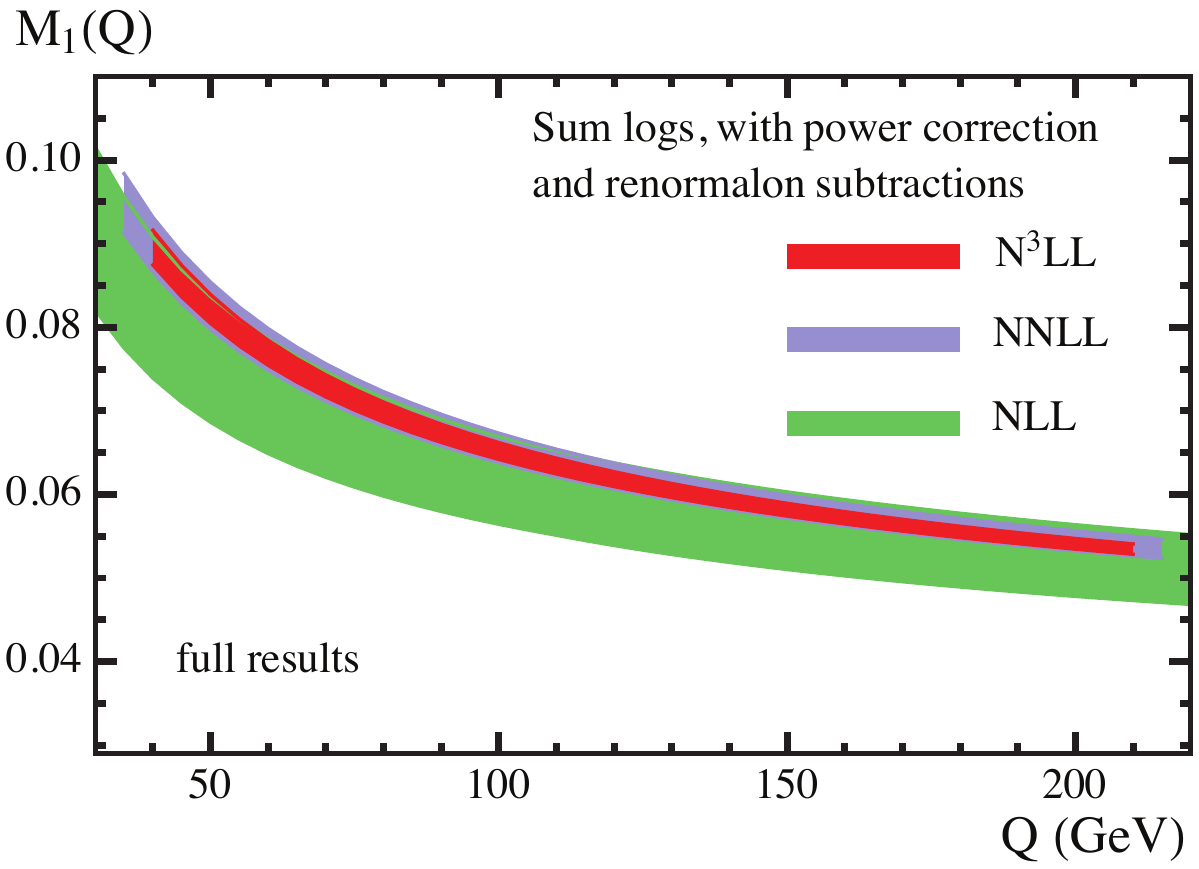}
\caption{Theory scan for uncertainties in pure QCD with massless quarks. The
  panels are fixed order (top-left), resummation without the nonperturbative
  correction (top-right), resummation with a nonperturbative function using the
  $\msbar$ scheme for $\overline\Omega_1$ (bottom-left), resummation with renormalon
  subtraction and a nonperturbative function in the Rgap scheme for $\Omega_1$
  (bottom-right).}
\label{fig:4-plots}
\end{figure*}

In Fig.~\ref{fig:Mnorm} the expansions of the partonic moments $\hat M_1$, $\hat
M_2$, and $\hat M_3$ are displayed in the fixed order expansion (blue circles)
and the log-resummed result with either the fixed order normalization (green
squares) or a properly normalized spectrum (red triangles).  We observe that the
fixed order expansion has rather small variations from scale variation, but
shows poor convergence indicating that its renormalization scale variation
underestimates the perturbative uncertainty. For $\hat M_1$ the fixed order and
log-resummed expressions with a common fixed-order normalization (blue circles
and green squares) agree well at each order, indicating that, as expected, large
logarithms do not play a significant role for this moment. On the other hand,
the expansion based on the properly normalized log-resummed spectrum exhibits
excellent convergence, and also has larger perturbative uncertainties at the
lowest order. In particular, for the red triangles the higher order results are
always within the 1-$\sigma$ uncertainties of the previous order.  The result
shows that using the normalized log-resummed spectrum for thrust, which
converges nicely for all $\tau$, also leads to better convergence properties of
the moments. At third order all the fixed order and resummed partonic moments
are consistent with each other.  Since the log-resummed moments exhibit more
realistic estimates of perturbative uncertainties at each order, we will use the
normalized resummed moments for our fit analysis.\footnote{At N$^3$LL in our
  most complete theory set up the norm of the distribution and total hadronic
  cross section are fully compatible within uncertainties, so it does not matter
  which is used. Following Ref.~\cite{Abbate:2010xh}, at N$^3$LL we choose to
   normalize the distribution with the fixed-order total hadronic cross section since it is faster.}

\begin{figure}[t!]
\vspace{0pt}
\includegraphics[width=1\linewidth]{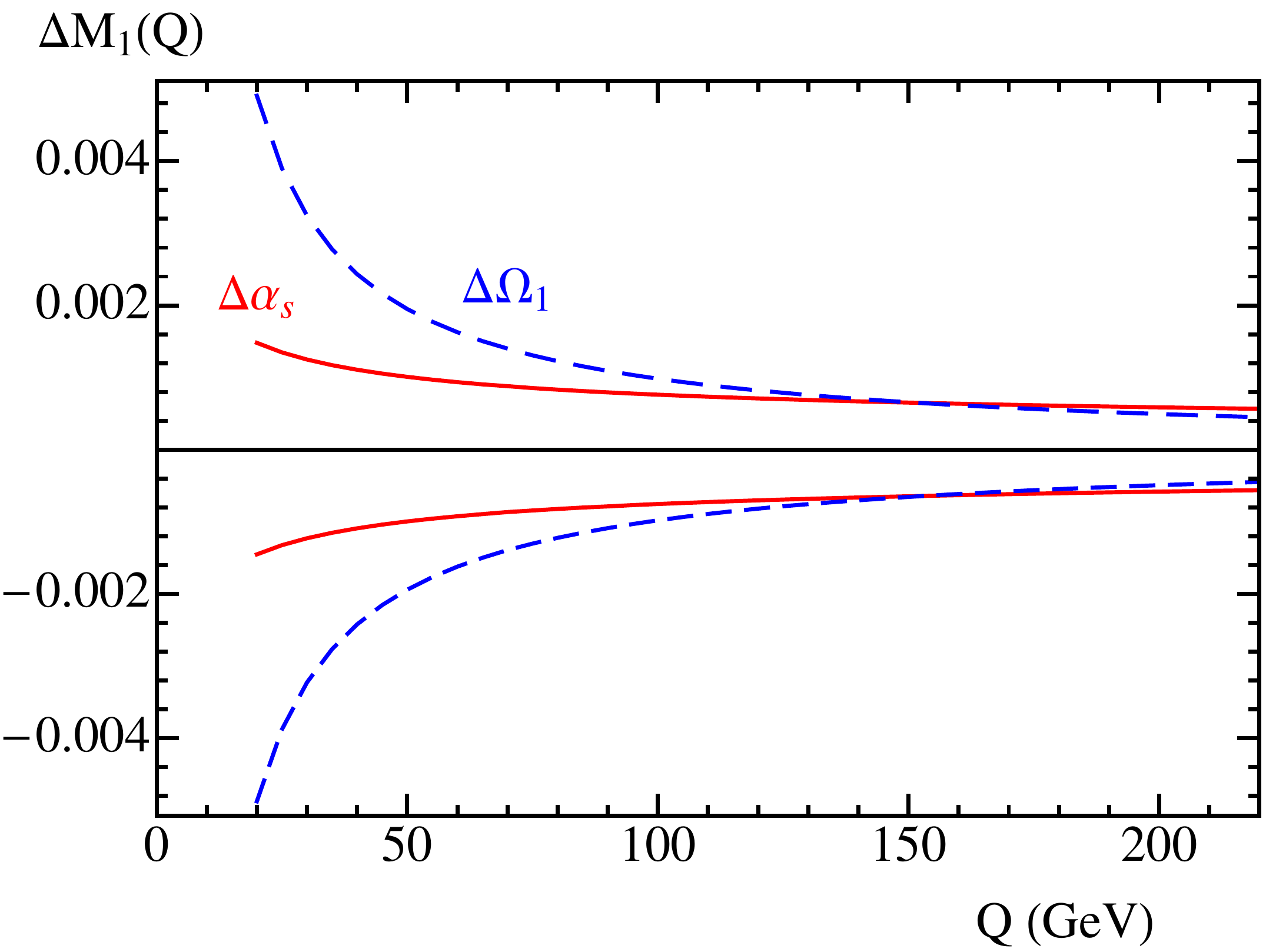}
\caption{  Difference between theoretical predictions with default parameters
for the first moment
as function of $Q$ when varying one parameter at a time. The red solid line
corresponds to varying
$\Delta\alpha_s(m_Z)=\pm 0.001$ and the blue dashed lines to varying
$\Delta\Omega_1=\pm 0.1$, with respect to
the pure QCD best-fit values. There is a strong degeneracy of the two parameters
in the region $Q>100$~GeV,
which is obviously broken when considering values of $Q$ below $70$~GeV.}
\label{fig:degeneracy}
\end{figure}

In Fig.~\ref{fig:4-plots} we show how the inclusion of various ingredients
(fixed order contributions, log resummation, power corrections, renormalon
subtraction) affects the convergence and uncertainty of our theoretical
prediction for the first moment of the thrust distribution as a function of $Q$.
From these plots we can observe four points: 
i) Fixed order perturbation theory does not converge very well.
ii) Resummation of large logarithms in the distribution, when normalized with
the integral of the resummed distribution, improves convergence for every
center of mass energy.
iii) The inclusion of power corrections has the effect of a $1/Q$-modulated
vertical shift on the value of the first moment.
iv) The subtraction of the renormalon ambiguity reduces the theoretical
uncertainty.
This picture for the first moment is consistent with
the results of Ref.~\cite{Abbate:2010xh} for the thrust distribution.

Another important element of our analysis is that we perform global fits,
simultaneously using data at a wide range of center of mass energies $Q$. This
is motivated by the fact that for each $Q$ there is a complete degeneracy
between changing $\alpha_s(m_Z)$ and changing $\Omega_1$, which can be lifted
only through a global analysis.  Fig.~\ref{fig:degeneracy} shows the difference
between the theoretical prediction of $M_1$ as a function of $Q$, when
$\alpha_s(m_Z)$ or $\Omega_1$ are varied by $\pm\,0.001$ and $\pm\,0.1$ GeV,
respectively. We see that the effect of a variation in $\alpha_s(m_Z)$ can be
compensated with an appropriate variation in $\Omega_1$ at a given center of
mass energy (or in a small $Q$ range).  This degeneracy is broken if we
perform a global fit including the wide range of $Q$ values shown in the figure.

\begin{figure*}[t!]
\vspace{0pt}
{\center
\includegraphics[width=0.9\linewidth]{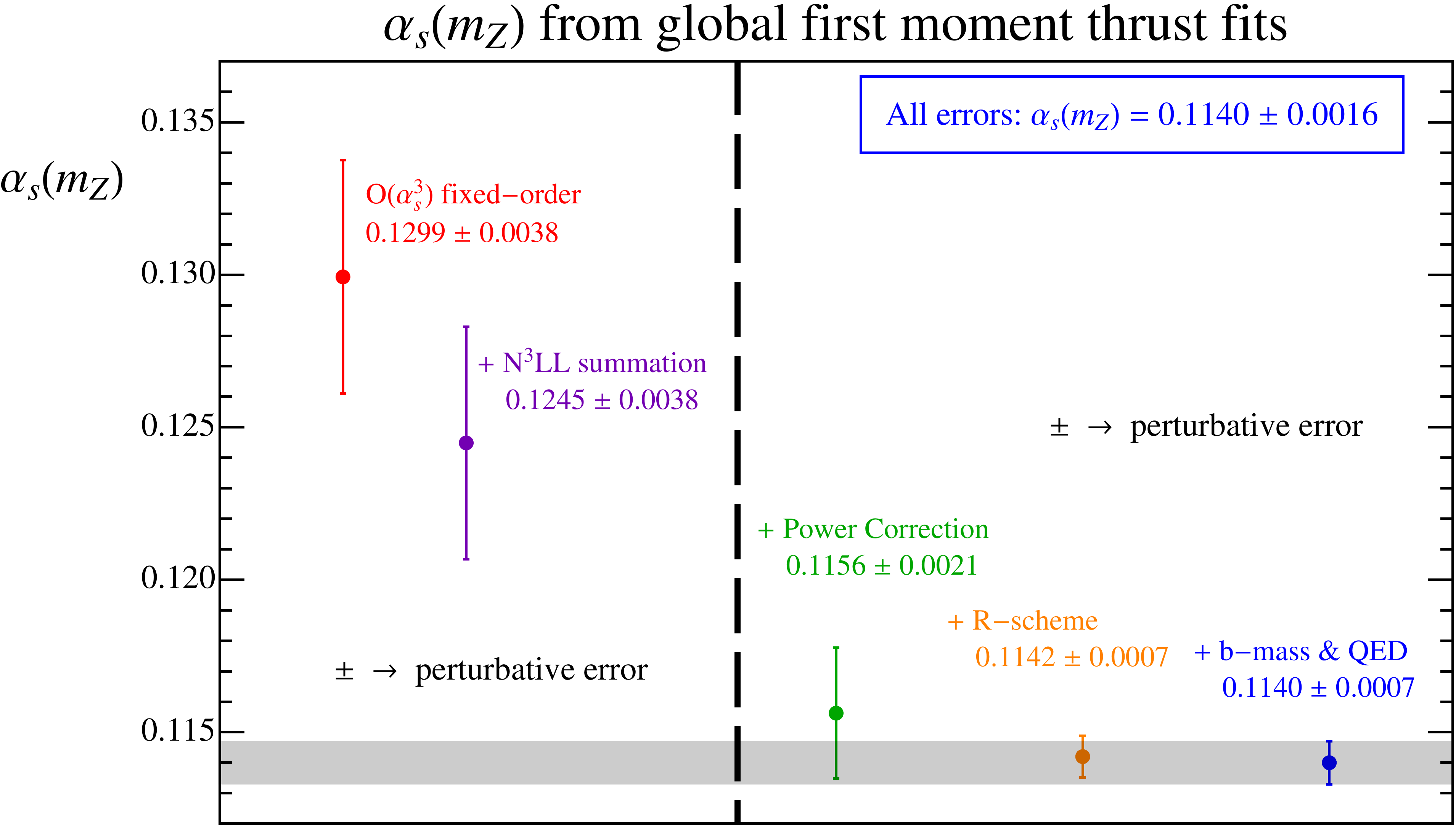}
\caption{Evolution of the best-fit values for $\alpha_s(m_Z)$ from
thrust first moment fits when including various levels of improvement
with respect to fixed order QCD. Only points at the right of the
vertical dashed line include nonperturbative effects.}
\label{fig:alpha-evolution}
}
\end{figure*}

Finally, in Fig.~\ref{fig:alpha-evolution} we show $\alpha_s(m_Z)$ extracted
from fits to the first moment of the thrust distribution at three-loop accuracy
including sequentially the different effects our code has implemented:
O$(\alpha_s^3)$ fixed order, \ntll resummation, power corrections, renormalon
subtraction, b-quark mass and QED. The error bars of the first two points at the
left hand side do not contain an estimate of uncertainties associated with the power
correction.  Though smaller, the resummed result is compatible at the
1-$\sigma$ level with the fixed order result.  The inclusion of the power correction
is the element which has the greatest impact on $\alpha_s(m_Z)$; for the
$\msbar$ definition of $\Omega_1$ it reduces the central value by 7\%. The
subtraction of the renormalon ambiguity in the Rgap scheme reduces the
theoretical uncertainty by a factor of 3, while b-quark mass and QED effects
give negligible contributions with current uncertainties.

\begin{table}[t!]
\begin{tabular}{ccc}
order &$\alpha_s(m_Z)$ (with $\overline\Omega_1^{\msbar}$) & $\alpha_s(m_Z)$
(with $\Omega_1^{\rm Rgap}$)\\
\hline
NLL 
  & $0.1173(82)(13)$ & $0.1172(82)(13)$ \\
NNLL
  & $0.1159(41)(14)$ & $0.1139(15)(13)$ \\
\ntll(full)
  &  $0.1153(21)(14)$ & $\mathbf{0.1140(07)(14)}$ \\
\hline
\ntll$\!\!${\tiny(QCD+$m_b$)}
  &  $0.1160(20)(14)$ & $0.1146(07)(14)$ \\
\ntll$\!\!${\tiny (pure QCD)}
  & $0.1156(21)(14)$ & $0.1142(07)(14)$ \\
\end{tabular}
\caption{Central values for $\alpha_s(m_Z)$ at various 
  orders with theory uncertainties from the parameter scan (first value in 
  parentheses), and experimental and hadronic error added in quadrature (second 
  value in parentheses). The bold \ntll value above the line is 
  our final  result,
  while  values below the line show the effect of leaving
  out the QED and $b$-mass corrections.}
\label{tab:results}
\end{table}
\begin{table}[t!]
\begin{tabular}{ccc}
order & \hspace{5mm}$\overline\Omega_1$ ($\msbar$) [GeV]\hspace{4mm} 
 & \hspace{1mm}$\Omega_1$ (Rgap) [GeV]\hspace{1mm} \\
\hline
NLL
  & $0.504(157)(45)$ & $0.500(153)(45)$ \\
NNLL
  & $0.405(82)(47)$ & $0.413(43)(44)$ \\
\ntll(full)
  & $0.318(75)(49)$ & $\mathbf{0.377(39)(44)}$ \\
\hline
\ntll$\!\!${\tiny (QCD+$m_b$)}
  & $0.310(74)(49)$ & $0.369(34)(44)$ \\
\ntll$\!\!${\tiny (pure QCD)}
  & $0.350(67)(49)$ & $0.402(35)(44)$ \\
\end{tabular}
\caption{Central values for  $\Omega_1$ at the 
  reference scales $R_\Delta=\mu_\Delta=2$~GeV and for $\overline\Omega_1$ and at various
  orders.  The parentheses show theory uncertainties from the parameter scan,
  and experimental and hadronic uncertainty added in quadrature, respectively. 
  The bold value above the  line is our final result,
  while the \ntll values below the horizontal line show the effect of leaving
  out the QED and $b$-mass corrections.}
\label{tab:O1results}
\end{table}
\begin{table}[t!]
\begin{tabular}{l|c c}
& $\alpha_s(m_Z)$ & $\chi^2/({\rm dof})$ \\
\hline
\ntll with $\Omega_1^{\rm Rgap}$&
$0.1140(07)(14)$&
$1.33$\\
\ntll with $\overline\Omega_1^{\msbar}$&
$0.1153(21)(14)$&
$1.33$\\
\ntll no power corr.\,\,&
$0.1236(39)(03)$&
$2.03$\\
\!\!\parbox{20ex}{${\cal O}(\alpha_s^3)$ fixed order\\[2pt]
 no power corr.}
 &
$0.1305(39)(04)$&
$2.52$ 
\end{tabular}
\caption{
Comparison of first moment fit results for analyses with full results and
$\Omega_1=\Omega_1^{\rm Rgap}$, with $\overline\Omega_1$ and no renormalon 
subtractions, without  power corrections, 
  and at fixed order without power corrections or log resummation. 
  The first number in parentheses corresponds to the theory uncertainty, whereas
  the second corresponds
  to the experimental and hadronic uncertainty added in quadrature for the first two rows,
and experimental uncertainty for the last two rows.
\label{tab:nomodel}} 
\end{table}

\subsection{Uncertainty Analysis}\label{sec:erroranalysis}

\begin{figure*}[t!]
\includegraphics[width=0.485\textwidth]{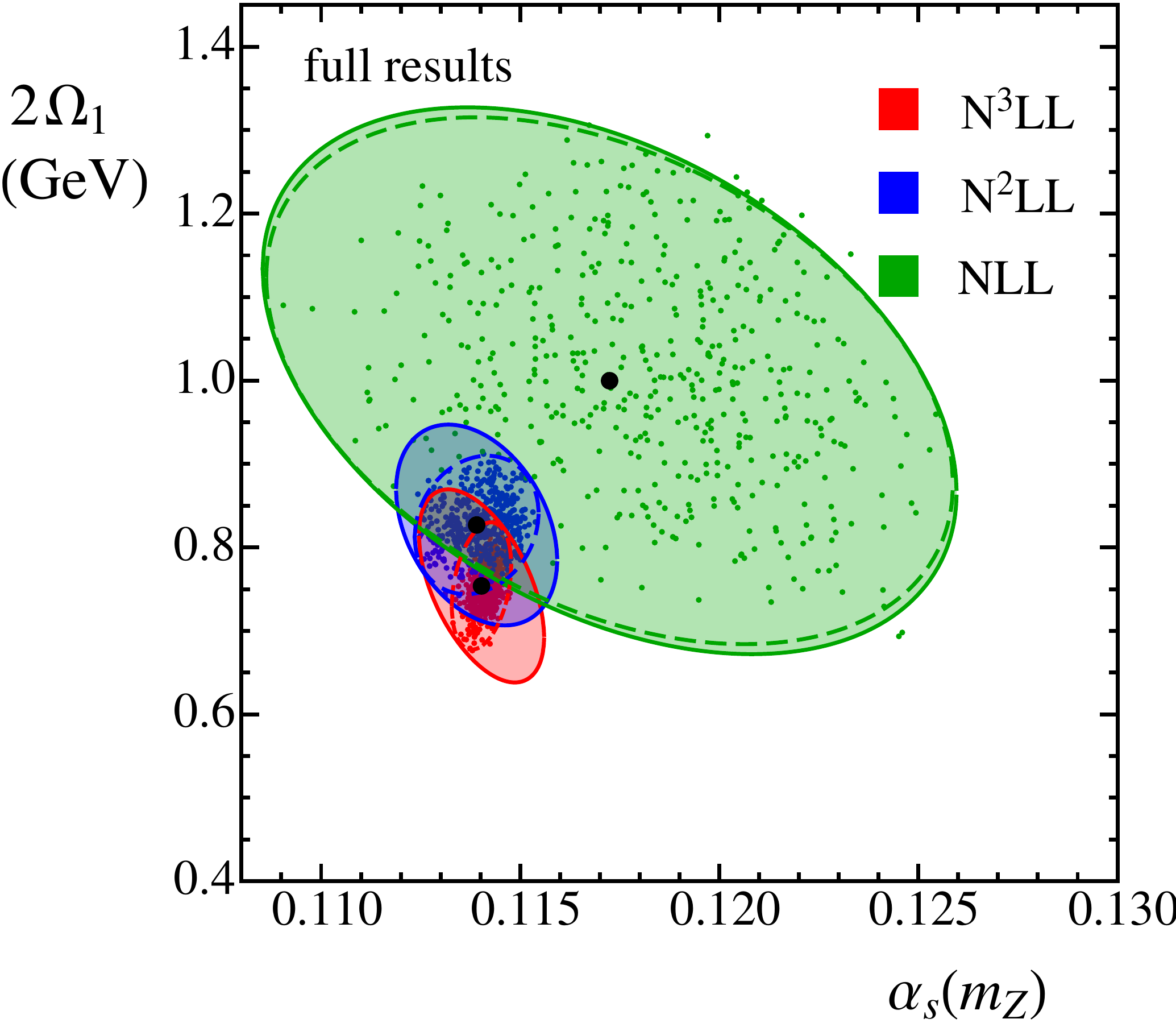}
\includegraphics[width=0.485\textwidth]{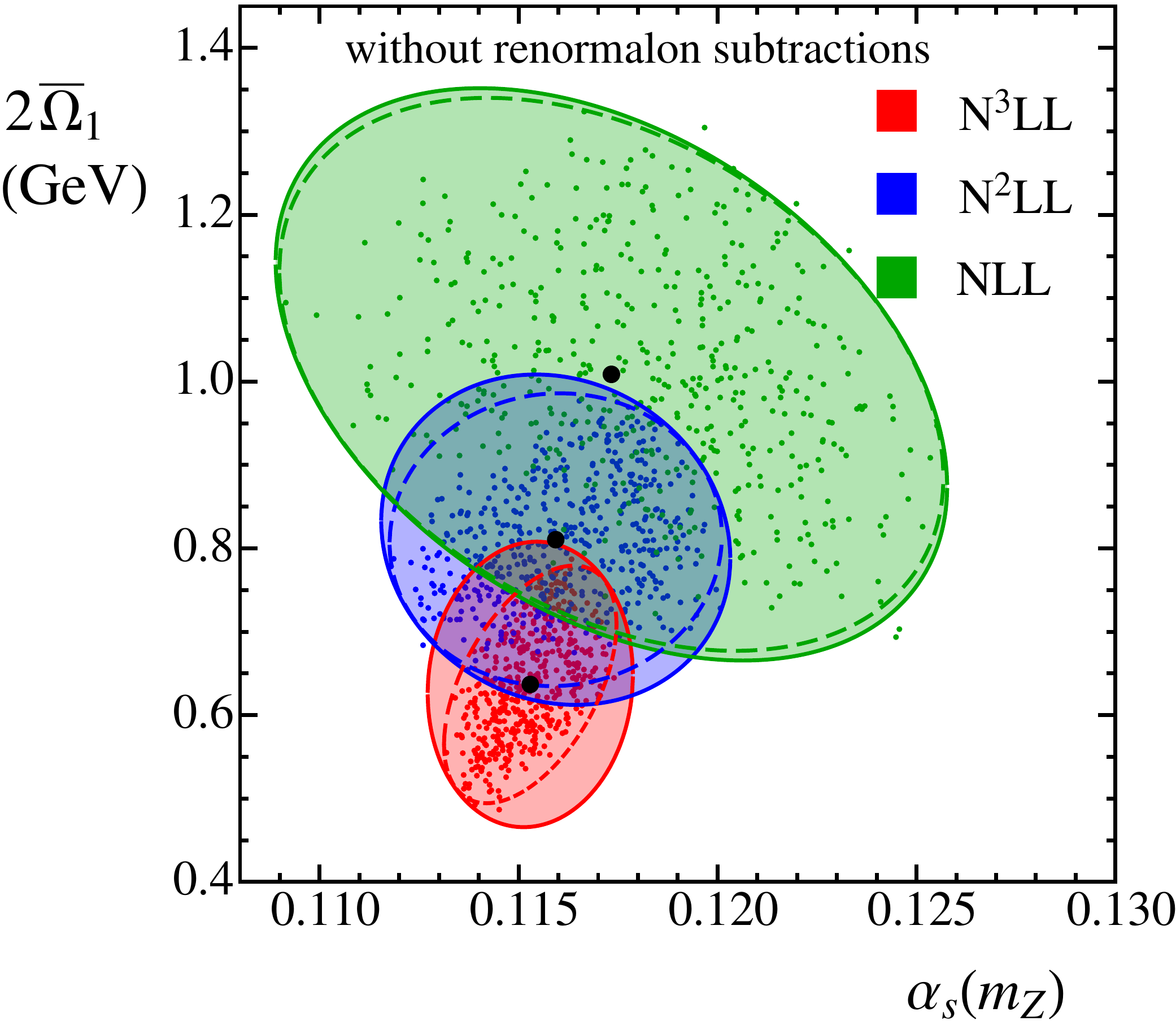}
\vspace{-0.2cm}
\caption{Distribution of best-fit points in the $\alpha_s(m_Z)$-$2\,\Omega_1$
  and $\alpha_s(m_Z)$-$2\,\overline\Omega_1$ planes. The left panel shows
  results including perturbation theory, resummation of the logs, the soft
  nonperturbative function, and $\Omega_1$ defined in the Rgap scheme with
  renormalon subtractions. The right panel shows the same results, but with
  $\overline\Omega_1$ defined in the $\msbar$ scheme, and without renormalon
  subtractions.  In both panels the dashed lines corresponds to an ellipse fit
  to the contour of the best-fit points to determine the theoretical
  uncertainty. The respective total (experimental\,+\,theoretical) 39\% CL
  standard error ellipses are displayed (solid lines), which correspond to
  $1$-$\sigma$ (68\% CL) for either one-dimensional projection.
  \label{fig:M1alpha} }
\end{figure*}

\begin{figure}[t]
\vspace{0pt}
\includegraphics[width=1\linewidth]{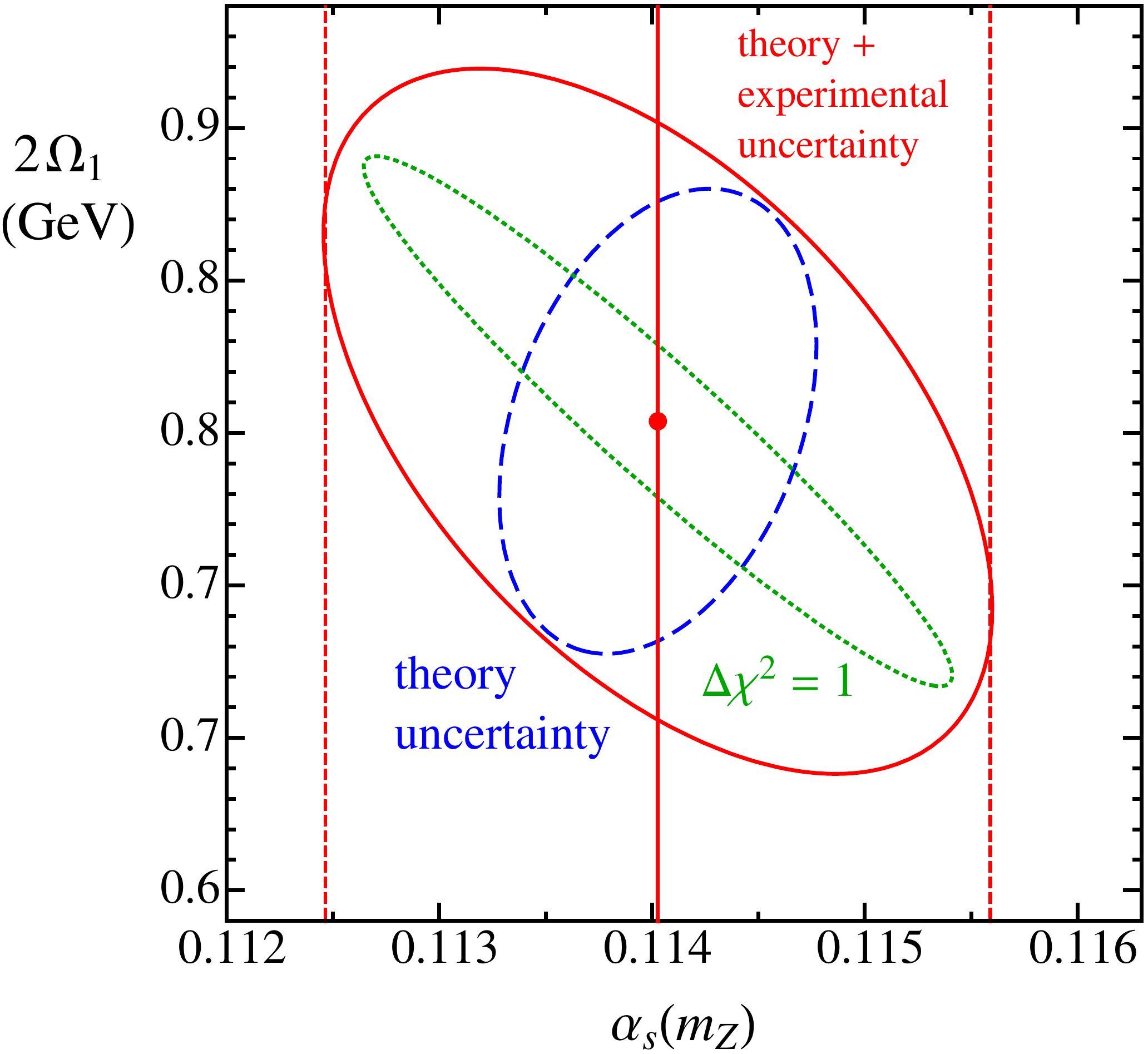}
\caption{ Experimental $\Delta\chi^2=1$ standard error ellipse (dotted green) at
  N${}^3$LL accuracy with renormalon subtractions, in the
  $\alpha_s$-$2\,\Omega_1$ plane.  The dashed blue ellipse represents the theory
  uncertainty which is obtained by fitting an ellipse to the contour of the
  distribution of the best-fit points. This ellipse should be interpreted as
  the $1$-$\sigma$ theory uncertainty for $1$-parameter (39\% confidence for
  $2$-parameters). The solid red ellipse represents the total (combined
  experimental and perturbative) uncertainty ellipse.}
\label{fig:ellipses}
\end{figure}

In Fig.~\ref{fig:M1alpha} we show the result of our theory scan to determine the
perturbative uncertainties. At each order we carried out 500 fits, with theory
parameters randomly chosen in the ranges given in Table \ref{tab:theoryerr} of
App.~\ref{app:scan} (where further details may be found). The left panel of
Fig.~\ref{fig:M1alpha} shows results with renormalon subtractions using the
Rgap scheme for $\Omega_1$, and the right-panel shows results in the $\msbar$
scheme without renormalon subtractions.  Each point in the plot represents the
result of a single fit. As described in App.~\ref{app:scan}, in order to
estimate perturbative uncertainties, we fit an ellipse to the contour of
best-fit points in the $\alpha_s$-$2\,\Omega_1$ plane, and we interpret this as
\mbox{1-$\sigma$} theoretical error ellipse.  This is represented by the dashed lines in
Fig.~\ref{fig:M1alpha}. The solid lines represent the combined (theoretical and
experimental) standard error ellipses.  These are obtained by adding the
theoretical and experimental error matrices which determined the individual
ellipses. The central values of the fits, collected in Tables~\ref{tab:results}
and \ref{tab:O1results}, are determined from the average of the maximal and
minimal values of the theory scan, and are very close to the central values
obtained when running with our default parameters.
The minimal $\chi^2$ values for these fits are quoted in Table
\ref{tab:nomodel} as well.  The best fit based on our full code has $\chi^2/{\rm
  dof}=1.325 \pm\,0.002 $ where the range incorporates the variation from the
displayed scan points at N$^3$LL.  The fit results show a substantial reduction
of the theoretical uncertainties with increasing perturbative order.  Removal of
the $\mathcal O(\Lambda_{\rm QCD})$ renormalon improves the perturbative
convergence and leads to a reduction of the theoretical uncertainties at the
highest order by a factor of 2 in $\Omega_1$, and factor of 3 in $\alpha_s(m_Z)$

To analyze in detail the experimental and the total uncertainties of our
results, we refer now to Fig.~\ref{fig:ellipses}.  Here we show the error
ellipses for our highest order fit, which includes resummation, power
corrections, renormalon subtraction, QED and b-quark mass contributions. The
green dotted, blue dashed, and the solid red lines represent the standard error
ellipses for, respectively, experimental, theoretical, and combined theoretical
and experimental uncertainties. The experimental and theory error ellipses are
defined by $\Delta \chi^2=1$ since we are most interested in the 1-dimensional
projection onto $\alpha_s$.  The correlation matrix of the experimental, theory,
and total error ellipses are ($i,j=\alpha_s, 2\,\Omega_1$)
\begin{align} \label{Vijresult}
V_{ij}&  =\, 
\left( \begin{array}{cc}
\sigma_{\alpha_s}^2 
   & \,\, 2 \sigma_{\alpha_s} \sigma_{\Omega_1}\rho_{\alpha\Omega}\\
2\sigma_{\alpha_s} \sigma_{\Omega_1}\rho_{\alpha\Omega} 
   & \,\,4 \sigma_{\Omega_1}^2
\end{array}\right) ,
 \\
V^{\rm exp}_{ij}& =
\left( \begin{array}{cc}
1.93(15)\cdot 10^{-6}  & \,\, -1.18(13)\cdot 10^{-4}\,\mbox{GeV}\\ 
-1.18(13)\cdot 10^{-4}\,\mbox{GeV} & \,\, 0.79(13)\cdot 10^{-2}\,\mbox{GeV}^2
\end{array}\right) \! , \nn\\
V^{\rm theo}_{ij}& =
\left( \begin{array}{cc}
5.56\cdot 10^{-7}  & \,\, 1.85\cdot 10^{-5}~\mbox{GeV}\\ 
1.85\cdot 10^{-5}~\mbox{GeV} & \,\, 5.82\cdot 10^{-3}~\mbox{GeV}^2
\end{array}\right), \nn\\
V^{\rm tot}_{ij}& =
\left( \begin{array}{cc}
2.49(15)\cdot 10^{-6}  & \,\, -0.99(13)\cdot 10^{-4}\,\mbox{GeV}\\ 
-0.99(13)\cdot 10^{-4}\,\mbox{GeV} & \,\, 1.37(13)\cdot 10^{-2}\,\mbox{GeV}^2
\end{array}\right)\! , \nn
\end{align} 

\noindent where the experimental correlation coefficient is significant and
reads
\begin{align} \label{eq:rhoaO}
  \rho^{\rm exp}_{\alpha\Omega}\,=\,-\,0.96(14) \,.
\end{align}
Adding the theory scan uncertainties reduces the correlation coefficient in
\eq{rhoaO} to
\begin{align}\label{eq:rhoaOtot}
\rho_{\alpha\Omega}^{\rm total} \,=\, -\,0.54(8).
\end{align}
In both \eqs{rhoaO}{rhoaOtot} the numbers in parentheses capture the range of
values obtained from the theory scan.  From $V_{ij}^{\rm exp}$ in Eq.~(\ref{Vijresult}) it is possible
to extract the experimental uncertainty for $\alpha_s$ and $\Omega_1$ and the
uncertainty due to variations of $\Omega_1$ and $\alpha_s$, respectively:
\begin{align}
\sigma_{\alpha_s}^{\rm exp} 
  & = \,\sigma_{\alpha_s}\,\sqrt{1-\rho^2_{\alpha\Omega}}
  =  \,0.0004 \,,
\\
\sigma_{\Omega_1}^{\rm exp} 
  & = \,\sigma_{\Omega_1}\,\sqrt{1-\rho^2_{\alpha\Omega}}
  =  \,0.013~\mbox{GeV} \,,
\nonumber\\
\sigma_{\alpha_s}^{\rm \Omega_1} 
  & = \,\sigma_{\alpha_s}\, |\rho_{\alpha\Omega}|\,
  =  \,0.0014 \,,
\nonumber\\
\sigma_{\Omega_1}^{\rm \alpha_s} 
  & = \,\sigma_{\Omega_1}\, |\rho_{\alpha\Omega}|\,
  =  \,0.044~\mbox{GeV}
\,.\nonumber
\end{align}
Fig.~\ref{fig:ellipses} shows the total uncertainty in our final result quoted in
\eq{asO1finalcor} below.

The correlation exhibited by the green dotted experimental error ellipse in
Fig.~\ref{fig:ellipses} is given by the line describing the semimajor axis
\begin{align}
  \frac{\Omega_1}{32.82\,{\rm GeV}} = 0.1255 - \alpha_s(m_Z) \,.
\end{align}
Note that extrapolating this correlation to the extreme case where we neglect
the nonperturbative corrections ($\Omega_1=0$) gives $\alpha_s(m_Z)\to 0.1255$.

\subsection{Effects of QED and the $b$-mass}\label{subsec:QED}

The experimental correction procedures applied to the AMY, JADE, SLC, DELPHI and
OPAL data sets were typically designed to eliminate initial state photon
radiation, while those of the TASSO, L3 and ALEPH collaborations eliminated
initial and final state photon radiation. It is
straightforward to test for the effect of these differences in the fits by using
our theory code with QED effects turned on or off depending on the data set.
Using our \ntll order code in the Rgap scheme we obtain the central values
$\alpha_s(m_Z)=0.1143$ and $\Omega_1=0.376$~GeV.  Comparing to our default
results given in Tabs.~\ref{tab:results} and \ref{tab:O1results}, which are
based on the theory code were QED effects are included for all data sets, we see
that the central value for $\alpha_s$ is larger by $0.0003$ and the one for
$\Omega_1$ is smaller by $0.001$~GeV. This shift is substantially smaller than
our perturbative uncertainty. Hence our choice to use the theory code with QED
effects included everywhere as the default for our analysis does not cause an
observable bias regarding experiments which remove final state photons.

By comparing the N$^3$LL (pure massless QCD) and N$^3$LL (QCD $+\,m_b$) entries in
Tabs.~\ref{tab:results} and \ref{tab:O1results} we see that including finite
$b$-mass corrections causes a very mild shift of $\simeq +0.0004$ to
$\alpha_s(m_Z)$, and a somewhat larger shift of $\simeq -0.033\,{\rm GeV}$ to
$\Omega_1$. In both cases these shifts are within the 1-$\sigma$ theory
uncertainties. In the N$^3$LL (pure massless QCD) analysis the $b$-quark is treated as a
massless flavor, hence this analysis differs from that done by
JADE~\cite{Pahl:2008uc} where primary $b$ quarks were removed using MC generators.

\subsection{Final Results}\label{subsec:finalresult}

\begin{figure}[t!]
\vspace{0pt}
\includegraphics[width=1\linewidth]{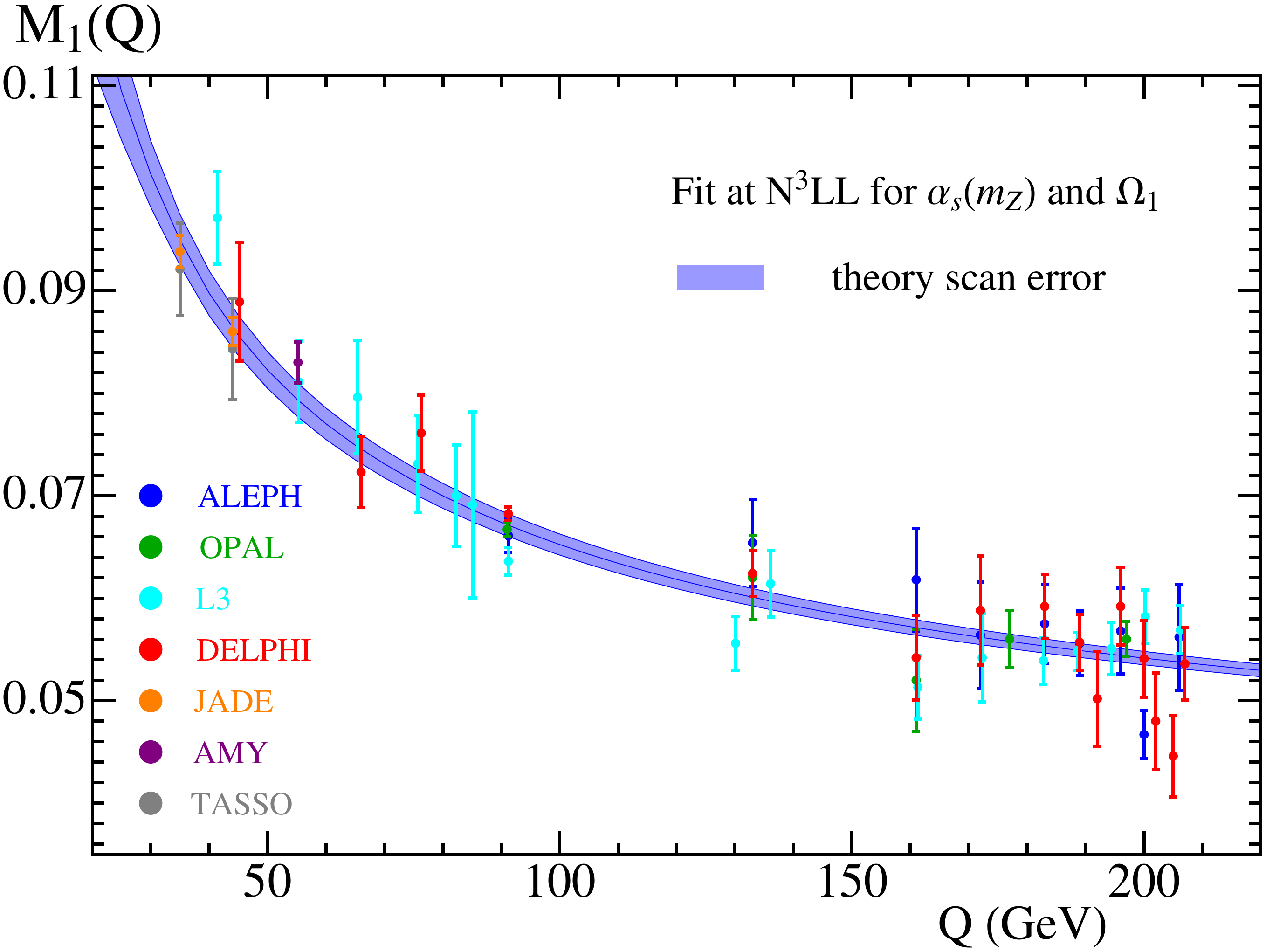}
\caption{ First moment of the thrust distribution as a function of the center of
  mass energy $Q$, using the best-fit values
  for $\alpha_s(m_Z)$ and $\Omega_1$ in the Rgap scheme as given in
  Eq.~(\ref{eq:asO1finalcor}). The blue band represents the perturbative
  uncertainty determined by our theory scan. Data is from ALEPH, OPAL, L3,
  DELPHI, JADE, AMY and TASSO. }
\label{fig:theovsexp}
\end{figure}

As our final result for $\alpha_s(m_Z)$ and $\Omega_1$, obtained at \ntll order
in the Rgap scheme for $\Omega_1(R_\Delta,\mu_\Delta)$, including bottom quark
mass and QED corrections we obtain
\begin{align} \label{eq:asO1finalcor}
\alpha_s(m_Z) & \, = \, 
 0.1140 \,\pm\, (0.0004)_{\rm exp} 
\\[2mm] & \,\pm\, (0.0013)_{\rm hadr} \,\pm \, (0.0007)_{\rm pert},
\nonumber\\[4mm]
\Omega_1(R_\Delta,\mu_\Delta) & \, = \,
 0.377 \,\pm\, (0.013)_{\rm exp} 
\nonumber\\[2mm] &        \,\pm\, (0.042)_{\rm \alpha_s(m_Z)} 
\,\pm \, (0.039)_{\rm pert}~\mbox{GeV},
\nonumber
\end{align} 
where $R_\Delta=\mu_\Delta=2$~GeV and we quote individual \mbox{$1$-$\sigma$}
uncertainties for each parameter.  Here $\chi^2/\rm{dof}=1.33$.
Eq.~(\ref{eq:asO1finalcor}) is the main result of this work.

In Fig.~\ref{fig:theovsexp} we show the first moment of the thrust distribution
as a function of the center of mass energy $Q$, including QED and $m_b$
corrections. We use here the best-fit values given in
Eq.~(\ref{eq:asO1finalcor}).  The band displays the theoretical uncertainty and
has been determined with a scan on the parameters included in our theory, as
explained in App.~\ref{app:scan}. The fit result is shown in comparison with
data from ALEPH, OPAL, L3, DELPHI, JADE, AMY and TASSO. Good agreement is observed for all
$Q$ values.

\begin{figure}[t!]
\vspace{0pt}
\includegraphics[width=1\linewidth]{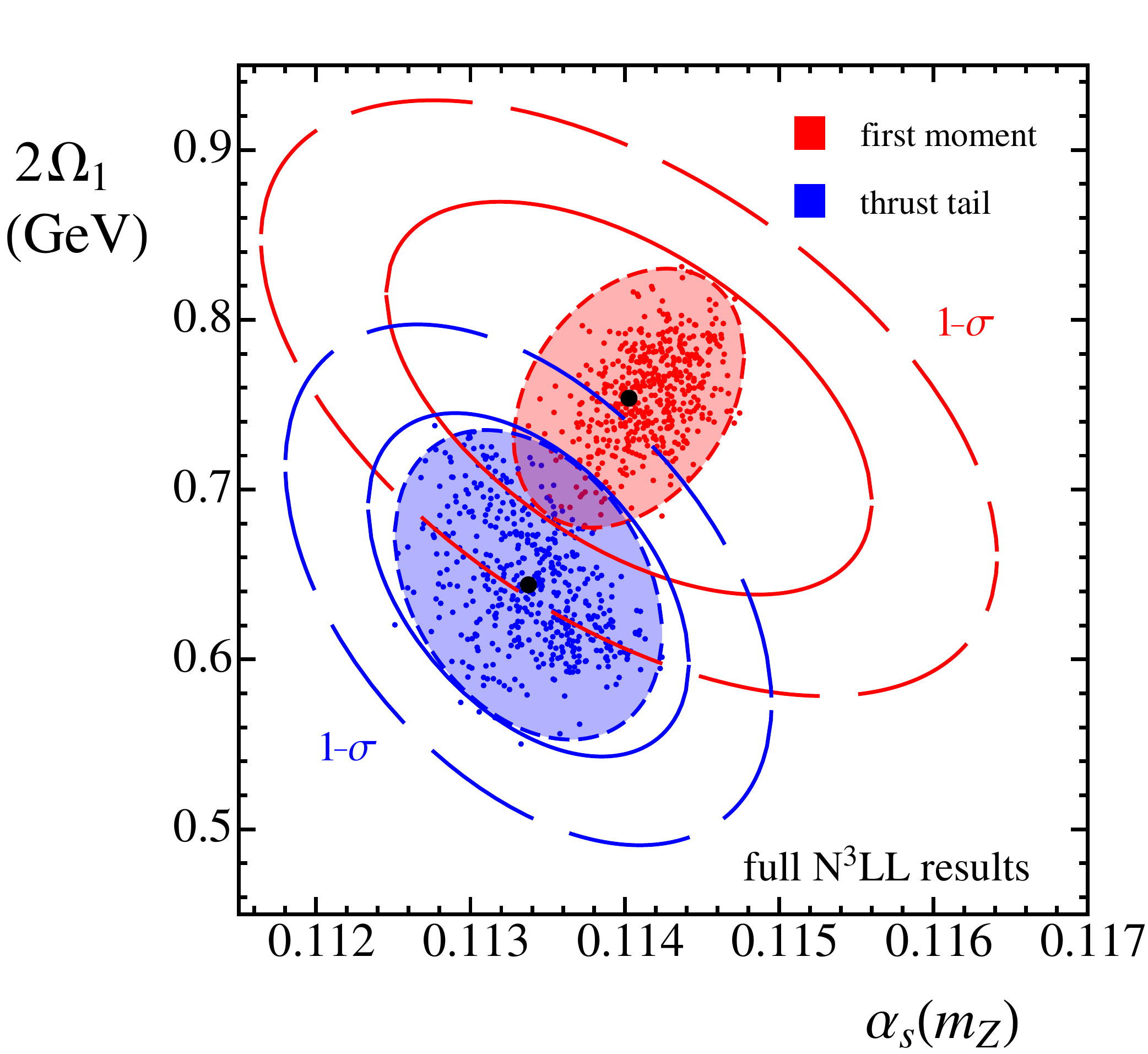}
\caption{Comparison of $\alpha_s(m_Z)$ and $\Omega_1$ determinations from thrust
  first moment data (red upper right ellipses) and thrust tail data (blue lower
  left ellipses).  The plot corresponds
  to fits with N${}^3$LL accuracy and in the Rgap scheme. The tail fits are
  performed with our improved code which uses a new nonsingular two-loop
  function, and the now known two-loop soft function. Dashed lines correspond to
  theory uncertainties, solid lines correspond to $\Delta\chi^2=1$ combined
  theoretical and experimental error ellipses, and wide-dashed lines correspond to
  $\Delta\chi^2=2.3$ combined error ellipses (corresponding to 1-$\sigma$
  uncertainty in two dimensions).}
\label{fig:moment-tail-comparison}
\end{figure}

It is interesting to compare the result of this analysis with the result of our
earlier fit of thrust tail distributions in Ref.~\cite{Abbate:2010xh}.  This is
shown in Fig.~\ref{fig:moment-tail-comparison}. Here the red upper shaded area
and corresponding ellipses show the results from fits to the first moment of the
thrust distribution, while the blue lower shaded area and ellipses show the
result from fits of its tail region. Both analyses show the theory
(dashed lines) and combined theoretical and experimental (solid lines) standard
error ellipses, as well as the ellipses which correspond to \mbox{$\Delta
  \chi^2=2.3$} (68\% CL for a two-parameter fit, wide-dashed lines). We see that
the two analyses are compatible.

\section{Fixed Order Analysis of $M_1$}\label{sec:FO}

It is interesting to compare the result of our best fit with an analysis where
we do not perform resummation in the thrust distribution, but where power corrections
and renormalon subtractions are still considered. This is achieved by setting
the scales $\mu_H$, $\mu_S$, $\mu_J$, $\mu_{\rm ns}$ in our theoretical
prediction all to a common scale $\mu\sim Q$.  We use $R$ for
the scale of the renormalon subtractions and renormalization group evolved power
correction.  Finally we will neglect QED and $b$-mass corrections in this
subsection.  Up to the treatment of power corrections and perturbative
subtractions, the fixed order results used for this analysis are thus equivalent
to those used in Ref.~\cite{Gehrmann:2009eh}.

The OPE formula for the first moment in the Rgap scheme for this situation is
given by
\begin{align} \label{eq:GAP-OPE}
M_1&= {\hat M}_1^{\rm Rgap}(R,\mu)
  +\dfrac{2\,\Omega_1(R,\mu)}{Q}\,,\\
\Omega_1(R,\mu)&=\Omega_1
  +\bar{\Delta}(R,\mu)-\bar{\Delta}(R_\Delta,\mu_\Delta)\,,\nonumber
\end{align}
In Eq.~(\ref{eq:GAP-OPE}), the $\Omega_1$ with no arguments is the value
determined by the fits, which is in the Rgap scheme at the reference scale
$\mu_\Delta=R_\Delta=2\,{\rm GeV}$. Here $\bar{\Delta}(R,\mu)$ is the running
gap parameter, and 
$\bar{\Delta}(R,\mu)-\bar{\Delta}(R_\Delta,\mu_\Delta)$
is used to sum logarithms from $(R_\Delta,\mu_\Delta)$ to
$(R,\mu)$ in \eq{GAP-OPE}. The analytic expression for
$\bar{\Delta}(R,\mu)-\bar{\Delta}(R_\Delta,\mu_\Delta)$ can be found in Eq.~(41)
of Ref.~\cite{Abbate:2010xh} (see also \cite{Hoang:2008fs}). The perturbative ${\hat M}_1^{\rm Rgap}$ is
related to the perturbative $\msbar$ result by
\begin{align} \label{eq:MhatRgap}
{\hat M}_1^{\rm Rgap}(R,\mu) 
 &={\hat M}_1^{\rm\overline{MS}}(\mu)+\dfrac{2\,\delta(R,\mu)}{Q}
 \,,\\
\delta(R,\mu)&=e^{\gamma_E}R \sum_{i=1}^3\alpha_s(\mu)^i\delta_i(R,\mu)\,,
\nonumber
\end{align}
where the subtractions terms are~\cite{Hoang:2008fs,Abbate:2010xh}
\begin{align} \label{eq:d123}
\delta_1(R,\mu) &= -0.848826 L_R \,,  \\
\delta_2(R,\mu) &= -0.156279 - 0.46663 L_R
 - 0.517864 L_R^2 \,, \nn \\
\delta_3(R,\mu)&= -\,0.552986
- 0.622467 L_R  - 0.777219 L_R^2
  \notag\\&\quad - 0.421261 L_R^3 
  \,,\nn
\end{align}
with $L_R=\ln(\mu/R)$.  In \eq{MhatRgap} $\delta(R,\mu)$ cancels the ${\cal
  O}(\Lambda_{\rm QCD})$ renormalon in ${\hat M}_1^{\rm\overline{MS}}(\mu)$, and
it is crucial that the coupling expansions in both these objects are done at the
same scale, $\alpha_s(\mu)$, for this cancellation to take place. The relation
to the $\msbar$ scheme power correction is $\overline \Omega_1 = \Omega_1 +
\delta(R_\Delta,\mu_\Delta)$, and the OPE in the $\msbar$ scheme at this level
is
\begin{align} \label{eq:MSbar-OPE}
  M_1&= {\hat M}_1^{\overline{\rm MS}}(\mu)
  +\dfrac{2\,{\overline \Omega}_1}{Q}\,.
\end{align}
In the $\msbar$ result there are no perturbative renormalon subtractions (and
thus no log resummation related to the renormalon subtractions) and the
parameter $\overline\Omega_1$ has a $\Lambda_{\rm QCD}$ renormalon ambiguity.

We will perform fits to the experimental data following the same procedure
discussed in the previous section.  Using Eq.~(\ref{eq:GAP-OPE}) we consider two
cases, i) $R\sim Q$ where $\Omega_1$ is renormalization group evolved to $R$ and
there are no large logarithms in the renormalon subtractions, and ii) fixing $R$
at the reference scale, $R=2\,{\rm GeV}$, in which case large logarithms are
present in the renormalon subtractions.  We will also consider a third case,
iii), using the $\msbar$-OPE of \eq{MSbar-OPE}. 
\begin{table}[t!]
\begin{tabular}{lcc}
  order 
    & \hspace{0.6cm}  ${\cal O}(\alpha_s^2)$  \hspace{0.6cm}
    &  ${\cal O}(\alpha_s^3)$   \\
\hline
(i) Rgap R-RGE
  & $0.1159(27)(14)$ & $0.1146(06)(14)$ \\
(ii) Rgap FO Subt.
  & $0.1185(63)(15)$ & $0.1138(20)(14)$ \\
(iii) $\msbar$ for ${\overline \Omega}_1$
  &  $0.1278(124)(19)$ & $0.1186(38)(14)$ \\
\end{tabular}
\caption{$\msbar$ scheme values for $\alpha_s(m_Z)$ obtained from various fixed order
  analyses. The first value in parentheses is the uncertainty from higher order
  perturbative corrections (obtained by the method described in the text), while
  the second value is the combined experimental and hadronization uncertainty.}
\label{tab:FOresults}
\end{table}
\begin{table}[t!]
\begin{tabular}{lcc}
  order 
    & \hspace{0.6cm}  ${\cal O}(\alpha_s^2)$  \hspace{0.6cm}
    & ${\cal O}(\alpha_s^3)$   \\
\hline
(i) Rgap R-RGE
  & $0.407(8)(45)$ & $0.400(8)(45)$ \\
(ii) Rgap FO Subt.
  & $0.216(126)(133)$ & $0.359(42)(62)$ \\
(iii) $\msbar$ for ${\overline \Omega}_1$
  &  $0.388(62)(47)$ & $0.350(54)(44)$ \\
\end{tabular}
\caption{$\Omega_1$ or $\overline\Omega_1$ 
  values obtained from fixed order analyses at various 
  orders. The first value in parentheses is the uncertainty from higher order
  perturbative corrections (obtained by the method described in the text), while
  the second value is the combined experimental and hadronization uncertainty.}
\label{tab:FOO1results}
\end{table}
Results for these fits are shown in Tabs.~\ref{tab:FOresults} and
\ref{tab:FOO1results}.  For all cases $\chi^2/\rm{dof} \simeq 1.32$.  

For case i) we take $R\sim \mu\sim Q$, so there are no large logarithms in the
$\delta(R,\mu)$ of \eq{GAP-OPE}, and all large logarithms associated with
renormalon subtractions are summed in
$\bar\Delta(R,\mu)-\bar\Delta(R_\Delta,\mu_\Delta)$.  Here we estimate the
perturbative uncertainty in $\alpha_s(m_Z)$ and $\Omega_1$ by varying the
renormalization scale $\mu$ and the scale $R$ independently in the range
$\{2\,Q,Q/2\}$. We use one-half the maximum minus minimum variation as the
uncertainty, and the average for the central value. The results for both
$\alpha_s(m_Z)$ and $\Omega_1$ are fully compatible at 1-$\sigma$ to our final
results shown in \eq{asO1finalcor}. The agreement is even closer to the central
values for the fits without QED or $b$-mass corrections in
Tabs.~\ref{tab:results} and \ref{tab:O1results}, namely
$\alpha_s(m_Z)=0.1142(07)(14)$ and $\Omega_1=0.402(35)(44)$. The one difference
is that the perturbative uncertainty for $\Omega_1$ in
Tab.~\ref{tab:FOO1results} is a factor of three smaller.  The case i) results in
the table also exhibit nice order-by-order convergence, and if one plots $M_1$
versus $Q$ (analogous to Fig.~\ref{fig:Mnorm}) the uncertainty bands are
entirely contained within one another.  In order to be conservative, we take our
resummation analysis in \eq{asO1finalcor} as our final results (with its larger
perturbative uncertainty and inclusion of QED and $b$-mass corrections).

For case ii) we take $R\sim 2\,{\rm GeV}$ and $\mu\sim Q$ as typical values, so
there are large logarithms, $\ln(R/Q)$, in the $\delta(R,\mu)$ renormalon
subtractions. The central value for $\alpha_s(m_Z)$ at ${\cal O}(\alpha_s^3)$ is
again fully compatible with that in \eq{asO1finalcor}. Here we estimate the
perturbative uncertainty in $\alpha_s(m_Z)$ by varying $\mu\in \{2\,Q,Q/2\}$ and
$R=2\pm 1\,{\rm GeV}$. Due to the large logarithms the perturbative uncertainty
in $\alpha_s(m_Z)$ for case ii), shown in Tab.~\ref{tab:FOresults}, is three
times larger than for case i).  It is also compatible with the difference
between central values at ${\cal O}(\alpha_s^2)$ and ${\cal O}(\alpha_s^3)$.  To
estimate the uncertainty for $\Omega_1$ we only vary $\mu$, which leads to the
rather large error estimate for $\Omega_1$ shown in
Tab.~\ref{tab:FOO1results}. The contrast between the precision of the results in
case i), to the results in case ii), illustrates the importance of summing large
logarithms in the renormalon subtractions.

For case iii), where the ${\overline\Omega}_1$ power correction is defined in
$\msbar$ we do not have renormalon subtractions (and hence no large logs in
subtractions). Due to the poor convergence of the fixed order prediction for the
first moment, seen from the blue fixed order points in Fig.~\ref{fig:Mnorm}, it
is not clear whether varying $\mu$ in the range $\{2\,Q,Q/2\}$ gives a realistic
perturbative uncertainty estimate.  Hence we determine the perturbative
uncertainty for case iii) in Tabs.~\ref{tab:FOresults} and \ref{tab:FOO1results}
by varying $\mu$ in the range $\{2\,Q,Q/2\}$ and multiply the result by a factor
of two. The perturbative uncertainties for $\alpha_s(m_Z)$ are a factor of two
larger than in case ii). The central values for $\alpha_s(m_Z)$ in case iii) are
also larger, but are compatible with those in case ii) and \eq{asO1finalcor}
within 1-$\sigma$.

It is interesting to compare our results to those of
Ref.~\cite{Gehrmann:2009eh}, which also performs a fixed order analysis at
${\cal O}(\alpha_s^3)$, and incorporates subtractions based on the dispersive
model.\footnote{On the experimental side, Ref.~\cite{Gehrmann:2009eh} uses only
  the new JADE data from \cite{Pahl:2008uc} and OPAL data. In our analysis the
  new JADE was excluded, but we utilized a larger dataset that includes ALEPH,
  OPAL, L3, DELPHI, AMY, TASSO, and older JADE data. This may have a
  non-negligible impact on the outcome of the comparison.} Here the subtractions
contain logarithms, $\ln(\mu_I/\mu)$, where $\mu_I\sim 2\,{\rm GeV}$ and
$\mu\sim Q$, that are not resummed. From a fit to $M_1$ in thrust they obtained
$\alpha_s(m_Z)\,=\,0.1166\,\pm\,0.0015_{\rm exp}\,\pm\,0.0032_{\rm th}$ where
the first uncertainty is experimental and the second is theoretical. Our
corresponding result is the one in case ii), and the central values and
uncertainties for $\alpha_s(m_Z)$ are fully compatible.  The perturbative
uncertainty they obtain is a factor of $1.6$ larger than ours. It arises from
varying the renormalization scale $\mu\in \{2\,Q,Q/2\}$, the ${\cal
  O}(\alpha_s^2)$ Milan factor $\cal M$ by 20\%, and the infrared scale
$\mu_I=2\pm 1\,{\rm GeV}$ in the dispersive model. In our analysis there is no
precise analog of the Milan factor because our subtractions and Rgap scheme for
$\Omega_1$ fully account for two and three gluon infrared effects up to ${\cal
  O}(\alpha_s^3)$ that are associated to thrust. Other than this, the difference
can be simply attributed to the differences in subtraction schemes which have an
impact on the $\mu$ scale uncertainty. Finally, note that we have implemented
the analytic results of Ref.~\cite{Gehrmann:2009eh} and confirmed their $\mu$
and $\mu_I$ uncertainties.

\section{JADE Datasets}\label{sec:dataset}

\begin{figure}[t!]
\includegraphics[width=1\linewidth]{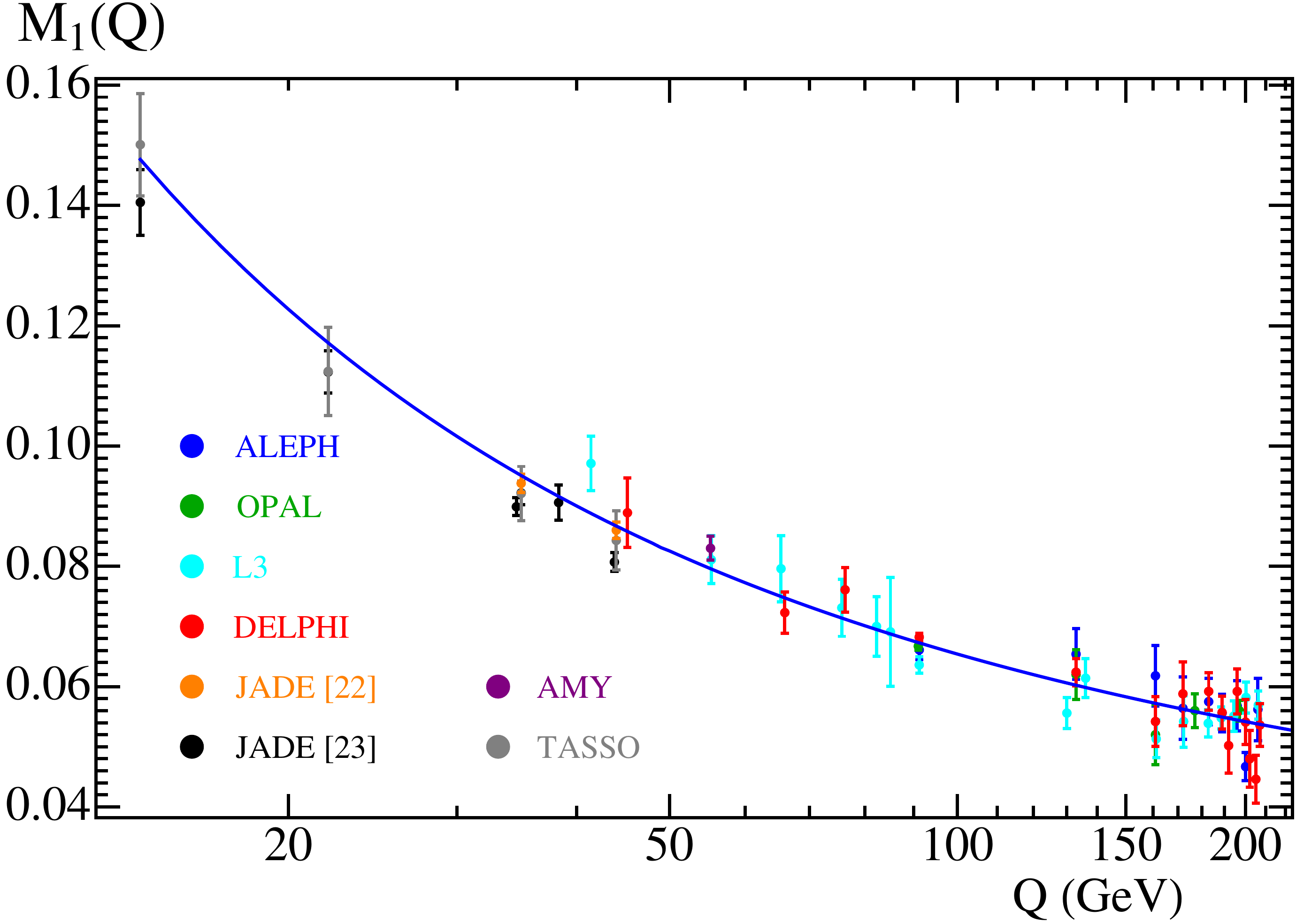}
\caption{Experimental data for the first moment of thrust. The solid line
  corresponds to the result from the first row of Tab.~\ref{tab:FOresults},
  and uses a fixed order code with power corrections in a renormalon-free
  scheme, but no resummation (neither QED nor bottom mass corrections).}
\label{fig:JADE-data}
\end{figure}

As discussed in Sec.~\ref{sec:intro} our global dataset includes thrust
moment results from ALEPH, OPAL, L3, DELPHI, AMY, TASSO and the JADE data from
Ref.~\cite{MovillaFernandez:1997fr}. In this section we discuss the impact on
the results in Secs.~\ref{sec:results} and \ref{sec:FO} of replacing the JADE
data from Ref.~\cite{MovillaFernandez:1997fr} with moment results from an
updated analysis carried out in Ref.~\cite{Pahl:2008uc}, which removes the
contributions from primary $b\bar b$ pair production and provides in addition
measurements at $Q=14$ and $22$~GeV. In 
Fig.~\ref{fig:JADE-data} we show the data for $M_1$, including the JADE results
from Refs.~\cite{MovillaFernandez:1997fr} and~\cite{Pahl:2008uc}.
The most significant difference occurs at $Q=44\,{\rm GeV}$.  Our
analysis will treat these datasets on the same footing without attempting to
account for the effect of removing the $b\bar b$'s.

\begin{figure}[t!]
\includegraphics[width=1\linewidth]{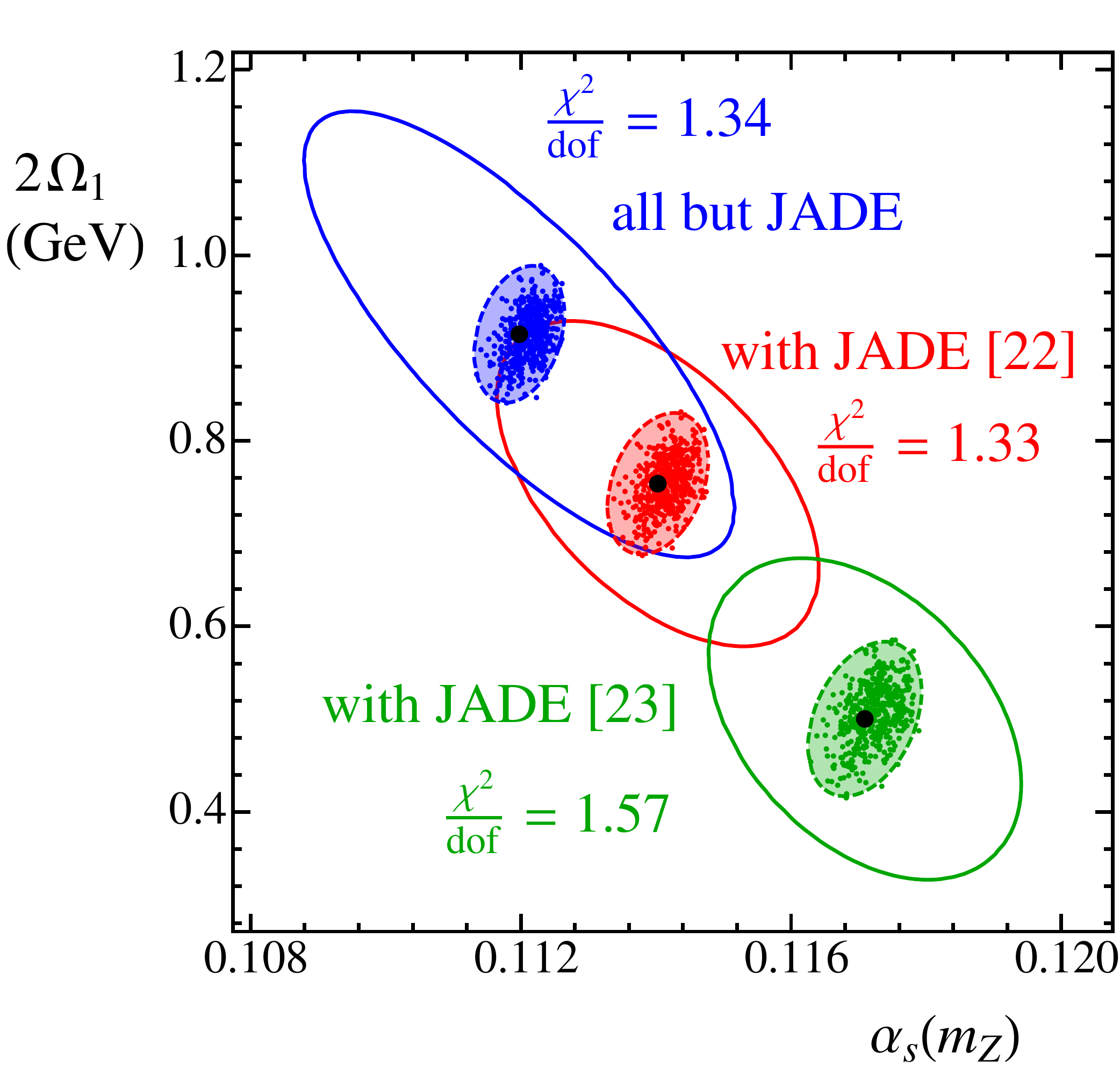}
\caption{Fit results when using ALEPH, DELPHI, OPAL, L3, AMY, TASSO, but no JADE data
  (upper blue ellipse), when also including JADE data from
  Ref.~\cite{MovillaFernandez:1997fr} (red central ellipse) [\,our default data
  set\,], and when instead including the JADE data from Ref.~\cite{Pahl:2008uc}
  (green lower ellipse). The ellipses here correspond to $1$-$\sigma$ for two
  parameters (68\% CL).}
\label{fig:JADE-effect}
\end{figure}

For our analysis here, with theory results at N$^3$LL\,+\,${\cal
  O}(\alpha_s^3)$, we continue to exclude center of mass energies $Q \le 22$~GeV
as in Sec.~\ref{sec:results}.  The dependence of the global fit result on the
data set for $M_1$ is shown in Fig.~\ref{fig:JADE-effect}. Theoretical
uncertainties are analyzed again by the scan method giving the central dots and
three inner ellipses, while the outer three ellipses show the respective
combined 1-$\sigma$ total experimental and theoretical uncertainties. Using all
experimental data but excluding JADE measurements entirely gives the fit result
shown by the upper blue ellipse. This result is compatible at 1-$\sigma$ with
the central red ellipse which shows our default analysis, using the
Ref.~\cite{MovillaFernandez:1997fr} JADE $M_1$ measurements.  Replacing these
two JADE data points by the four $Q>22\,{\rm GeV}$ JADE $M_1$ results from
Ref.~\cite{Pahl:2008uc} yields the lower green ellipse (whose center is $\simeq
1.5$-$\sigma$ from the central ellipse). For this fit the $\chi^2/{\rm dof}$
increases from $1.33$ to $1.52$ demonstrating that there is less compatibility
between the data. For this reason, together with the concern about the impact of
removing primary $b\bar b$ events with MC simulations, we have used only JADE
data from Ref.~\cite{MovillaFernandez:1997fr} in our main analysis.

A similar pattern is observed using the fixed order fits of $M_1$ discussed in
Sec.~\ref{sec:FO}. In this case it is also straightforward to include the
$Q=14,22\,{\rm GeV}$ JADE data from Ref.~\cite{Pahl:2008uc}. If these two points
are added to our default dataset (which contains $Q=35$ and $45$~GeV 
as the lowest $Q$ results for $M_1$) then we find $\alpha_s(m_Z)=0.1155\pm0.0012$
and $\Omega_1=0.361\pm0.035\,{\rm GeV}$ with $\chi^2/{\rm dof}=1.3$. This is
compatible at 1-$\sigma$ with our final pure QCD result in
Tab.~\ref{tab:results}. If we include the entire set of JADE data from
Ref.~\cite{Pahl:2008uc} instead of those from
Ref.~\cite{MovillaFernandez:1997fr}
then we find $\alpha_s(m_Z)=0.1166\pm 0.0012$ and
$\Omega_1=0.306\pm 0.033\,{\rm GeV}$ with $\chi^2/{\rm dof}=1.6$, very similar
to the values observed for the green lower ellipse in
Fig.~\ref{fig:JADE-effect}. Hence, overall the fixed order analysis does not
change the comparison of fits with the two different JADE datasets.


\section{Higher Moment Analysis}
\label{sec:comparison}

In this section we consider higher moments, $M_{n\ge 2}$, which have been
measured experimentally up to $n=5$. From \eq{MnOPE} we see that these moments
have power corrections $\propto 1/Q^k$ for $k\ge 1$. Since for the perturbative
moments we have $\hat M_{n}/\hat M_{n+1} \simeq 4$--$9$, we estimate that the
$1/Q^2$ power corrections are suppressed by $9\Lambda_{\rm QCD}/Q$ which varies
from $1/8$ to $1/44$ for the $Q$-values in our dataset, $Q\ge 35\,{\rm GeV}$.
Hence, for the analysis in this section we can safely drop the $1/Q^2$ and
higher power corrections and use the form
\begin{align}
M_n &= \hat M_n+ \frac{2\,n\,\Omega_1}{Q}\,\hat M_{n-1} \,.
\end{align}

\begin{figure}[t!]
\includegraphics[width=1\linewidth]{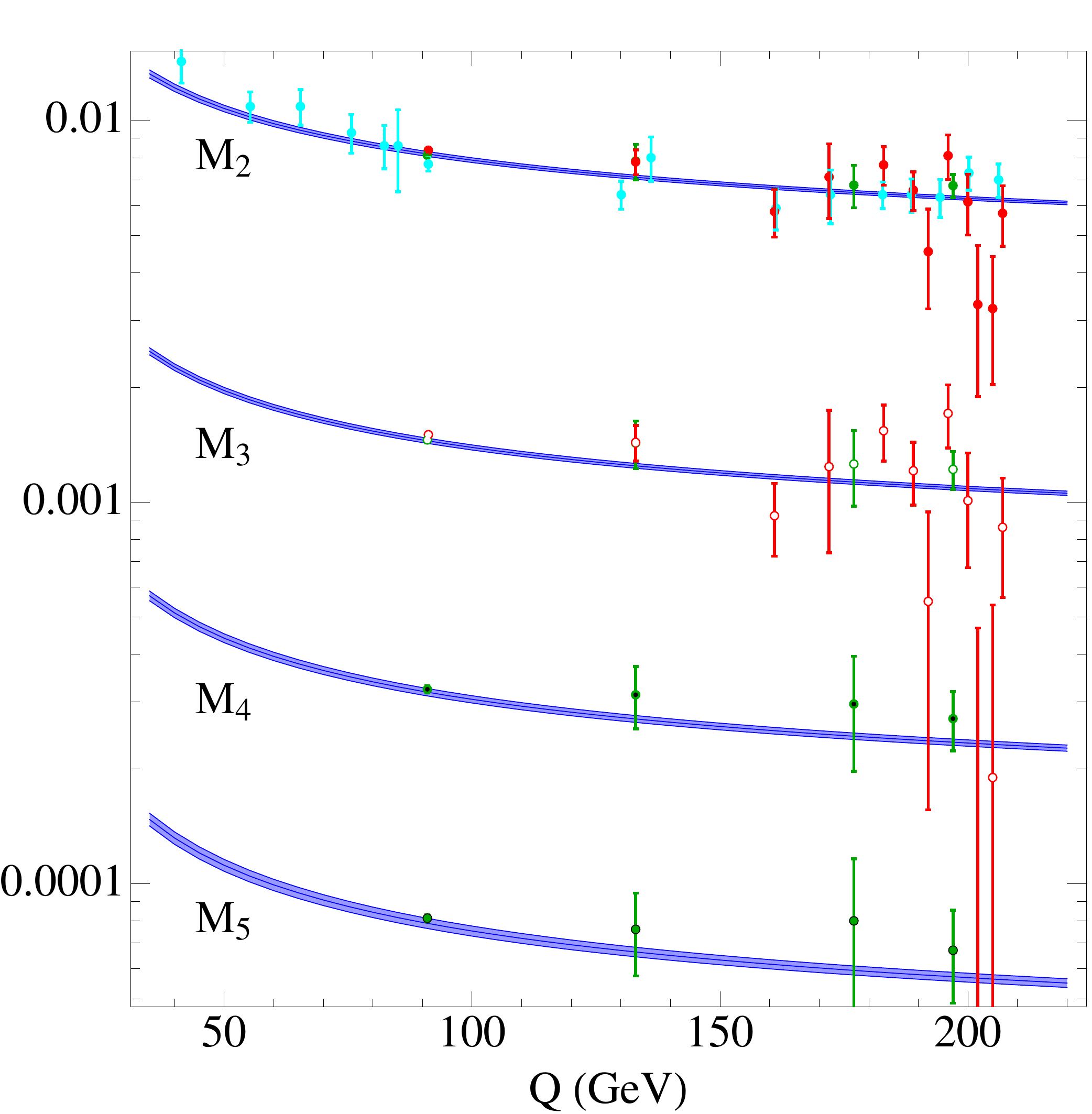}
\caption{Predictions for the higher moments $M_2$, $M_3$, $M_4$, $M_5$ using the
  best fit values from \eq{asO1finalcor}, and our full N$^3$LL\,+\,${\cal
    O}(\alpha_s^3)$ code in the Rgap scheme, but with QED and mass effects
  turned off. The central points use different symbols for different moments. }
\label{fig:mnlogplot}
\end{figure}

By using our fit results for $\alpha_s(m_Z)$ and $\Omega_1$ from
\eq{asO1finalcor} we can directly make predictions for the moments
$M_{2,3,4,5}$. This tests how well the theory does at calculating the
perturbative contributions $\hat M_{2,3,4,5}$. The results for these moments are
shown in Fig.~\ref{fig:mnlogplot} and correspond to $\chi^2/{\rm dof} = 1.3,
2.5, 0.8, 1.1$ for $n=2,3,4,5$ respectively, indicating that our formalism does
quite well at reproducing these moments.  The larger $\chi^2/{\rm dof}$ for $n=3$
is related to a quite significant spread in the experimental data for this moment at
$Q\gtrsim 190\,{\rm GeV}$. Note that we also see that the relation
$M_{n}/M_{n+1} \simeq 4$--$9$ is satisfied by the experimental moments.

\begin{figure}[t!]
\includegraphics[width=0.7\linewidth]{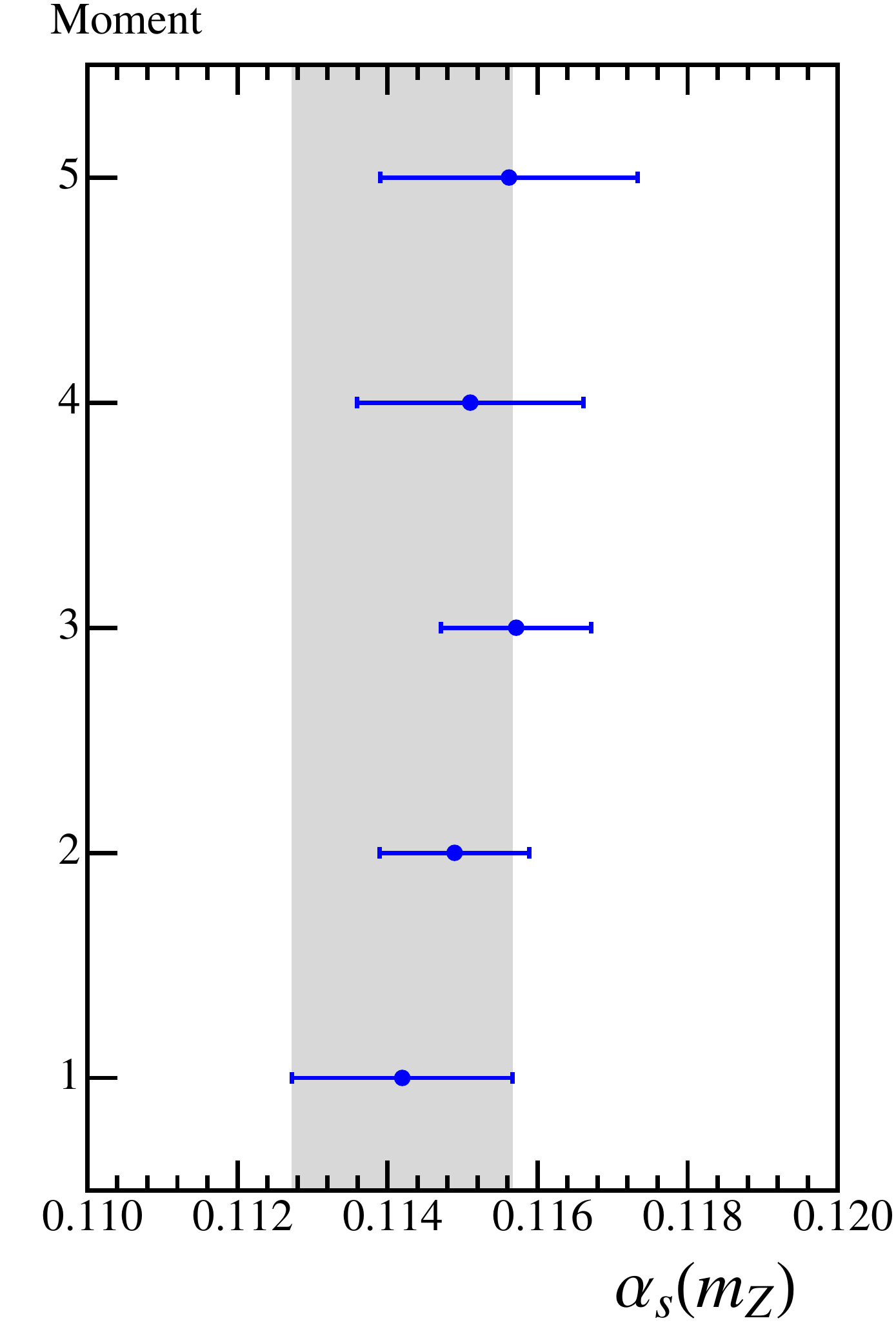}
\caption{One-parameter fits for $\alpha_s(m_Z)$ to the first five moments. We
  use our full set up with power corrections and renormalon subtractions, but with
  QED and mass corrections turned off. The value of $\Omega_1$ is fixed from
  \eq{asO1finalcor}. The error bars include theoretical and experimental errors
  added in quadrature (not including uncertainty in $\Omega_1$).}
\label{fig:higher-fits}
\end{figure}

An alternate way to test the higher moments is to perform a fit to this data.
Since we have excluded the new JADE data in Ref.~\cite{Pahl:2008uc}, we do not
have a significant dataset at smaller $Q$ values for the higher moments.  With
our higher moment dataset the degeneracy between $\alpha_s(m_Z)$ and $\Omega_1$
is not broken for $n\ge 2$, and one finds very large experimental errors for a two-parameter
fit already at $n=2$. However we can still fit for $\alpha_s(m_Z)$ from data for
each individual $M_{n\ge 2}$ by fixing the value of $\Omega_1$ to the best fit
value in \eq{asO1finalcor} from our fit to $M_1$.  For this exercise we use our
full N$^3$LL\,+\,${\cal O}(\alpha_s^3)$ code, but with QED and mass effects turned
off.  The outcome is shown in Fig.~\ref{fig:higher-fits} and
Tab.~\ref{tab:higher-fits}.  We find only a little dependence of $\alpha_s$ on $n$,
and all values are compatible with the fit to the first moment within less than
1-$\sigma$. This again confirms that our value for $\Omega_1$ and perturbative
predictions for $\hat M_{n\ge 2}$ are consistent with the higher moment data.

\begin{table}[t!]
\begin{centering}
\begin{tabular}{ccccc}
\vspace{0.1cm} $n$ \hspace{0.1cm} 
 &\hspace{0.1cm} $\alpha_{s}(m_{Z})$ \hspace{0.1cm}
 &\hspace{0.1cm} $\Delta_{{\rm th}}[\alpha_{s}]$ \hspace{0.1cm}
 &\hspace{0.1cm} $\Delta_{{\rm exp}}[\alpha_{s}]$ \hspace{0.1cm}
 &\hspace{0.1cm} ${\chi^{2}}/{{\rm dof}}$ \hspace{0.1cm} \\
\hline 
$2$ & $0.1149$ & $0.0009$ & $0.0005$ & $1.24$\\
$3$ & $0.1157$ & $0.0009$ & $0.0005$ & $1.87$\\
$4$ & $0.1151$ & $0.0011$ & $0.0010$ & $0.39$\\
$5$ & $0.1156$ & $0.0015$ & $0.0010$ & $0.23$\\
\end{tabular}
\end{centering}
\caption{Numerical results for $\alpha_s$ from one-parameter fits to
the $M_{n}$ moments.
The second column gives
the central values for $\alpha_s(m_Z)$, the
third and fourth show the theoretical and experimental errors, respectively. 
Since $\Omega_1$ was fixed for this analysis we do
not quote a hadronization error.
\label{tab:higher-fits}}
\end{table}

In Ref.~\cite{Gehrmann:2009eh} a two-parameter fit to higher thrust moments was
carried out using OPAL data and the latest low energy JADE data. For $n=2$ to
$n=5$ the results increase linearly from $\alpha_s(m_Z)=0.1202\pm (0.0018)_{\rm
  exp}\pm (0.0046)_{\rm th}$ to $\alpha_s(m_Z)=0.1294\pm (0.0027)_{\rm exp}\pm
(0.0070)_{\rm th}$ respectively, and the weighted average for the first five
moments of thrust is $\alpha_s(m_Z)\,=\,0.1208\,\pm\,0.0018_{\rm exp}\,\pm\,
0.0045_{\rm th}$. The results are fully compatible within the uncertainties, and
there is an indication of a trend towards larger $\alpha_s(m_Z)$ extracted from
higher moments.  In our analysis we do not observe this trend, but our results
should not be directly compared since we have only performed a one parameter fit.
After further averaging over results obtained from event shapes other than
thrust Ref.~\cite{Gehrmann:2009eh} obtained as their final result
$\alpha_s(m_Z)\,=\,0.1153\,\pm\,0.0017_{\rm exp}\,\pm\,0.0023_{\rm th}$. This is
again perfectly compatible with our result in \eq{asO1finalcor}.

\section{Higher power corrections from Cumulant Moments}\label{sec:power-data}

In this section we use cumulant moments as defined in
Eq.~(\ref{eq:OPE-subleading}) to discuss the presence of higher power
corrections and their constraints from experimental data.  There are two types
of power corrections that are relevant for the cumulants, those defined
rigorously by QCD matrix elements which come from the leading thrust
factorization theorem, $\Omega_n^\prime$, and those from our simple
parameterization of higher order power corrections in
Eq.~(\ref{eq:sublead_fact_thm}), $\Omega_{n,j\ge 1}$. For the latter a
systematic matching onto QCD matrix elements has not been carried out and the
corresponding perturbative coefficients have not been determined.

For the second cumulant $M_2'$ both types of power correction contribute to the
leading $1/Q^2$ term in the combination
\begin{align} \label{eq:tO2}
  \tilde \Omega_2^\prime &= \Omega_2^\prime + {\overline M}_{1,1}\, \Omega_{1,1}
  \,.
\end{align}
Without a calculation of the perturbative coefficient ${\overline M}_{1,1}$ we
cannot argue that either one dominates, and hence we keep both of them.  In
terms of this parameter the OPE with its leading power correction for the second
cumulant becomes simply
\begin{align}  \label{eq:M2pOPE}
  M_2' = \hat M_2' + \frac{4\,\tilde\Omega_2^\prime}{Q^2} \,,
\end{align}
where $\hat M_2'$ is computed from our leading order factorization theorem, see
\eq{Mnhat}. For the third cumulant $M_3'$ the power correction from the leading
thrust factorization theorem is $1/Q^3$, while that from the subleading
factorization theorem is $1/Q^2$, so
\begin{align} \label{eq:M3pOPE}
  M_3' = \hat M_3'
  + \frac{6\,{\overline M}_{2,1}\,\Omega_{1,1}}{Q^2}  
  + \frac{8\,\Omega_3^\prime}{Q^3} \,.
\end{align}
where we keep both of these power corrections.  

For our analysis we assume
that the perturbative coefficients ${\overline M}_{1,1}$ and ${\overline
  M}_{2,1}$ get contributions at tree-level, and hence that their logarithmic
dependence on $Q$ is $\alpha_s$-suppressed.  Thus for fits to $M_2'$ and $M_3'$ we 
consider the three parameters $\tilde\Omega_2^\prime$, ${\overline
  M}_{2,1}\,\Omega_{1,1}$, and $\Omega_3^\prime$.  Our theoretical
expectations are that $(\Omega_n^\prime)^{1/n} \sim \Lambda_{\rm QCD}$ and
$(\Omega_{1,1})^{1/2} \sim (\Omega_n^\prime)^{1/n}$.

\begin{figure}[t!]
\includegraphics[width=1\linewidth]{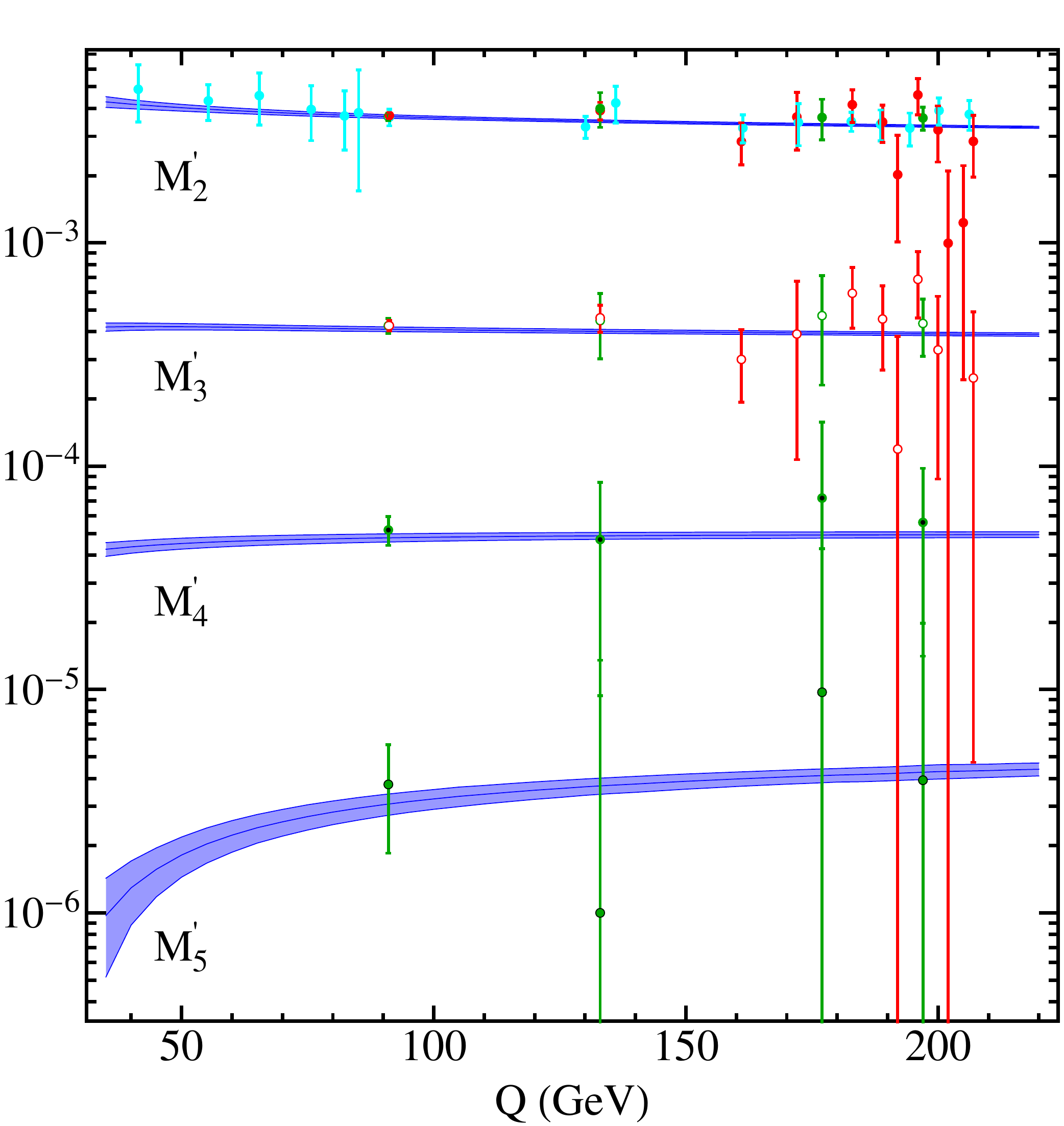}
\caption{Prediction of cumulants using our best-fit values for
$\alpha_s(m_Z)$ and $\Omega_1$ from the fit to
the first thrust moment. The band includes only the theoretical uncertainty from
the random scan. The theory prediction includes
QED and mass corrections, in contrast to our numerical analysis which has no
QED and $b$-mass effects and uses our default model, which translates into the
following values for higher nonperturbative
power corrections: $\Omega_2^\prime=\Omega_1^2/4$,
$\Omega_3^\prime=\Omega_1^3/8$, $\Omega_4^\prime=3\,\Omega_1^4/32$,
$\Omega_5^\prime=3\,\Omega_1^5/32$.}
\label{fig:cumulant-moments-prediction}
\end{figure}
Since most of the experimental collaborations provide measurements only for
moments we computed the cumulants using Eq.~(\ref{eq:variance-skewness}). To
propagate the errors to the $n$-th cumulant one needs the correlations between
the first $n$ moments, both statistical and systematical.  Following
experimental procedures we estimate the statistical correlation matrix from
Monte Carlo simulations.  These matrices are provided in Ref.~\cite{Pahl:thesis}
for $Q=14,\,91.3,\,206.6\,$GeV.\footnote{We thank Christoph Pahl for providing
  details on the use of correlation matrices for moments.} The computation of
these matrices does not depend on the simulation of the detector and hence can
be a priory employed on the data provided by any experimental collaboration.  It
was found that statistical correlation matrices depend very mildly on the center
of mass energy, and our approach is to use the matrix computed at $14\,$GeV for
$Q<60\,$GeV, the one computed at $91.3$ for $60\,{\rm GeV}\leq Q<120\,{\rm GeV}$
and the one at $206.6\,$GeV for $Q\geq 120\,$GeV.  The systematic correlation
matrix for the moments is estimated using the minimal overlap model based on the
systematic uncertainties, and then converted to uncertainties for the cumulants.
We use this method even for the few cases in which experimental collaborations
provide uncertainties for the cumulants directly, since we want to treat all
data on the same footing. In these cases we have checked that the results are
very similar.

To some extent the prescription we employ lies in between two extreme
situations: a) moments are completely uncorrelated, and b) cumulants are
completely uncorrelated. Situation a) corresponds to the naive assumption that
the moments are independent. Situation b) is motivated by considering that
properties like the location of the peak of the distribution ($\sim M_1$), the
width of the peak ($\sim M_2'$), etc.\ are independent pieces of information.
By assuming moments are 
uncorrelated one overestimates the errors of the cumulants. This would translate
into larger experimental errors for our fit results and very small $\chi^2/{\rm
  dof}$.  Assuming that cumulants are uncorrelated induces very strong positive
correlations between moments, which then leads to small uncertainties for the
cumulants, especially for the variance, and larger $\chi^2/{\rm dof}$ values.
With the adopted prescription we use one finds a weaker positive correlation
among moments, which translates into a situation between these two
extremes.\footnote{One might also construct the correlation matrices using the
  statistical and systematic errors from the thrust distributions themselves.
  Bins in distributions are statistically independent and systematic
  correlations are estimated using the minimal overlap model. Unfortunately this
  can introduce biases, and we thank Christoph Pahl for clarifying this point.}

\begin{table}[t!]
\begin{tabular}{lcccc}
 & \phantom{00}central\phantom{00}& \phantom{00}$\Delta_{{\rm th}}$ \phantom{00}&
\phantom{00}$\Delta_{{\rm exp}}$ \phantom{00}& $\frac{\chi^{2}}{{\rm
dof}}$\\
\hline
$(\tilde \Omega^\prime_2)^{1/2}$ & $0.74\phantom{0}$ & $0.09\phantom{0}$ &
$0.11\phantom{0}$ & $0.72$
\\
$(\Theta_2)^{1/2}$ & $1.21\phantom{0}$ & $0.10\phantom{0}$ & $0.22\phantom{0}$ &\multirow{2}{*}{$0.93$}
\\
$(\Theta_3)^{1/3}$ & $-2.61\phantom{0}$ & $0.15\phantom{0}$ & $1.51\phantom{0}$
& \\
\end{tabular}
\caption{Determination of power corrections from fits to $M_2^\prime$ and
  $M_3^\prime$. All values in the table are in GeV. 
  Columns two to four correspond to central value, theoretical uncertainty, and
  experimental uncertainty, respectively (the latter includes both statistical and 
  systematic errors added in quadrature).  The values displayed correspond to the linear
  combinations in Eq.~(\ref{eq:linear-combinations}), which for $M_3^\prime$ 
  diagonalize the experimental error matrix. 
  \label{tab:omega-fits}}
\end{table}
\begin{figure*}[t!]
\includegraphics[width=0.48\textwidth]{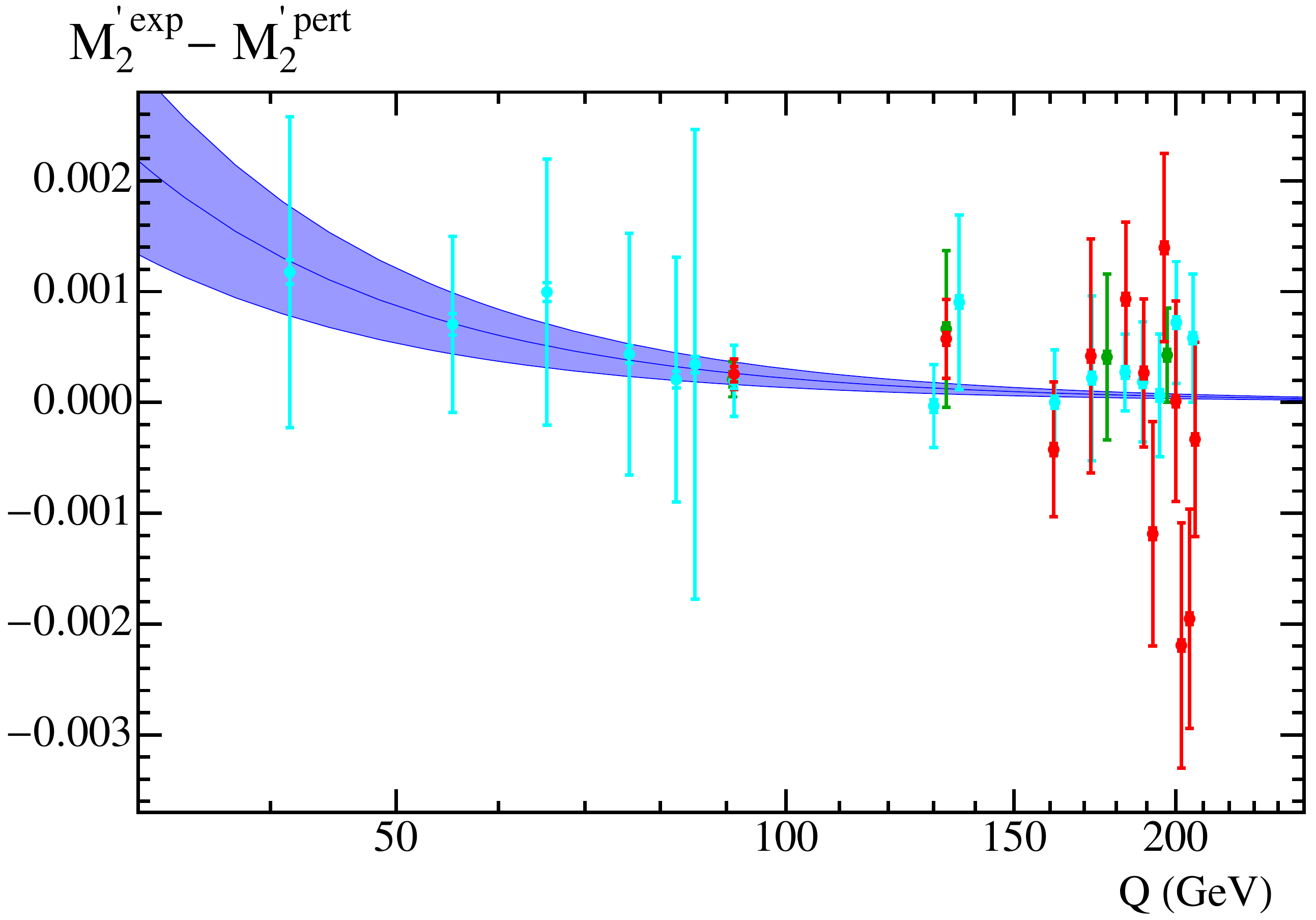}
\includegraphics[width=0.48\textwidth]{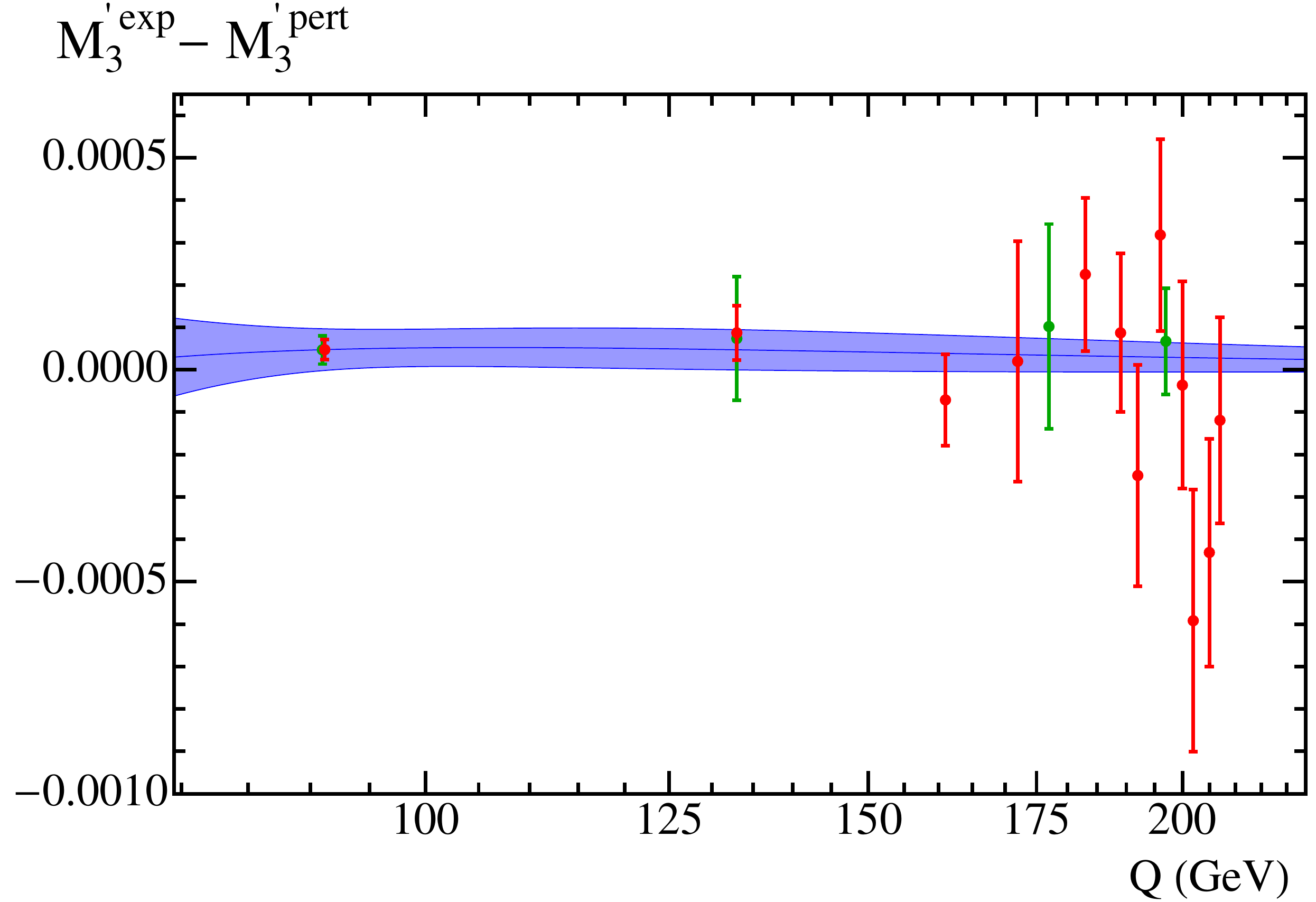}
\includegraphics[width=0.48\textwidth]{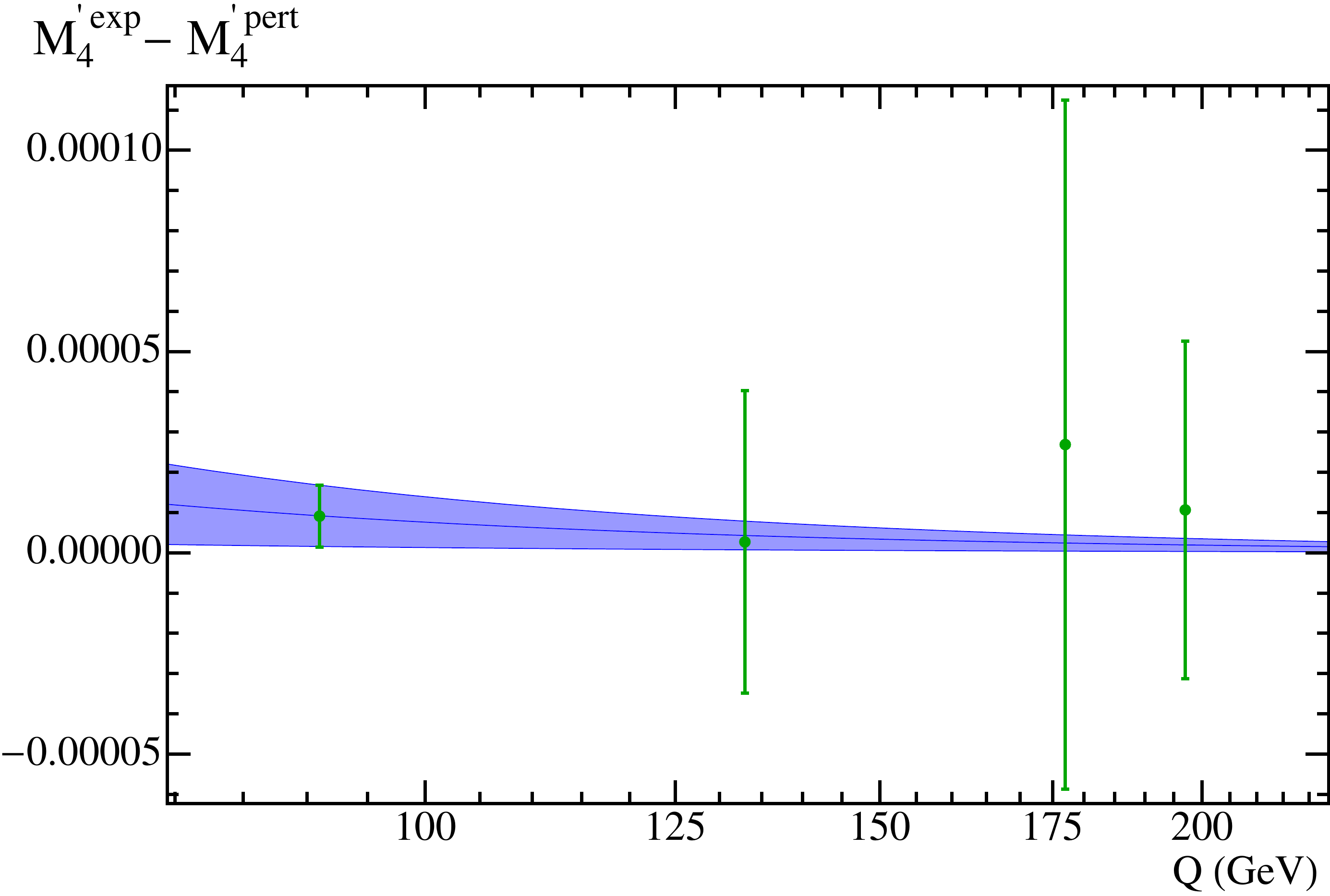}
\includegraphics[width=0.48\textwidth]{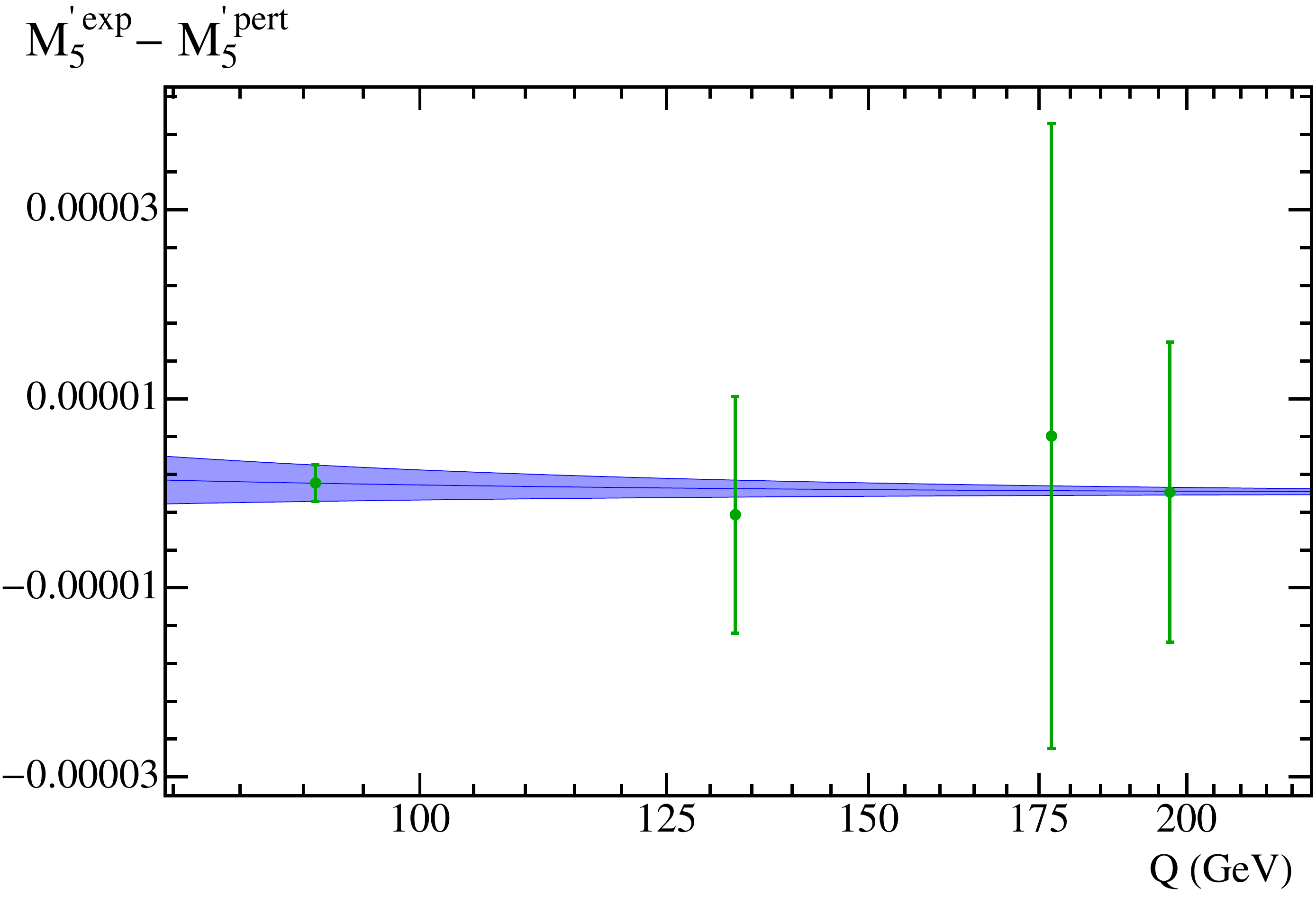}
\caption{Determination of power corrections from fits to data. On the vertical
  axes we display the $n$-th experimental cumulant with the perturbative part
  subtracted $M_n^\prime-{\hat M}_n^\prime$.  The error bars shown are
  experimental (statistical and systematic combined) added in quadrature with
  perturbative errors from the random scan over the profile parameters.  The
  top-left panel shows the fit to $\tilde\Omega_2^\prime/Q^2$, and the top-right
  panel shows the fit to ${\overline M}_{2,1}\,\Omega_{1,1}/Q^2$ and
  $\Omega_3^\prime/Q^3$ through the linear combinations in $\Theta_{2,3}$.  The
  bottom two panels for $n=4,5$ show a simple fit to $\overline
  M_{3,1}\Omega_{1,1}$ and $\overline M_{4,1}\Omega_{1,1}$ taking
  $\Omega_4^\prime=\Omega_5^\prime=0$.}
\label{fig:omega-plots}
\end{figure*}

For our analysis we use our highest order code as described in \sec{results},
and take the value $\alpha_s(m_Z)=0.1142$
obtained in our fit to the first moment data with this code (see
Tab.~\ref{tab:results}). Since we are analyzing 
cumulants $M_{n\ge 2}^\prime$ the value of $\Omega_1$ is not required,
and there is no distinction between having this parameter in $\msbar$ or the
Rgap scheme.  Hence in order to fit for higher power corrections we use our
purely perturbative code in the $\msbar$ scheme.
Thus all of the power correction parameters extracted in this section are in the
$\msbar$ scheme. The perturbative error is estimated as in
\sec{results}, by a 500 point scan of theory parameters (see
App.~\ref{app:scan}).

Before we fit for the higher power corrections, we will check how well our
factorization theorem predicts the experimental cumulants using a simple
exponential model for the nonperturbative soft function (the model with only one
coefficient $c_0=1$ from Refs.~\cite{Abbate:2010xh,Ligeti:2008ac}). This model
has higher power corrections that are determined by its one parameter
$\Omega_1$: $\Omega_2^\prime=\Omega_1^2/4$, $\Omega_3^\prime=\Omega_1^3/8$,
$\Omega_4^\prime=3\,\Omega_1^4/32$, $\Omega_5^\prime=3\,\Omega_1^5/32$.  Results
are shown in Fig.~\ref{fig:cumulant-moments-prediction}, where good agreement
between theory and data is observed. 

For the $M_n'$ in Fig.~\ref{fig:cumulant-moments-prediction} we also observe
that $M_{n+1}^\prime /M_n^\prime \sim 1/10$, so the $(n+1)$-th order cumulant is
generically one order of magnitude smaller than the $n$-th order cumulant.

Next we will fit for the power correction parameters $\tilde\Omega_2^\prime$,
${\overline M}_{2,1}\,\Omega_{1,1}$, and $\Omega_3^\prime$. For this analysis
we neglect QED and $b$-mass effects. To facilitate this we
consider the difference between the experimental cumulants $M_n'$ and the
perturbative theoretical cumulants $\hat M_n'$, namely $M_2'-\hat M_2'$ and
$M_3'-\hat M_3'$. From \eqs{M2pOPE}{M3pOPE} these differences are determined
entirely by the power correction parameters we wish to fit. The results are
shown in Tab.~\ref{tab:omega-fits} and the upper two panels of
Fig.~\ref{fig:omega-plots}. From the $M_2'-\hat M_2'$ fit a fairly precise
result is obtained for $(\tilde \Omega_2^\prime)^{1/2}$. Its central value of
$740\,{\rm MeV}$ is compatible with $\sim 2\Lambda_{\rm QCD}$, and hence agrees
with naive dimensional analysis.  Interestingly, we have checked that including
a constant and $1/Q$ term in the second cumulant fit one finds that their
coefficients are compatible with zero, in support of the theoretically expected
$1/Q^2$-dependence.

For the fit to $M_3'-\hat M_3'$ there is a strong correlation between
$\Omega_3^\prime$ and ${\overline M}_{2,1}\,\Omega_{1,1}$ even though they occur
at different orders in $1/Q$. Since the $\chi^2$ is quadratic in these two
parameters we can determine the linear combinations that exactly diagonalize
their correlation matrix:
\begin{align}\label{eq:linear-combinations}
  \Theta_2 & \equiv \bigg[\frac{6\,{\overline
      M}_{2,1}}{0.07}\bigg] \frac{\Omega_{1,1}}{4} + (0.3105\,{\rm GeV}^{-1})\,
  \Omega_3^\prime
  \,,\\
  \Theta_3 & \equiv \Omega_3^\prime 
   -(0.3105\,{\rm GeV})\, \bigg[\frac{6\,{\overline
      M}_{2,1}}{0.07}\bigg] \frac{\Omega_{1,1}}{4}
  \,.\nonumber
\end{align}
Note that these combinations arise solely from experimental data. We have
presented the coefficients of these combinations grouping together a factor of
$(6{\overline M}_{2,1}/0.07)$, which is close to unity if $6{\overline
  M}_{2,1}\simeq \hat M_1$. The results in Tab.~\ref{tab:omega-fits} exhibit a
reasonable uncertainty for $\Theta_2$, but a large uncertainty for $\Theta_3$.
Hence, at this time it is not possible to determine the original parameters
$\Omega_3^\prime$ and ${\overline M}_{2,1}\,\Omega_{1,1}$ independently. As in
the previous case, the fit does not exhibit any evidence for a $1/Q$ correction,
confirming the theoretical prediction for this cumulant.

In Fig.~\ref{fig:omega-plots} we also show results for cumulant differences
$M_n^\prime-\hat M_n^{\prime}$ versus $Q$ for $n=4$ and $n=5$. In all cases
$n=2,3,4,5$ the perturbative cumulants $\hat M_n^{\prime}$ are the largest
component of the cumulant moments $M_n^\prime$, as can be verified by the
reduction of the values by a factor of $2$--$3$ in Fig.~\ref{fig:omega-plots}
compared to the values in Fig.~\ref{fig:cumulant-moments-prediction}. We also
observe an order of magnitude suppression between the $(n+1)$'th and $n$'th
terms, $(M_{n+1}'-\hat M_{n+1}')/(M_n'-\hat M_n')\sim 1/10$. For $n=4,5$ the OPE
formula in \eq{OPE-subleading} involves both $2^n\Omega_n^\prime/Q^n$ terms and
terms with non-trivial perturbative coefficients: $(2\,n\,\overline M_{n-1,1}
\Omega_{1,1})/Q^2+\ldots$ (where here the ellipses are terms at $1/Q^3$ and
beyond). If the former dominated we would expect a suppression by
$2\,\Lambda_{\rm QCD}/Q$ for the $(n+1)$'th versus $n$'th term.  The observed
suppression by $1/10$ is less strong and is instead consistent with domination
by the $1/Q^2$ power correction terms in the $n=4,5$ cumulant differences. This
would imply $[(n+1) \overline M_{n,1}] / [n \overline M_{n-1,1}]\sim 1/10$ and
could in principle be verified by an explicit computation of these coefficients.
In Fig.~\ref{fig:omega-plots} we show fits to a $1/Q^2$ power correction, which
are essentially dominated by the lowest energy point at the Z-pole. The results
are $\sqrt{8\,\overline{M}_{3,1}\,\Omega_{1,1}}=0.20\pm0.08$ from fits to
$M_4^\prime$ and $\sqrt{10\,\overline{M}_{4,1}\,\Omega_{1,1}}=0.07\pm0.06$ from
fits to $M_5^\prime$. These values agree with our expectation of the $\sim 1/10$
suppression between the two $\overline M_{n,1}$ perturbative coefficients.

In this section we have determined the $1/Q^2$ power correction parameter
$\tilde \Omega_2^\prime$ with $25\%$ accuracy, and find it is $3.8\,\sigma$
different from zero. For the higher moments there are important contributions
from a $\Omega_{1,1}/Q^2$ power correction, which appears to even dominate for
$n\ge 4$.  Clearly experimental data supports the pattern expected from the OPE relation in
Eq.~(\ref{eq:OPE-subleading}).

\section{Conclusions}
\label{eq:conclusions}

In this work we have used a full $\tau$-distribution factorization formula
developed by the authors in a previous publication~\cite{Abbate:2010xh} to study
moments and cumulant moments (cumulants) of the thrust distribution.  Perturbatively it
incorporates $\mathcal{O}(\alpha_s^3)$ matrix elements and nonsingular terms, a
resummation of large logarithms, $\ln^k\tau$, to $\ntll$ accuracy, and the
leading QED and bottom mass corrections. It also describes the dominant
nonperturbative corrections, is free of the leading renormalon ambiguity, and
sums up large logs appearing in perturbative renormalon subtractions.

Theoretically there are no large logs in the perturbative expression of the
thrust moments, and when normalized in the same way the perturbative result from
the full $\tau$ code with resummation agrees very well with the fixed order
results. Nevertheless, when the code is properly self normalized it
significantly improves the order-by-order perturbative convergence towards the
${\cal O}(\alpha_s^3)$ result. In particular, the results remain within the
perturbative error band of the previous order, in contrast to what is observed
using fixed order expressions. This lends support to the theoretical uncertainty
analysis from the code with resummation.

From fits to the first moment of the thrust distribution, $M_1$, we find the
results for $\alpha_s(m_Z)$ and the leading power correction parameter
$\Omega_1$ given in \eq{asO1finalcor}. They are in nice agreement with values
from the fit to the tail of the thrust distribution in
Ref.~\cite{Abbate:2010xh}. The moment results have larger experimental
uncertainties, and these dominate over theoretical uncertainties, in contrast
with the situation in the tail region analysis of Ref.~\cite{Abbate:2010xh}.
Repeating the $M_1$ fit using a fixed order code with no $\ln\tau$ resummation,
but still retaining the summation of large logs in the perturbative renormalon
subtractions, yields fully compatible results for $\alpha_s(m_Z)$ and
$\Omega_1$.

Using a Fourier space operator product expansion we have parameterized higher
order power corrections which are beyond the leading factorization formula, and
analyzed the OPE both for moments $M_n$ and cumulants $M_n'$. In
the moments $M_n$ the $\Omega_1/Q$ power correction from the leading
factorization theorem enters with a perturbative suppression in its coefficient,
and dominates numerically over higher $1/Q$ corrections. In contrast, the
cumulants $M_{n\ge 2}'$ depend on higher order cumulant power corrections
$\Omega_n'/Q^n$ from the leading factorization theorem, and are independent of
$\Omega_1/Q$, \ldots, $\Omega^\prime_{n-1}/Q^{n-1}$.  Data on these cumulants
appear to indicate that they receive important contributions from a
$1/Q^2$ power correction that enters at a level beyond the leading thrust
factorization theorem. Thus the OPE reveals that cumulants are appealing
quantities for exploring subleading power corrections.  We performed a fit to
the second cumulant and determined a non-vanishing $\tilde
\Omega_2^\prime/Q^2$ power correction with a precision of $25\%$.

It would be interesting to extend the analysis performed here, based on OPE
formulas related to factorization theorems, to other event shape moments and
cumulants.  Examples of interest include the heavy jet mass event
shape~\cite{Clavelli:1979md,Chandramohan:1980ry,Clavelli:1981yh,Catani:1991bd,Chien:2010kc},
angularities~\cite{Berger:2003iw,Hornig:2009vb}, as well as more exclusive event shapes like jet
broadening~\cite{Catani:1992jc,Dokshitzer:1998kz,Chiu:2011qc,Becher:2011pf,Chiu:2012ir}. Other event shape
moments were considered at ${\cal O}(\alpha_s^3)$ in Ref.~\cite{Gehrmann:2009eh} in the context
of the dispersive model for the $1/Q$ power corrections.

\begin{acknowledgments}
  
  This work was supported in part by the Office of Nuclear Physics of the U.S.\ 
  Department of Energy under the Contracts DE-FG02-94ER40818, DE-FG02-06ER41449,
  the European Community's Marie-Curie Research Networks under contract
  MRTN-CT-2006-035482 (FLAVIAnet), MRTN-CT-2006-035505 (HEPTOOLS) and
  PITN-GA-2010-264564 (LHCphenOnet), and the U.S.\ National Science Foundation,
  grant NSF-PHY-0969510 (LHC Theory Initiative).  VM has been supported in part
  by a Marie Curie Fellowship under contract PIOF-GA-2009-251174, and IS in part
  by a Friedrich Wilhelm Bessel award from the Alexander von Humboldt
  foundation. VM, IS and AHH are also supported in part by MISTI global seed
  funds.  We thank C.~Pahl for useful discussions concerning the treatment of
  JADE experimental data.  MF thanks S.~Fleming for discussions.

\end{acknowledgments}

\appendix

\section{Theory parameter scan}
\label{app:scan}

\begin{table}[t] 
\begin{tabular}{ccc}
   parameter\ & \ default value\ & \ range of values \ \\
  \hline 
  $\mu_0$ & $2$\,{\rm GeV} & $1.5$ to $2.5$\, {\rm GeV}\\ 
  $n_1$ & $5$ & $2$ to $8$\\
  $t_2$ & $0.25$ & $0.20$ to $0.30$\\
  $e_J$ & $0$ & $-1$, $0$, $1$\\
  $e_H$ & $1$ & $0.5$ to $2.0$\\
  $n_s$ & $0$ & $-1$, $0$, $1$\\
\hline
  $\Gamma^{\rm cusp}_3$ & $1553.06$ & $-1553.06$ to $+4659.18$ \\
  $j_3$ & $0$ & $-3000$ to $+3000$ \\
  $s_3$ & $0$ & $-500$ to $+500$ \\
\hline
  $\epsilon_2$ & $0$ & $-1$, $0$, $1$ \\
  $\epsilon_3$ & $0$ & $-1$, $0$, $1$ \\ 
\end{tabular}
\caption{Theory parameters relevant for estimating the theory uncertainty, their
default values and range of values used for the theory scan during the fit
procedure.}
\label{tab:theoryerr}
\end{table}

\begin{figure*}[t]
\vspace{0pt}
\includegraphics[width=1\linewidth]{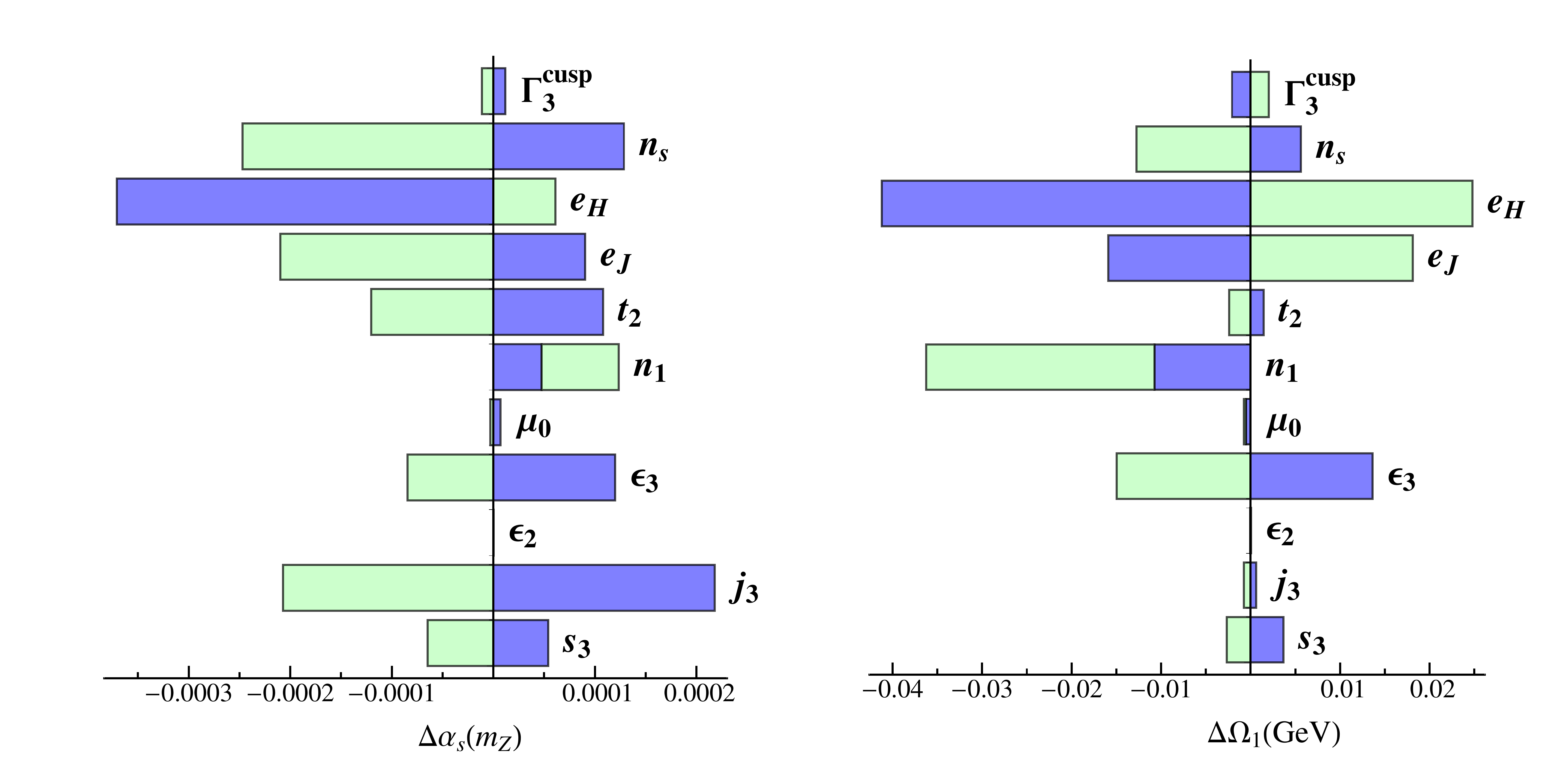}
\caption{Impact on parameters of the $M_1$ fit from variations of the best-fit values for $\alpha_s(m_Z)$ and
  $\Omega_1$ values in the ranges given in Table~\ref{tab:theoryerr}. The dark
  shaded blue regions represent values of the parameters larger than their
  default values, the light shaded green regions where the parameters are
  smaller than their default values.  }
\label{fig:barplots}
\end{figure*}
In this Appendix we describe the method we use to estimate uncertainties in our analysis.
We will briefly review the profile functions and the theoretical parameters which
determine the theory uncertainty. We will also describe the scan over those parameters
and the effects they have on the fit results.

The profile functions used in Ref.~\cite{Abbate:2010xh}, to which we refer for a
more extensive description, are $\tau$-dependent factorization scales which
allow us to smoothly interpolate between the theoretical constraints the hard, jet
and soft scale must obey in different regions of the thrust distribution:
\begin{align}
\label{eq:profile123}
&  \text{1) peak:} & &
\muh\sim Q \,,\ \ 
\muj\sim \sqrt{\Lambda_{\rm QCD}\,Q}\,,\  
& \mus &\gtrsim \Lambda_{\rm QCD}\,, 
\nn\\
&  \text{2) tail:} & & 
\muh\sim Q \,,\ \
\muj\sim Q\sqrt{\tau} \,,\
&\mus &\sim Q\,\tau \,,
\nn\\
& \text{3) far-tail:} & & 
\muh=\muj = \mus \sim Q \,.
\end{align}
The factorization theorem derived for thrust in Ref.~\cite{Abbate:2010xh} is
formally invariant under ${\cal O}(1)$ changes of the profile function scales.
The residual dependence on the choice of profile functions constitutes one part
of the theoretical uncertainties and provides a method to estimate higher order
perturbative corrections. We adopt a set of six parameters that can be varied in
our theory error analysis which encode this residual freedom while still
satisfying the constraints in Eq.~(\ref{eq:profile123}).

For the profile function at the hard scale, we adopt
\begin{align}
\muh=&\,e_H\,Q,
\end{align}
where $e_H$ is a free parameter which we vary from $1/2$ to $2$ in our theory
error analysis.

For the soft profile function we use the form
\begin{align}
\mus(\tau)=\left\{\begin{array}{ll}\mu_0+\frac{b}{2 t_1} \tau^2,\hskip1.6cm
& 0\leq \tau\leq t_1,\\
b \, \tau+d,   & t_1\leq \tau\leq  t_2,\\
\muh-\frac{b}{1-2 t_2}(\frac{1}{2}-\tau)^2, &
t_2\leq \tau\leq \frac{1}{2}.\end{array} \right.  
\end{align}
Here, $t_1$ and $t_2$ represent the borders between the peak, tail and far-tail
regions. $\mu_0$ is the value of $\mu_S$ at \mbox{$\tau=0$}. Since the thrust
value where the peak region ends and the tail region begins is $Q$ dependent,
$t_1\simeq 1/Q$, we define the \mbox{Q-independent} parameter $n_1$ by
$t_1=n_1/(Q/1\,\mbox{GeV})$. To ensure that $\mus(\tau)$ is a smooth function,
the quadratic and linear forms are joined by demanding continuity of the
function and its first derivative at $\tau=t_1$ and $\tau=t_2$, which fixes
$b=2\,\big(\muh-\mu_0\big)/\big(t_2-t_1+\frac{1}{2}\big)$ and
$d=\big[\mu_0(t_2+\frac{1}{2})-\muh \,t_1\big]/\big(t_2-t_1+\frac{1}{2}\big)$.
In our theory error analysis we vary the free parameters $n_1$, $t_2$ and
$\mu_0$.

The profile function for the jet scale is determined by the natural relation
between the hard, jet, and soft scales
\begin{align}
\muj(\tau)=\bigg(1+
e_J\Big(\frac{1}{2}-\tau\Big)^2\bigg)\,\sqrt{\muh\,\mus(\tau)}\,.
\end{align}
The term involving the free ${\cal O}(1)$-parameter $e_J$ implements a
modification to this relation and vanishes in the multijet region where
$\tau=1/2$. We use a variation of $e_J$ to include the effect of such
modifications in our estimation of the theoretical uncertainties.

In our theory error analysis we vary $\mu_{\rm ns}$ to account for our ignorance
on the resummation of logarithms of $\tau$ in the nonsingular corrections.  We
consider three possibilities
\begin{align}
\mu_{\rm ns}(\tau)=\left\{\begin{array}{ll}
    \muh,\hskip3.08cm  & n_s=1,\\
    \muj(\tau), & n_s=0, \\
    \frac{1}{2}[\,\muj(\tau)+\mus(\tau)\,],   & n_s=-1.
\end{array}\right. 
\end{align}

The complete set of theoretical parameters and the their ranges of variation are summarized in Table~\ref{tab:theoryerr}.

Besides the parameters associated with the profile functions, the other theory parameters are $\Gamma_3^{\rm cusp}$,
$j_3$, $s_3$, and $\epsilon_{2,3}$. The cusp anomalous dimension at $\mathcal O(\alpha_s^4)$, $\Gamma_3^{\rm cusp}$ is
estimated via Pad\'e approximants and we assign a 200\% uncertainty to this approximation. $j_3$ and $s_3$ represent the nonlogarithmic 3-loop
term in the position-space hemisphere jet and soft functions, respectively. 
These two parameters and their variations  are estimated via Pad\'e approximations.
The last two parameters $\epsilon_2$ and $\epsilon_3$ allow us to include the statistical errors in the numerical determination of
the nonsingular distribution at two (from EVENT2~\cite{Catani:1996jh,Catani:1996vz}) and three
(from EERAD3~\cite{GehrmannDeRidder:2007bj}) loops, respectively.

At each order we randomly scan the parameter space summarized in
Table~\ref{tab:theoryerr} with a uniform measure, extracting 500 points.  Each
of the points in Fig.~\ref{fig:M1alpha} is the result of the fit performed with
a single choice of a point in the parameter space. The contour of the area in
the $\alpha_s$-$2\,\Omega_1$ plane covered by the fit results at each given
order is fitted to an ellipse, which is interpreted as a 1-$\sigma$ theoretical
uncertainty. The ellipse is determined as follows: in a first step we determine
the outermost points on the \mbox{$\alpha_s$-$2\,\Omega_1$} plane (defined by
the outermost convex polygon). We then perform a fit to these points using a
$\chi^2$ which is the square of the formula for an ellipse:
\begin{align} \label{eq:ellipse-fitter}
  \chi^2_{\rm ellipse} 
 & = \sum_i \big[a\,(\alpha_i-\alpha_0)^2+ 4\,b\, (\Omega_i - \Omega_0)^2 
  \\
  & +\,2\, c \,(\alpha_i-\alpha_0)(\Omega_i - \Omega_0)\,-\,1\big]^2\,.\nonumber
\end{align}
Here the sum is over the outermost points. The coordinates for the center of the
ellipse, $\alpha_0$ and $\Omega_0$, are fixed ahead of time to the average of the
maximum and minimum values of $\alpha_s(m_Z)$ and $\Omega_1$ in the scan. We
then minimize $\chi^2_{\rm ellipse}$ to determine the parameters $a,b,c$ of the
ellipse.

One could further express the coefficients $a$ and $b$ by 
\begin{align} \label{eq:ab}
 a & = \frac{1\,+\,\sqrt{1\,+\,4\,c^2 \,\Delta \alpha ^2 \,\Delta \Omega
     ^2}}{2\, \Delta \alpha ^2}\,, \\
 b & = \frac{1\,+\,\sqrt{1\,+4\,\,c^2 \,\Delta \alpha ^2 \,\Delta \Omega ^2}}{8\, \Delta \Omega ^2}\,, \nonumber
\end{align}
where $\Delta \alpha$ and $\Delta \Omega$ are just the half of the difference
of the maximum and minimum values of $\alpha_s(m_Z)$ and $\Omega_1$,
respectively, on the ellipse. Setting $\Delta \alpha$ and $\Delta \Omega$
to the corresponding values obtained from the fit points of the scan (i.e.\ the
perturbative errors) the coefficients $a$ and $b$ can be fixed and only $c$
remains as a free parameter. The
minimization of $\chi^2_{\rm 
  ellipse}$ in Eq.~(\ref{eq:ellipse-fitter}) gives almost identical results
regardless of whether or not Eqs.~(\ref{eq:ab}) are imposed.

In Fig.~\ref{fig:barplots} we vary a single parameter of
Table~\ref{tab:theoryerr} keeping all the others fixed at their respective
default values, and we plot the change of $\alpha_s(m_Z)$ and $\Omega_1$ as
compared to the values obtained from the first moment thrust fit with the
default setup.  In the figure, the dark shaded blue area represents a variation
where the parameter is larger than the default value, and the light shaded green
one where the parameter is smaller.  The largest uncertainty is associated with
the variation of the hard scale, $e_H$.  The value of $\alpha_s(m_Z)$ is
similarly affected by the uncertainty of the profile function parameters, the
statistical error from the numerical determination of the 3-loop nonsingular
distribution from EERAD3~\cite{GehrmannDeRidder:2007bj}, and by the parameter
$j_3$. It is rather insensitive to the variation of the 4-loop cusp anomalous
dimension and the statistical error from the determination of the 2-loop
nonsingular contribution to the thrust distribution.
The value of $\Omega_1$ is mainly sensitive to the profile function
parameters and $\epsilon_3$, but is quite insensitive to $j_3$.

\bibliography{thrust3}

\begin{thebibliography}{78}
\expandafter\ifx\csname natexlab\endcsname\relax\def\natexlab#1{#1}\fi
\expandafter\ifx\csname bibnamefont\endcsname\relax
  \def\bibnamefont#1{#1}\fi
\expandafter\ifx\csname bibfnamefont\endcsname\relax
  \def\bibfnamefont#1{#1}\fi
\expandafter\ifx\csname citenamefont\endcsname\relax
  \def\citenamefont#1{#1}\fi
\expandafter\ifx\csname url\endcsname\relax
  \def\url#1{\texttt{#1}}\fi
\expandafter\ifx\csname urlprefix\endcsname\relax\def\urlprefix{URL }\fi
\providecommand{\bibinfo}[2]{#2}
\providecommand{\eprint}[2][]{\url{#2}}

\bibitem[{\citenamefont{Kluth}(2006)}]{Kluth:2006bw}
\bibinfo{author}{\bibfnamefont{S.}~\bibnamefont{Kluth}},
  \bibinfo{journal}{Rept. Prog. Phys.} \textbf{\bibinfo{volume}{69}},
  \bibinfo{pages}{1771} (\bibinfo{year}{2006}), \eprint{hep-ex/0603011}.

\bibitem[{\citenamefont{{Gehrmann-De Ridder}
  et~al.}(2007{\natexlab{a}})\citenamefont{{Gehrmann-De Ridder}, Gehrmann,
  Glover, and Heinrich}}]{GehrmannDeRidder:2007bj}
\bibinfo{author}{\bibfnamefont{A.}~\bibnamefont{{Gehrmann-De Ridder}}},
  \bibinfo{author}{\bibfnamefont{T.}~\bibnamefont{Gehrmann}},
  \bibinfo{author}{\bibfnamefont{E.~W.~N.} \bibnamefont{Glover}},
  \bibnamefont{and} \bibinfo{author}{\bibfnamefont{G.}~\bibnamefont{Heinrich}},
  \bibinfo{journal}{Phys. Rev. Lett.} \textbf{\bibinfo{volume}{99}},
  \bibinfo{pages}{132002} (\bibinfo{year}{2007}{\natexlab{a}}),
  \eprint{0707.1285}.

\bibitem[{\citenamefont{{Gehrmann-De Ridder}
  et~al.}(2007{\natexlab{b}})\citenamefont{{Gehrmann-De Ridder}, Gehrmann,
  Glover, and Heinrich}}]{GehrmannDeRidder:2007hr}
\bibinfo{author}{\bibfnamefont{A.}~\bibnamefont{{Gehrmann-De Ridder}}},
  \bibinfo{author}{\bibfnamefont{T.}~\bibnamefont{Gehrmann}},
  \bibinfo{author}{\bibfnamefont{E.~W.~N.} \bibnamefont{Glover}},
  \bibnamefont{and} \bibinfo{author}{\bibfnamefont{G.}~\bibnamefont{Heinrich}},
  \bibinfo{journal}{JHEP} \textbf{\bibinfo{volume}{12}}, \bibinfo{pages}{094}
  (\bibinfo{year}{2007}{\natexlab{b}}), \eprint{0711.4711}.

\bibitem[{\citenamefont{Weinzierl}(2008)}]{Weinzierl:2008iv}
\bibinfo{author}{\bibfnamefont{S.}~\bibnamefont{Weinzierl}},
  \bibinfo{journal}{Phys. Rev. Lett.} \textbf{\bibinfo{volume}{101}},
  \bibinfo{pages}{162001} (\bibinfo{year}{2008}), \eprint{0807.3241}.

\bibitem[{\citenamefont{Weinzierl}(2009{\natexlab{a}})}]{Weinzierl:2009ms}
\bibinfo{author}{\bibfnamefont{S.}~\bibnamefont{Weinzierl}},
  \bibinfo{journal}{JHEP} \textbf{\bibinfo{volume}{06}}, \bibinfo{pages}{041}
  (\bibinfo{year}{2009}{\natexlab{a}}), \eprint{0904.1077}.

\bibitem[{\citenamefont{Becher and Schwartz}(2008)}]{Becher:2008cf}
\bibinfo{author}{\bibfnamefont{T.}~\bibnamefont{Becher}} \bibnamefont{and}
  \bibinfo{author}{\bibfnamefont{M.~D.} \bibnamefont{Schwartz}},
  \bibinfo{journal}{JHEP} \textbf{\bibinfo{volume}{07}}, \bibinfo{pages}{034}
  (\bibinfo{year}{2008}), \eprint{0803.0342}.

\bibitem[{\citenamefont{Chien and Schwartz}(2010)}]{Chien:2010kc}
\bibinfo{author}{\bibfnamefont{Y.-T.} \bibnamefont{Chien}} \bibnamefont{and}
  \bibinfo{author}{\bibfnamefont{M.~D.} \bibnamefont{Schwartz}},
  \bibinfo{journal}{JHEP} \textbf{\bibinfo{volume}{08}}, \bibinfo{pages}{058}
  (\bibinfo{year}{2010}), \eprint{1005.1644}.

\bibitem[{\citenamefont{Abbate et~al.}(2011)\citenamefont{Abbate, Fickinger,
  Hoang, Mateu, and Stewart}}]{Abbate:2010xh}
\bibinfo{author}{\bibfnamefont{R.}~\bibnamefont{Abbate}},
  \bibinfo{author}{\bibfnamefont{M.}~\bibnamefont{Fickinger}},
  \bibinfo{author}{\bibfnamefont{A.~H.} \bibnamefont{Hoang}},
  \bibinfo{author}{\bibfnamefont{V.}~\bibnamefont{Mateu}}, \bibnamefont{and}
  \bibinfo{author}{\bibfnamefont{I.~W.} \bibnamefont{Stewart}},
  \bibinfo{journal}{Phys. Rev.} \textbf{\bibinfo{volume}{D83}},
  \bibinfo{pages}{074021} (\bibinfo{year}{2011}), \eprint{1006.3080}.

\bibitem[{\citenamefont{Farhi}(1977)}]{Farhi:1977sg}
\bibinfo{author}{\bibfnamefont{E.}~\bibnamefont{Farhi}},
  \bibinfo{journal}{Phys. Rev. Lett.} \textbf{\bibinfo{volume}{39}},
  \bibinfo{pages}{1587} (\bibinfo{year}{1977}).

\bibitem[{\citenamefont{Bauer et~al.}(2001{\natexlab{a}})\citenamefont{Bauer,
  Fleming, and Luke}}]{Bauer:2000ew}
\bibinfo{author}{\bibfnamefont{C.~W.} \bibnamefont{Bauer}},
  \bibinfo{author}{\bibfnamefont{S.}~\bibnamefont{Fleming}}, \bibnamefont{and}
  \bibinfo{author}{\bibfnamefont{M.~E.} \bibnamefont{Luke}},
  \bibinfo{journal}{Phys. Rev. D} \textbf{\bibinfo{volume}{63}},
  \bibinfo{pages}{014006} (\bibinfo{year}{2001}{\natexlab{a}}),
  \eprint{hep-ph/0005275}.

\bibitem[{\citenamefont{Bauer et~al.}(2001{\natexlab{b}})\citenamefont{Bauer,
  Fleming, Pirjol, and Stewart}}]{Bauer:2000yr}
\bibinfo{author}{\bibfnamefont{C.~W.} \bibnamefont{Bauer}},
  \bibinfo{author}{\bibfnamefont{S.}~\bibnamefont{Fleming}},
  \bibinfo{author}{\bibfnamefont{D.}~\bibnamefont{Pirjol}}, \bibnamefont{and}
  \bibinfo{author}{\bibfnamefont{I.~W.} \bibnamefont{Stewart}},
  \bibinfo{journal}{Phys. Rev. D} \textbf{\bibinfo{volume}{63}},
  \bibinfo{pages}{114020} (\bibinfo{year}{2001}{\natexlab{b}}),
  \eprint{hep-ph/0011336}.

\bibitem[{\citenamefont{Bauer and Stewart}(2001)}]{Bauer:2001ct}
\bibinfo{author}{\bibfnamefont{C.~W.} \bibnamefont{Bauer}} \bibnamefont{and}
  \bibinfo{author}{\bibfnamefont{I.~W.} \bibnamefont{Stewart}},
  \bibinfo{journal}{Phys. Lett. B} \textbf{\bibinfo{volume}{516}},
  \bibinfo{pages}{134} (\bibinfo{year}{2001}), \eprint{hep-ph/0107001}.

\bibitem[{\citenamefont{Bauer et~al.}(2002{\natexlab{a}})\citenamefont{Bauer,
  Pirjol, and Stewart}}]{Bauer:2001yt}
\bibinfo{author}{\bibfnamefont{C.~W.} \bibnamefont{Bauer}},
  \bibinfo{author}{\bibfnamefont{D.}~\bibnamefont{Pirjol}}, \bibnamefont{and}
  \bibinfo{author}{\bibfnamefont{I.~W.} \bibnamefont{Stewart}},
  \bibinfo{journal}{Phys. Rev.} \textbf{\bibinfo{volume}{D65}},
  \bibinfo{pages}{054022} (\bibinfo{year}{2002}{\natexlab{a}}),
  \eprint{hep-ph/0109045}.

\bibitem[{\citenamefont{Bauer et~al.}(2002{\natexlab{b}})\citenamefont{Bauer,
  Fleming, Pirjol, Rothstein, and Stewart}}]{Bauer:2002nz}
\bibinfo{author}{\bibfnamefont{C.~W.} \bibnamefont{Bauer}},
  \bibinfo{author}{\bibfnamefont{S.}~\bibnamefont{Fleming}},
  \bibinfo{author}{\bibfnamefont{D.}~\bibnamefont{Pirjol}},
  \bibinfo{author}{\bibfnamefont{I.~Z.} \bibnamefont{Rothstein}},
  \bibnamefont{and} \bibinfo{author}{\bibfnamefont{I.~W.}
  \bibnamefont{Stewart}}, \bibinfo{journal}{Phys. Rev. D}
  \textbf{\bibinfo{volume}{66}}, \bibinfo{pages}{014017}
  (\bibinfo{year}{2002}{\natexlab{b}}), \eprint{hep-ph/0202088}.

\bibitem[{\citenamefont{Hoang and Stewart}(2008)}]{Hoang:2007vb}
\bibinfo{author}{\bibfnamefont{A.~H.} \bibnamefont{Hoang}} \bibnamefont{and}
  \bibinfo{author}{\bibfnamefont{I.~W.} \bibnamefont{Stewart}},
  \bibinfo{journal}{Phys. Lett.} \textbf{\bibinfo{volume}{B660}},
  \bibinfo{pages}{483} (\bibinfo{year}{2008}), \eprint{0709.3519}.

\bibitem[{\citenamefont{Hoang and Kluth}(2008)}]{Hoang:2008fs}
\bibinfo{author}{\bibfnamefont{A.~H.} \bibnamefont{Hoang}} \bibnamefont{and}
  \bibinfo{author}{\bibfnamefont{S.}~\bibnamefont{Kluth}}
  (\bibinfo{year}{2008}), \eprint{0806.3852}.

\bibitem[{\citenamefont{Hoang et~al.}(2008)\citenamefont{Hoang, Jain, Scimemi,
  and Stewart}}]{Hoang:2008yj}
\bibinfo{author}{\bibfnamefont{A.~H.} \bibnamefont{Hoang}},
  \bibinfo{author}{\bibfnamefont{A.}~\bibnamefont{Jain}},
  \bibinfo{author}{\bibfnamefont{I.}~\bibnamefont{Scimemi}}, \bibnamefont{and}
  \bibinfo{author}{\bibfnamefont{I.~W.} \bibnamefont{Stewart}},
  \bibinfo{journal}{Phys. Rev. Lett.} \textbf{\bibinfo{volume}{101}},
  \bibinfo{pages}{151602} (\bibinfo{year}{2008}), \eprint{0803.4214}.

\bibitem[{\citenamefont{Hoang et~al.}(2010)\citenamefont{Hoang, Jain, Scimemi,
  and Stewart}}]{Hoang:2009yr}
\bibinfo{author}{\bibfnamefont{A.~H.} \bibnamefont{Hoang}},
  \bibinfo{author}{\bibfnamefont{A.}~\bibnamefont{Jain}},
  \bibinfo{author}{\bibfnamefont{I.}~\bibnamefont{Scimemi}}, \bibnamefont{and}
  \bibinfo{author}{\bibfnamefont{I.~W.} \bibnamefont{Stewart}},
  \bibinfo{journal}{Phys.Rev.} \textbf{\bibinfo{volume}{D82}},
  \bibinfo{pages}{011501} (\bibinfo{year}{2010}), \eprint{0908.3189}.

\bibitem[{\citenamefont{Bethke}(2009)}]{Bethke:2009jm}
\bibinfo{author}{\bibfnamefont{S.}~\bibnamefont{Bethke}},
  \bibinfo{journal}{Eur. Phys. J.} \textbf{\bibinfo{volume}{C64}},
  \bibinfo{pages}{689} (\bibinfo{year}{2009}), \eprint{0908.1135}.

\bibitem[{\citenamefont{Bethke}(2012)}]{PDG:2012}
\bibinfo{author}{\bibfnamefont{S.}~\bibnamefont{Bethke}},
  \bibinfo{journal}{Nucl. Phys. B Proc. Supp.} \textbf{\bibinfo{volume}{(to
  appear)}} (\bibinfo{year}{2012}).

\bibitem[{\citenamefont{Bethke et~al.}(2011)\citenamefont{Bethke, Hoang, Kluth,
  Schieck, Stewart et~al.}}]{Bethke:2011tr}
\bibinfo{author}{\bibfnamefont{S.}~\bibnamefont{Bethke}},
  \bibinfo{author}{\bibfnamefont{A.~H.} \bibnamefont{Hoang}},
  \bibinfo{author}{\bibfnamefont{S.}~\bibnamefont{Kluth}},
  \bibinfo{author}{\bibfnamefont{J.}~\bibnamefont{Schieck}},
  \bibinfo{author}{\bibfnamefont{I.~W.} \bibnamefont{Stewart}},
  \bibnamefont{et~al.} (\bibinfo{year}{2011}), \bibinfo{note}{long author list
  - awaiting processing}, \eprint{1110.0016}.

\bibitem[{\citenamefont{{Movilla Fernandez} et~al.}(1998)\citenamefont{{Movilla
  Fernandez}, Biebel, Bethke, Kluth, and
  Pfeifenschneider}}]{MovillaFernandez:1997fr}
\bibinfo{author}{\bibfnamefont{P.~A.} \bibnamefont{{Movilla Fernandez}}},
  \bibinfo{author}{\bibfnamefont{O.}~\bibnamefont{Biebel}},
  \bibinfo{author}{\bibfnamefont{S.}~\bibnamefont{Bethke}},
  \bibinfo{author}{\bibfnamefont{S.}~\bibnamefont{Kluth}}, \bibnamefont{and}
  \bibinfo{author}{\bibfnamefont{P.}~\bibnamefont{Pfeifenschneider}}
  (\bibinfo{collaboration}{JADE}), \bibinfo{journal}{Eur. Phys. J.}
  \textbf{\bibinfo{volume}{C1}}, \bibinfo{pages}{461} (\bibinfo{year}{1998}),
  \eprint{hep-ex/9708034}.

\bibitem[{\citenamefont{Pahl et~al.}(2009{\natexlab{a}})\citenamefont{Pahl,
  Bethke, Kluth, Schieck, and collaboration}}]{Pahl:2008uc}
\bibinfo{author}{\bibfnamefont{C.}~\bibnamefont{Pahl}},
  \bibinfo{author}{\bibfnamefont{S.}~\bibnamefont{Bethke}},
  \bibinfo{author}{\bibfnamefont{S.}~\bibnamefont{Kluth}},
  \bibinfo{author}{\bibfnamefont{J.}~\bibnamefont{Schieck}}, \bibnamefont{and}
  \bibinfo{author}{\bibfnamefont{t.~J.} \bibnamefont{collaboration}},
  \bibinfo{journal}{Eur.Phys.J.} \textbf{\bibinfo{volume}{C60}},
  \bibinfo{pages}{181} (\bibinfo{year}{2009}{\natexlab{a}}),
  \eprint{0810.2933}.

\bibitem[{\citenamefont{Abbiendi et~al.}(2005)}]{Abbiendi:2004qz}
\bibinfo{author}{\bibfnamefont{G.}~\bibnamefont{Abbiendi}} \bibnamefont{et~al.}
  (\bibinfo{collaboration}{OPAL}), \bibinfo{journal}{Eur. Phys. J.}
  \textbf{\bibinfo{volume}{C40}}, \bibinfo{pages}{287} (\bibinfo{year}{2005}),
  \eprint{hep-ex/0503051}.

\bibitem[{\citenamefont{Ackerstaff et~al.}(1997)}]{Ackerstaff:1997kk}
\bibinfo{author}{\bibfnamefont{K.}~\bibnamefont{Ackerstaff}}
  \bibnamefont{et~al.} (\bibinfo{collaboration}{OPAL}), \bibinfo{journal}{Z.
  Phys.} \textbf{\bibinfo{volume}{C75}}, \bibinfo{pages}{193}
  (\bibinfo{year}{1997}).

\bibitem[{\citenamefont{Heister et~al.}(2004)}]{Heister:2003aj}
\bibinfo{author}{\bibfnamefont{A.}~\bibnamefont{Heister}} \bibnamefont{et~al.}
  (\bibinfo{collaboration}{ALEPH}), \bibinfo{journal}{Eur. Phys. J.}
  \textbf{\bibinfo{volume}{C35}}, \bibinfo{pages}{457} (\bibinfo{year}{2004}).

\bibitem[{\citenamefont{Abdallah et~al.}(2003)}]{Abdallah:2003xz}
\bibinfo{author}{\bibfnamefont{J.}~\bibnamefont{Abdallah}} \bibnamefont{et~al.}
  (\bibinfo{collaboration}{DELPHI}), \bibinfo{journal}{Eur. Phys. J.}
  \textbf{\bibinfo{volume}{C29}}, \bibinfo{pages}{285} (\bibinfo{year}{2003}),
  \eprint{hep-ex/0307048}.

\bibitem[{\citenamefont{Abdallah et~al.}(2004)}]{Abdallah:2004xe}
\bibinfo{author}{\bibfnamefont{J.}~\bibnamefont{Abdallah}} \bibnamefont{et~al.}
  (\bibinfo{collaboration}{DELPHI Collaboration}),
  \bibinfo{journal}{Eur.Phys.J.} \textbf{\bibinfo{volume}{C37}},
  \bibinfo{pages}{1} (\bibinfo{year}{2004}), \eprint{hep-ex/0406011}.

\bibitem[{\citenamefont{Abreu et~al.}(1999)}]{Abreu:1999rc}
\bibinfo{author}{\bibfnamefont{P.}~\bibnamefont{Abreu}} \bibnamefont{et~al.}
  (\bibinfo{collaboration}{DELPHI}), \bibinfo{journal}{Phys. Lett.}
  \textbf{\bibinfo{volume}{B456}}, \bibinfo{pages}{322} (\bibinfo{year}{1999}).

\bibitem[{\citenamefont{Acciarri et~al.}(2000)}]{Acciarri:2000hm}
\bibinfo{author}{\bibfnamefont{M.}~\bibnamefont{Acciarri}} \bibnamefont{et~al.}
  (\bibinfo{collaboration}{L3 Collaboration}), \bibinfo{journal}{Phys.Lett.}
  \textbf{\bibinfo{volume}{B489}}, \bibinfo{pages}{65} (\bibinfo{year}{2000}),
  \eprint{hep-ex/0005045}.

\bibitem[{\citenamefont{Achard et~al.}(2004)}]{Achard:2004sv}
\bibinfo{author}{\bibfnamefont{P.}~\bibnamefont{Achard}} \bibnamefont{et~al.}
  (\bibinfo{collaboration}{L3}), \bibinfo{journal}{Phys. Rept.}
  \textbf{\bibinfo{volume}{399}}, \bibinfo{pages}{71} (\bibinfo{year}{2004}),
  \eprint{hep-ex/0406049}.

\bibitem[{\citenamefont{Braunschweig et~al.}(1990)}]{Braunschweig:1990yd}
\bibinfo{author}{\bibfnamefont{W.}~\bibnamefont{Braunschweig}}
  \bibnamefont{et~al.} (\bibinfo{collaboration}{TASSO}), \bibinfo{journal}{Z.
  Phys.} \textbf{\bibinfo{volume}{C47}}, \bibinfo{pages}{187}
  (\bibinfo{year}{1990}).

\bibitem[{\citenamefont{Li et~al.}(1990)}]{Li:1989sn}
\bibinfo{author}{\bibfnamefont{Y.~K.} \bibnamefont{Li}} \bibnamefont{et~al.}
  (\bibinfo{collaboration}{AMY}), \bibinfo{journal}{Phys. Rev.}
  \textbf{\bibinfo{volume}{D41}}, \bibinfo{pages}{2675} (\bibinfo{year}{1990}).

\bibitem[{\citenamefont{{Gehrmann-De Ridder}
  et~al.}(2009)\citenamefont{{Gehrmann-De Ridder}, Gehrmann, Glover, and
  Heinrich}}]{GehrmannDeRidder:2009dp}
\bibinfo{author}{\bibfnamefont{A.}~\bibnamefont{{Gehrmann-De Ridder}}},
  \bibinfo{author}{\bibfnamefont{T.}~\bibnamefont{Gehrmann}},
  \bibinfo{author}{\bibfnamefont{E.}~\bibnamefont{Glover}}, \bibnamefont{and}
  \bibinfo{author}{\bibfnamefont{G.}~\bibnamefont{Heinrich}},
  \bibinfo{journal}{JHEP} \textbf{\bibinfo{volume}{0905}}, \bibinfo{pages}{106}
  (\bibinfo{year}{2009}), \eprint{0903.4658}.

\bibitem[{\citenamefont{Weinzierl}(2009{\natexlab{b}})}]{Weinzierl:2009yz}
\bibinfo{author}{\bibfnamefont{S.}~\bibnamefont{Weinzierl}},
  \bibinfo{journal}{Phys. Rev.} \textbf{\bibinfo{volume}{D80}},
  \bibinfo{pages}{094018} (\bibinfo{year}{2009}{\natexlab{b}}),
  \eprint{0909.5056}.

\bibitem[{\citenamefont{Dokshitzer and Webber}(1995)}]{Dokshitzer:1995zt}
\bibinfo{author}{\bibfnamefont{Y.~L.} \bibnamefont{Dokshitzer}}
  \bibnamefont{and} \bibinfo{author}{\bibfnamefont{B.~R.}
  \bibnamefont{Webber}}, \bibinfo{journal}{Phys. Lett.}
  \textbf{\bibinfo{volume}{B352}}, \bibinfo{pages}{451} (\bibinfo{year}{1995}),
  \eprint{hep-ph/9504219}.

\bibitem[{\citenamefont{Akhoury and Zakharov}(1995)}]{Akhoury:1995sp}
\bibinfo{author}{\bibfnamefont{R.}~\bibnamefont{Akhoury}} \bibnamefont{and}
  \bibinfo{author}{\bibfnamefont{V.~I.} \bibnamefont{Zakharov}},
  \bibinfo{journal}{Phys. Lett.} \textbf{\bibinfo{volume}{B357}},
  \bibinfo{pages}{646} (\bibinfo{year}{1995}), \eprint{hep-ph/9504248}.

\bibitem[{\citenamefont{Akhoury and Zakharov}(1996)}]{Akhoury:1995fb}
\bibinfo{author}{\bibfnamefont{R.}~\bibnamefont{Akhoury}} \bibnamefont{and}
  \bibinfo{author}{\bibfnamefont{V.~I.} \bibnamefont{Zakharov}},
  \bibinfo{journal}{Nucl.Phys.} \textbf{\bibinfo{volume}{B465}},
  \bibinfo{pages}{295} (\bibinfo{year}{1996}), \eprint{hep-ph/9507253}.

\bibitem[{\citenamefont{Nason and Seymour}(1995)}]{Nason:1995hd}
\bibinfo{author}{\bibfnamefont{P.}~\bibnamefont{Nason}} \bibnamefont{and}
  \bibinfo{author}{\bibfnamefont{M.~H.} \bibnamefont{Seymour}},
  \bibinfo{journal}{Nucl. Phys.} \textbf{\bibinfo{volume}{B454}},
  \bibinfo{pages}{291} (\bibinfo{year}{1995}), \eprint{hep-ph/9506317}.

\bibitem[{\citenamefont{Korchemsky and Sterman}(1995)}]{Korchemsky:1994is}
\bibinfo{author}{\bibfnamefont{G.~P.} \bibnamefont{Korchemsky}}
  \bibnamefont{and} \bibinfo{author}{\bibfnamefont{G.}~\bibnamefont{Sterman}},
  \bibinfo{journal}{Nucl. Phys.} \textbf{\bibinfo{volume}{B437}},
  \bibinfo{pages}{415} (\bibinfo{year}{1995}), \eprint{hep-ph/9411211}.

\bibitem[{\citenamefont{Beneke}(1999)}]{Beneke:1998ui}
\bibinfo{author}{\bibfnamefont{M.}~\bibnamefont{Beneke}},
  \bibinfo{journal}{Phys. Rept.} \textbf{\bibinfo{volume}{317}},
  \bibinfo{pages}{1} (\bibinfo{year}{1999}), \eprint{hep-ph/9807443}.

\bibitem[{\citenamefont{Gardi}(2000)}]{Gardi:2000yh}
\bibinfo{author}{\bibfnamefont{E.}~\bibnamefont{Gardi}},
  \bibinfo{journal}{JHEP} \textbf{\bibinfo{volume}{0004}}, \bibinfo{pages}{030}
  (\bibinfo{year}{2000}), \eprint{hep-ph/0003179}.

\bibitem[{\citenamefont{Dokshitzer et~al.}(1996)\citenamefont{Dokshitzer,
  Marchesini, and Webber}}]{Dokshitzer:1995qm}
\bibinfo{author}{\bibfnamefont{Y.~L.} \bibnamefont{Dokshitzer}},
  \bibinfo{author}{\bibfnamefont{G.}~\bibnamefont{Marchesini}},
  \bibnamefont{and} \bibinfo{author}{\bibfnamefont{B.~R.}
  \bibnamefont{Webber}}, \bibinfo{journal}{Nucl. Phys.}
  \textbf{\bibinfo{volume}{B469}}, \bibinfo{pages}{93} (\bibinfo{year}{1996}),
  \eprint{hep-ph/9512336}.

\bibitem[{\citenamefont{Dokshitzer
  et~al.}(1998{\natexlab{a}})\citenamefont{Dokshitzer, Lucenti, Marchesini, and
  Salam}}]{Dokshitzer:1998pt}
\bibinfo{author}{\bibfnamefont{Y.~L.} \bibnamefont{Dokshitzer}},
  \bibinfo{author}{\bibfnamefont{A.}~\bibnamefont{Lucenti}},
  \bibinfo{author}{\bibfnamefont{G.}~\bibnamefont{Marchesini}},
  \bibnamefont{and} \bibinfo{author}{\bibfnamefont{G.}~\bibnamefont{Salam}},
  \bibinfo{journal}{JHEP} \textbf{\bibinfo{volume}{9805}}, \bibinfo{pages}{003}
  (\bibinfo{year}{1998}{\natexlab{a}}), \eprint{hep-ph/9802381}.

\bibitem[{\citenamefont{Gardi and Grunberg}(1999)}]{Gardi:1999dq}
\bibinfo{author}{\bibfnamefont{E.}~\bibnamefont{Gardi}} \bibnamefont{and}
  \bibinfo{author}{\bibfnamefont{G.}~\bibnamefont{Grunberg}},
  \bibinfo{journal}{JHEP} \textbf{\bibinfo{volume}{9911}}, \bibinfo{pages}{016}
  (\bibinfo{year}{1999}), \eprint{hep-ph/9908458}.

\bibitem[{\citenamefont{Biebel}(2001)}]{Biebel:2001dm}
\bibinfo{author}{\bibfnamefont{O.}~\bibnamefont{Biebel}},
  \bibinfo{journal}{Phys.Rept.} \textbf{\bibinfo{volume}{340}},
  \bibinfo{pages}{165} (\bibinfo{year}{2001}).

\bibitem[{\citenamefont{Pahl et~al.}(2009{\natexlab{b}})\citenamefont{Pahl,
  Bethke, Biebel, Kluth, and Schieck}}]{Pahl:2009aa}
\bibinfo{author}{\bibfnamefont{C.}~\bibnamefont{Pahl}},
  \bibinfo{author}{\bibfnamefont{S.}~\bibnamefont{Bethke}},
  \bibinfo{author}{\bibfnamefont{O.}~\bibnamefont{Biebel}},
  \bibinfo{author}{\bibfnamefont{S.}~\bibnamefont{Kluth}}, \bibnamefont{and}
  \bibinfo{author}{\bibfnamefont{J.}~\bibnamefont{Schieck}},
  \bibinfo{journal}{Eur.Phys.J.} \textbf{\bibinfo{volume}{C64}},
  \bibinfo{pages}{533} (\bibinfo{year}{2009}{\natexlab{b}}),
  \eprint{0904.0786}.

\bibitem[{\citenamefont{Gehrmann et~al.}(2010)\citenamefont{Gehrmann, Jaquier,
  and Luisoni}}]{Gehrmann:2009eh}
\bibinfo{author}{\bibfnamefont{T.}~\bibnamefont{Gehrmann}},
  \bibinfo{author}{\bibfnamefont{M.}~\bibnamefont{Jaquier}}, \bibnamefont{and}
  \bibinfo{author}{\bibfnamefont{G.}~\bibnamefont{Luisoni}},
  \bibinfo{journal}{Eur. Phys. J.} \textbf{\bibinfo{volume}{C67}},
  \bibinfo{pages}{57} (\bibinfo{year}{2010}), \eprint{0911.2422}.

\bibitem[{\citenamefont{Dokshitzer
  et~al.}(1998{\natexlab{b}})\citenamefont{Dokshitzer, Lucenti, Marchesini, and
  Salam}}]{Dokshitzer:1997iz}
\bibinfo{author}{\bibfnamefont{Y.~L.} \bibnamefont{Dokshitzer}},
  \bibinfo{author}{\bibfnamefont{A.}~\bibnamefont{Lucenti}},
  \bibinfo{author}{\bibfnamefont{G.}~\bibnamefont{Marchesini}},
  \bibnamefont{and} \bibinfo{author}{\bibfnamefont{G.}~\bibnamefont{Salam}},
  \bibinfo{journal}{Nucl.Phys.} \textbf{\bibinfo{volume}{B511}},
  \bibinfo{pages}{396} (\bibinfo{year}{1998}{\natexlab{b}}),
  \eprint{hep-ph/9707532}.

\bibitem[{\citenamefont{Dokshitzer and Webber}(1997)}]{Dokshitzer:1997ew}
\bibinfo{author}{\bibfnamefont{Y.~L.} \bibnamefont{Dokshitzer}}
  \bibnamefont{and} \bibinfo{author}{\bibfnamefont{B.}~\bibnamefont{Webber}},
  \bibinfo{journal}{Phys.Lett.} \textbf{\bibinfo{volume}{B404}},
  \bibinfo{pages}{321} (\bibinfo{year}{1997}), \eprint{hep-ph/9704298}.

\bibitem[{\citenamefont{Lee and Sterman}(2006)}]{Lee:2006fn}
\bibinfo{author}{\bibfnamefont{C.}~\bibnamefont{Lee}} \bibnamefont{and}
  \bibinfo{author}{\bibfnamefont{G.}~\bibnamefont{Sterman}}
  (\bibinfo{year}{2006}), \eprint{hep-ph/0603066}.

\bibitem[{\citenamefont{Lee and Sterman}(2007)}]{Lee:2006nr}
\bibinfo{author}{\bibfnamefont{C.}~\bibnamefont{Lee}} \bibnamefont{and}
  \bibinfo{author}{\bibfnamefont{G.}~\bibnamefont{Sterman}},
  \bibinfo{journal}{Phys. Rev.} \textbf{\bibinfo{volume}{D75}},
  \bibinfo{pages}{014022} (\bibinfo{year}{2007}), \eprint{hep-ph/0611061}.

\bibitem[{\citenamefont{Korchemsky and Sterman}(1999)}]{Korchemsky:1999kt}
\bibinfo{author}{\bibfnamefont{G.~P.} \bibnamefont{Korchemsky}}
  \bibnamefont{and} \bibinfo{author}{\bibfnamefont{G.}~\bibnamefont{Sterman}},
  \bibinfo{journal}{Nucl. Phys.} \textbf{\bibinfo{volume}{B555}},
  \bibinfo{pages}{335} (\bibinfo{year}{1999}), \eprint{hep-ph/9902341}.

\bibitem[{\citenamefont{Korchemsky and Tafat}(2000)}]{Korchemsky:2000kp}
\bibinfo{author}{\bibfnamefont{G.~P.} \bibnamefont{Korchemsky}}
  \bibnamefont{and} \bibinfo{author}{\bibfnamefont{S.}~\bibnamefont{Tafat}},
  \bibinfo{journal}{JHEP} \textbf{\bibinfo{volume}{10}}, \bibinfo{pages}{010}
  (\bibinfo{year}{2000}), \eprint{hep-ph/0007005}.

\bibitem[{\citenamefont{Ligeti et~al.}(2008)\citenamefont{Ligeti, Stewart, and
  Tackmann}}]{Ligeti:2008ac}
\bibinfo{author}{\bibfnamefont{Z.}~\bibnamefont{Ligeti}},
  \bibinfo{author}{\bibfnamefont{I.~W.} \bibnamefont{Stewart}},
  \bibnamefont{and} \bibinfo{author}{\bibfnamefont{F.~J.}
  \bibnamefont{Tackmann}}, \bibinfo{journal}{Phys. Rev.}
  \textbf{\bibinfo{volume}{D78}}, \bibinfo{pages}{114014}
  (\bibinfo{year}{2008}), \eprint{0807.1926}.

\bibitem[{\citenamefont{Lee and Stewart}(2005)}]{Lee:2004ja}
\bibinfo{author}{\bibfnamefont{K.~S.~M.} \bibnamefont{Lee}} \bibnamefont{and}
  \bibinfo{author}{\bibfnamefont{I.~W.} \bibnamefont{Stewart}},
  \bibinfo{journal}{Nucl. Phys.} \textbf{\bibinfo{volume}{B721}},
  \bibinfo{pages}{325} (\bibinfo{year}{2005}), \eprint{hep-ph/0409045}.

\bibitem[{\citenamefont{Bauer et~al.}(2003)\citenamefont{Bauer, Luke, and
  Mannel}}]{Bauer:2001mh}
\bibinfo{author}{\bibfnamefont{C.~W.} \bibnamefont{Bauer}},
  \bibinfo{author}{\bibfnamefont{M.~E.} \bibnamefont{Luke}}, \bibnamefont{and}
  \bibinfo{author}{\bibfnamefont{T.}~\bibnamefont{Mannel}},
  \bibinfo{journal}{Phys.Rev.} \textbf{\bibinfo{volume}{D68}},
  \bibinfo{pages}{094001} (\bibinfo{year}{2003}), \eprint{hep-ph/0102089}.

\bibitem[{\citenamefont{Bauer et~al.}(2002{\natexlab{c}})\citenamefont{Bauer,
  Luke, and Mannel}}]{Bauer:2002yu}
\bibinfo{author}{\bibfnamefont{C.~W.} \bibnamefont{Bauer}},
  \bibinfo{author}{\bibfnamefont{M.}~\bibnamefont{Luke}}, \bibnamefont{and}
  \bibinfo{author}{\bibfnamefont{T.}~\bibnamefont{Mannel}},
  \bibinfo{journal}{Phys.Lett.} \textbf{\bibinfo{volume}{B543}},
  \bibinfo{pages}{261} (\bibinfo{year}{2002}{\natexlab{c}}),
  \eprint{hep-ph/0205150}.

\bibitem[{\citenamefont{Leibovich et~al.}(2002)\citenamefont{Leibovich, Ligeti,
  and Wise}}]{Leibovich:2002ys}
\bibinfo{author}{\bibfnamefont{A.~K.} \bibnamefont{Leibovich}},
  \bibinfo{author}{\bibfnamefont{Z.}~\bibnamefont{Ligeti}}, \bibnamefont{and}
  \bibinfo{author}{\bibfnamefont{M.~B.} \bibnamefont{Wise}},
  \bibinfo{journal}{Phys.Lett.} \textbf{\bibinfo{volume}{B539}},
  \bibinfo{pages}{242} (\bibinfo{year}{2002}), \eprint{hep-ph/0205148}.

\bibitem[{\citenamefont{Bosch et~al.}(2004)\citenamefont{Bosch, Neubert, and
  Paz}}]{Bosch:2004cb}
\bibinfo{author}{\bibfnamefont{S.~W.} \bibnamefont{Bosch}},
  \bibinfo{author}{\bibfnamefont{M.}~\bibnamefont{Neubert}}, \bibnamefont{and}
  \bibinfo{author}{\bibfnamefont{G.}~\bibnamefont{Paz}},
  \bibinfo{journal}{JHEP} \textbf{\bibinfo{volume}{0411}}, \bibinfo{pages}{073}
  (\bibinfo{year}{2004}), \eprint{hep-ph/0409115}.

\bibitem[{\citenamefont{Beneke et~al.}(2005)\citenamefont{Beneke, Campanario,
  Mannel, and Pecjak}}]{Beneke:2004in}
\bibinfo{author}{\bibfnamefont{M.}~\bibnamefont{Beneke}},
  \bibinfo{author}{\bibfnamefont{F.}~\bibnamefont{Campanario}},
  \bibinfo{author}{\bibfnamefont{T.}~\bibnamefont{Mannel}}, \bibnamefont{and}
  \bibinfo{author}{\bibfnamefont{B.}~\bibnamefont{Pecjak}},
  \bibinfo{journal}{JHEP} \textbf{\bibinfo{volume}{0506}}, \bibinfo{pages}{071}
  (\bibinfo{year}{2005}), \eprint{hep-ph/0411395}.

\bibitem[{\citenamefont{Kelley et~al.}(2011)\citenamefont{Kelley, Schwartz,
  Schabinger, and Zhu}}]{Kelley:2011ng}
\bibinfo{author}{\bibfnamefont{R.}~\bibnamefont{Kelley}},
  \bibinfo{author}{\bibfnamefont{M.~D.} \bibnamefont{Schwartz}},
  \bibinfo{author}{\bibfnamefont{R.~M.} \bibnamefont{Schabinger}},
  \bibnamefont{and} \bibinfo{author}{\bibfnamefont{H.~X.} \bibnamefont{Zhu}},
  \bibinfo{journal}{Phys.Rev.} \textbf{\bibinfo{volume}{D84}},
  \bibinfo{pages}{045022} (\bibinfo{year}{2011}), \eprint{1105.3676}.

\bibitem[{\citenamefont{Hornig et~al.}(2011)\citenamefont{Hornig, Lee, Stewart,
  Walsh, and Zuberi}}]{Hornig:2011iu}
\bibinfo{author}{\bibfnamefont{A.}~\bibnamefont{Hornig}},
  \bibinfo{author}{\bibfnamefont{C.}~\bibnamefont{Lee}},
  \bibinfo{author}{\bibfnamefont{I.~W.} \bibnamefont{Stewart}},
  \bibinfo{author}{\bibfnamefont{J.~R.} \bibnamefont{Walsh}}, \bibnamefont{and}
  \bibinfo{author}{\bibfnamefont{S.}~\bibnamefont{Zuberi}},
  \bibinfo{journal}{JHEP} \textbf{\bibinfo{volume}{1108}}, \bibinfo{pages}{054}
  (\bibinfo{year}{2011}), \eprint{1105.4628}.

\bibitem[{\citenamefont{Monni et~al.}(2011)\citenamefont{Monni, Gehrmann, and
  Luisoni}}]{Monni:2011gb}
\bibinfo{author}{\bibfnamefont{P.~F.} \bibnamefont{Monni}},
  \bibinfo{author}{\bibfnamefont{T.}~\bibnamefont{Gehrmann}}, \bibnamefont{and}
  \bibinfo{author}{\bibfnamefont{G.}~\bibnamefont{Luisoni}},
  \bibinfo{journal}{JHEP} \textbf{\bibinfo{volume}{1108}}, \bibinfo{pages}{010}
  (\bibinfo{year}{2011}), \eprint{1105.4560}.

\bibitem[{\citenamefont{Pahl}(2007)}]{Pahl:thesis}
\bibinfo{author}{\bibfnamefont{C.}~\bibnamefont{Pahl}}, Ph.D. thesis,
  \bibinfo{school}{TU Munich} (\bibinfo{year}{2007}).

\bibitem[{\citenamefont{Clavelli}(1979)}]{Clavelli:1979md}
\bibinfo{author}{\bibfnamefont{L.}~\bibnamefont{Clavelli}},
  \bibinfo{journal}{Phys.Lett.} \textbf{\bibinfo{volume}{B85}},
  \bibinfo{pages}{111} (\bibinfo{year}{1979}).

\bibitem[{\citenamefont{Chandramohan and Clavelli}(1981)}]{Chandramohan:1980ry}
\bibinfo{author}{\bibfnamefont{T.}~\bibnamefont{Chandramohan}}
  \bibnamefont{and} \bibinfo{author}{\bibfnamefont{L.}~\bibnamefont{Clavelli}},
  \bibinfo{journal}{Nucl.Phys.} \textbf{\bibinfo{volume}{B184}},
  \bibinfo{pages}{365} (\bibinfo{year}{1981}).

\bibitem[{\citenamefont{Clavelli and Wyler}(1981)}]{Clavelli:1981yh}
\bibinfo{author}{\bibfnamefont{L.}~\bibnamefont{Clavelli}} \bibnamefont{and}
  \bibinfo{author}{\bibfnamefont{D.}~\bibnamefont{Wyler}},
  \bibinfo{journal}{Phys.Lett.} \textbf{\bibinfo{volume}{B103}},
  \bibinfo{pages}{383} (\bibinfo{year}{1981}).

\bibitem[{\citenamefont{Catani et~al.}(1991)\citenamefont{Catani, Turnock, and
  Webber}}]{Catani:1991bd}
\bibinfo{author}{\bibfnamefont{S.}~\bibnamefont{Catani}},
  \bibinfo{author}{\bibfnamefont{G.}~\bibnamefont{Turnock}}, \bibnamefont{and}
  \bibinfo{author}{\bibfnamefont{B.}~\bibnamefont{Webber}},
  \bibinfo{journal}{Phys.Lett.} \textbf{\bibinfo{volume}{B272}},
  \bibinfo{pages}{368} (\bibinfo{year}{1991}).

\bibitem[{\citenamefont{Berger et~al.}(2003)\citenamefont{Berger, K{\'u}cs, and
  Sterman}}]{Berger:2003iw}
\bibinfo{author}{\bibfnamefont{C.~F.} \bibnamefont{Berger}},
  \bibinfo{author}{\bibfnamefont{T.}~\bibnamefont{K{\'u}cs}}, \bibnamefont{and}
  \bibinfo{author}{\bibfnamefont{G.}~\bibnamefont{Sterman}},
  \bibinfo{journal}{Phys. Rev. D} \textbf{\bibinfo{volume}{68}},
  \bibinfo{pages}{014012} (\bibinfo{year}{2003}), \eprint{hep-ph/0303051}.

\bibitem[{\citenamefont{Hornig et~al.}(2009)\citenamefont{Hornig, Lee, and
  Ovanesyan}}]{Hornig:2009vb}
\bibinfo{author}{\bibfnamefont{A.}~\bibnamefont{Hornig}},
  \bibinfo{author}{\bibfnamefont{C.}~\bibnamefont{Lee}}, \bibnamefont{and}
  \bibinfo{author}{\bibfnamefont{G.}~\bibnamefont{Ovanesyan}},
  \bibinfo{journal}{JHEP} \textbf{\bibinfo{volume}{05}}, \bibinfo{pages}{122}
  (\bibinfo{year}{2009}), \eprint{0901.3780}.

\bibitem[{\citenamefont{Catani et~al.}(1992)\citenamefont{Catani, Turnock, and
  Webber}}]{Catani:1992jc}
\bibinfo{author}{\bibfnamefont{S.}~\bibnamefont{Catani}},
  \bibinfo{author}{\bibfnamefont{G.}~\bibnamefont{Turnock}}, \bibnamefont{and}
  \bibinfo{author}{\bibfnamefont{B.}~\bibnamefont{Webber}},
  \bibinfo{journal}{Phys.Lett.} \textbf{\bibinfo{volume}{B295}},
  \bibinfo{pages}{269} (\bibinfo{year}{1992}).

\bibitem[{\citenamefont{Dokshitzer
  et~al.}(1998{\natexlab{c}})\citenamefont{Dokshitzer, Lucenti, Marchesini, and
  Salam}}]{Dokshitzer:1998kz}
\bibinfo{author}{\bibfnamefont{Y.~L.} \bibnamefont{Dokshitzer}},
  \bibinfo{author}{\bibfnamefont{A.}~\bibnamefont{Lucenti}},
  \bibinfo{author}{\bibfnamefont{G.}~\bibnamefont{Marchesini}},
  \bibnamefont{and} \bibinfo{author}{\bibfnamefont{G.}~\bibnamefont{Salam}},
  \bibinfo{journal}{JHEP} \textbf{\bibinfo{volume}{9801}}, \bibinfo{pages}{011}
  (\bibinfo{year}{1998}{\natexlab{c}}), \eprint{hep-ph/9801324}.

\bibitem[{\citenamefont{Chiu et~al.}(2012{\natexlab{a}})\citenamefont{Chiu,
  Jain, Neill, and Rothstein}}]{Chiu:2011qc}
\bibinfo{author}{\bibfnamefont{J.-y.} \bibnamefont{Chiu}},
  \bibinfo{author}{\bibfnamefont{A.}~\bibnamefont{Jain}},
  \bibinfo{author}{\bibfnamefont{D.}~\bibnamefont{Neill}}, \bibnamefont{and}
  \bibinfo{author}{\bibfnamefont{I.~Z.} \bibnamefont{Rothstein}},
  \bibinfo{journal}{Phys.Rev.Lett.} \textbf{\bibinfo{volume}{108}},
  \bibinfo{pages}{151601} (\bibinfo{year}{2012}{\natexlab{a}}),
  \eprint{1104.0881}.

\bibitem[{\citenamefont{Becher et~al.}(2011)\citenamefont{Becher, Bell, and
  Neubert}}]{Becher:2011pf}
\bibinfo{author}{\bibfnamefont{T.}~\bibnamefont{Becher}},
  \bibinfo{author}{\bibfnamefont{G.}~\bibnamefont{Bell}}, \bibnamefont{and}
  \bibinfo{author}{\bibfnamefont{M.}~\bibnamefont{Neubert}},
  \bibinfo{journal}{Phys.Lett.} \textbf{\bibinfo{volume}{B704}},
  \bibinfo{pages}{276} (\bibinfo{year}{2011}), \bibinfo{note}{15 pages, 4
  figures}, \eprint{1104.4108}.

\bibitem[{\citenamefont{Chiu et~al.}(2012{\natexlab{b}})\citenamefont{Chiu,
  Jain, Neill, and Rothstein}}]{Chiu:2012ir}
\bibinfo{author}{\bibfnamefont{J.-Y.} \bibnamefont{Chiu}},
  \bibinfo{author}{\bibfnamefont{A.}~\bibnamefont{Jain}},
  \bibinfo{author}{\bibfnamefont{D.}~\bibnamefont{Neill}}, \bibnamefont{and}
  \bibinfo{author}{\bibfnamefont{I.~Z.} \bibnamefont{Rothstein}},
  \bibinfo{journal}{JHEP} \textbf{\bibinfo{volume}{1205}}, \bibinfo{pages}{084}
  (\bibinfo{year}{2012}{\natexlab{b}}), \eprint{1202.0814}.

\bibitem[{\citenamefont{Catani and Seymour}(1996)}]{Catani:1996jh}
\bibinfo{author}{\bibfnamefont{S.}~\bibnamefont{Catani}} \bibnamefont{and}
  \bibinfo{author}{\bibfnamefont{M.~H.} \bibnamefont{Seymour}},
  \bibinfo{journal}{Phys. Lett.} \textbf{\bibinfo{volume}{B378}},
  \bibinfo{pages}{287} (\bibinfo{year}{1996}), \eprint{hep-ph/9602277}.

\bibitem[{\citenamefont{Catani and Seymour}(1997)}]{Catani:1996vz}
\bibinfo{author}{\bibfnamefont{S.}~\bibnamefont{Catani}} \bibnamefont{and}
  \bibinfo{author}{\bibfnamefont{M.~H.} \bibnamefont{Seymour}},
  \bibinfo{journal}{Nucl. Phys.} \textbf{\bibinfo{volume}{B485}},
  \bibinfo{pages}{291} (\bibinfo{year}{1997}), \eprint{hep-ph/9605323}.

\end{thebibliography}
\end{document}